\documentclass[
 amsmath,amssymb,
 aps, twocolumn,
nofootinbib
]{revtex4-2}
\usepackage{float}
\usepackage{footnote}
\usepackage{placeins}
\usepackage{graphicx}
\usepackage{dcolumn}
\usepackage{bm}

\usepackage{orcidlink}
\usepackage[utf8]{inputenc}
\usepackage{layout}
\usepackage{amsmath}
\usepackage{amssymb}
\usepackage{amsbsy}
\usepackage{subcaption}
\usepackage{amsfonts}

\usepackage{fancyhdr}
\usepackage{mathrsfs}
\usepackage[english]{babel}
\usepackage{geometry}
\usepackage{float}
\usepackage{tcolorbox}
\usepackage{pythonhighlight}
\usepackage{graphicx}
\usepackage{dcolumn}
\usepackage{bm}
\usepackage{tikz}
\usetikzlibrary{decorations.markings}
\usepackage{hyperref}
\hypersetup{
colorlinks=false,
    urlcolor=blue,
}
\usepackage{physics}
\usepackage{simpler-wick}
\usepackage{tikz-feynman,contour}
\usepackage{units}

\usepackage{bbold}
\date{\today}
\usepackage{mathtools}

\newcommand{\inParen}[1]{\left( #1 \right)}
\newcommand{\inSqr}[1]{\left[ #1 \right]}

\newcommand{\mc}[1]{\mathcal{#1}}
\newcommand{\mf}[1]{\mathfrak{#1}}

\newcommand{\I}{\mathbb{1}}

\newcommand{\bp}{\bm{p}}
\newcommand{\bq}{\bm{q}}

\newcommand{\bpsi}{\bar{\psi}}

\usepackage{expl3}
\newcommand{\msc}[1]{\mathscr{#1}}
\title{Beyond Meanfield:\\Moment Closure}
\newcommand{\TR}{2}
\newcommand{\dR}{n}
\newcommand{\dL}{\delta \Lambda}

\definecolor{answer}{RGB}{100,200,100}
\date{\today}

\begin{document}
\title{Improved Approximations for Collective Neutrino Oscillations}

\author{Matthew Riccio \orcidlink{0000-0002-9659-2236}}
\email{mariccio@wisc.edu}
 \affiliation{Department of Physics, University of Wisconsin, Madison WI 53706, USA}

 \author{A.B. Balantekin \orcidlink{0000-0002-2999-0111}}
 \email{baha@physics.wisc.edu}
 \affiliation{Department of Physics, University of Wisconsin, Madison WI 53706, USA}
 
\preprint{APS/123-QED}
\begin{abstract}
A one- and two-body  $\mf{su}(n)$ Hamiltonian governing the dynamics of many systems, including collective neutrino oscillations, is investigated. We start by analyzing the algebraic structure($\mf{u}(n^N)$), formulate a product structure of the algebra, and utilize this to construct generic expressions for operator expectation values, R\'{e}nyi Entropy, and Wigner Functions. Performing BBGKY hierarchy truncation we develop a systematic methodology for going beyond the mean field with polynomial scaling on a classical computer. 
\end{abstract}
\maketitle
\section{Introduction}

Essentially the entire gravitational binding energy of a core-collapse supernova is released as neutrinos. Mergers of neutron stars also exhibit a similar behavior. The very high neutrino densities that result necessitate taking into account neutrino-neutrino interactions in neutrino transport on top of neutrino interactions with the background particles. These collective neutrino oscillations are inherently a many-body phenomenon. Neutrino-neutrino interactions are already present in the Standard Model although there could be beyond the Standard Model contributions. The impact of collective neutrino oscillations on various aspects of astrophysics is well reviewed in the literature \cite{Duan:2010bg,Balantekin:2018mpq,Tamborra:2020cul,Johns:2025mlm,Volpe:2023met}. Collective neutrino oscillations represent emergent many-body dynamics, related to a diverse set of problems in astrophysics, high-energy physics, condensed matter physics, and quantum information.  

Because of its many-body nature, collective neutrino oscillations are typically treated with approximate methods. The most commonly used  approximation is the mean-field approach where each neutrino separately interacts with a mean-field created by all the other neutrinos. For a simplified Hamiltonian where each neutrino only forward scatter, one can show that stationary phase approximation to the path integral for the many-body problem yields the mean-field limit \cite{Balantekin:2006tg}. However, going beyond the stationary-phase approximation becomes increasingly difficult.

In recent years, there has been increased interest in exploring the full quantum dynamics of the system beyond the mean-field limit using other methods. This has led to the development of many-body models to account for the full quantum dynamics of the system, either using classical computers \cite{Cervia:2019res,Cervia:2022pro,Lacroix:2022krq,Illa:2022zgu,Martin:2021bri,Martin:2023gbo,Carlson:2026mir,Lacroix:2024pbb,Neill:2024klc,Chernyshev:2024pqy,Bhaskar:2024myw,Brokemeier:2024lhq,Kiss:2025jgt,Laraib:2025uza,Laraib:2025ziz,Hite:2026fsj,Froustey:2026slw} or with quantum computers \cite{Yeter-Aydeniz:2021olz,Amitrano:2022yyn,Hall:2021rbv,Siwach:2023wzy,Balantekin:2023qvm,Turro:2024shh,Spagnoli:2025etu,Heimsoth}. However exact treatment of even simplified cases with classical  computers has proven to be computationally challenging, requiring resources that scale exponentially with system size. Quantum computers remain too technologically and algorithmically demanding at present to provide simulations with beyond a few neutrinos. 

As a result, it is necessary to develop a framework that overcomes these challenges, enabling the study of the quantum dynamics in large, interacting systems without requiring exponential computational resources. Classical approaches include the use tensor network matrix product states \cite{Cervia:2022pro,Roggero:2021asb}. However, this approach remains bond-dimension limited \cite{Cervia:2022pro}. Another approach, the phase-space approximation \cite{Lacroix:2024pbb, Mangin-Brinet:2025sau}, samples initial quantum fluctuations via the Husimi quasi-probability distribution and evolves the resulting ensemble using independent mean-field trajectories. This framework offers a highly scalable alternative enabling simulations of much larger numbers of neutrinos while still capturing the gross features of phase-space delocalization. However, because the individual trajectories are propagated via the mean field equations of motion one sacrifices some of the dynamical quantum interference that occurs during the system's evolution. 

These existing methodologies provide solutions that allow us to explore different regimes of the problem, but it is very beneficial to explore new approaches. Here we explore an approach utilizing systematic truncation of the Bogoliubov-Born-Green-Kirkwood-Yvon (BBGKY) hierarchy.  BBGKY hierarchy was introduced to the neutrino physics by Volpe and her collaborators \cite{Volpe:2013uxl,Vaananen:2013qja,Serreau:2014cfa,Volpe:2023met}. Multi neutrino correlations were also explored in \cite{Illa:2022zgu}.
Our method is based on systematic truncation of many-neutrino cumulants. 
As we detail in the body of the paper this approach requires only polynomial classical resources.  

In Sec.~\ref{Algebra} we outline the necessary mathematical background and aspects of quantum information, Sec.~\ref{Cumulant} we discuss collective neutrino oscillations specifically, and outline our proposed methodology of approximation.  In Sec.~\ref{Num} we discuss numerical complexities and methodologies used to solve the problem and in Sec.~\ref{Results} we outline and discuss the results of the simulations for the both the two and three-flavor cases. Note that in this paper, unless otherwise specified, the word entropy refers to the R\'enyi entropy, except for one case where R\'enyi and von Neumann entropies are compared.

\section{$n$-Level Algebra and its consequences \label{Algebra}}
\subsection{$n$ Level Systems and $\mf{u}(n^N)$}
{We consider a system with $n$-degrees of freedom. The state of the system is described by a vector $\ket{i}$, where set $i\in\{0, \cdots, n-1\}$ which lives within a $n$-dimensional Hilbert space $\msc{H}_{n}$, spanned by the orthonormal basis $\{\ket{0}\cdots\ket{n-1}\}$. For such a system the evolution operator must be an element of the unitary group $U(n)$. Consequently, 
the Hamiltonian is an element of $\mf{u}(n)$ algebra. We will neglect the $\mf{u}(1)$ phase and thus $H\in \mf{su}(n)$. 
It is spanned by the elements of the $\mf{su}(n)$ algebra defined by:
\begin{align}
    \Tr(\lambda_a \lambda_b) &= 2 \delta_{ab}, \label{trace}\\
    [\lambda_{a},\lambda_{b}] &= i f_{ab}{}^{c}\lambda_c, \label{commutator}\\
    \{\lambda_{a},\lambda_{b}\} &= \frac{4}{\dR}\delta_{ab}\I + d_{ab}{}^{c}\lambda_{c}, \label{anticommutator}
\end{align}
where $a,b \in \{1, ... n^2-1\}$, 
$d_{ab}{}^c = \frac{1}{2} 
\Tr(\{\lambda_a,\lambda_b\}\lambda^c)$ are the totally symmetric structure constants, $f_{ab}{}^c = \frac{-i}{2} \Tr([\lambda_a ,\lambda_b]\lambda^c)$ are the totally antisymmetric structure constants. 
Equations \ref{trace}, \ref{commutator}, \ref{anticommutator} imply 
\begin{equation}
\label{lambdaproduct}
 \lambda_{a}\lambda_{b} = \frac{2}{n}\delta_{ab}\I + (d_{ab}{}^{c}+ i f_{ab}{}^{c})\lambda_{c}, 
\end{equation}
We write a one-body Hamiltonian as 
\begin{equation}
    H = C^{a}\lambda_{a}
\end{equation}
where $C^a$ are real for the Hamiltonian to be Hermitian. 
We can write down the elements of the fundamental representation of $\mf{su}(n)$ algebra in the generalized Gell-Mann basis using the indexing schema outlined in ~\cite{GellRepr}:
\begin{align*}
\lambda_{i^2+2(j-i)} = \ketbra{i}{j} + \text{h.c.}\\
\lambda_{i^2+2(j-i)+1} = -i \ketbra{i}{j} + \text{h.c.}\\
\lambda_{i^2 - 1} = \sqrt{\frac{2}{(i-1)i}} \left( \sum_{j<i} \ket{j}\bra{j} - (i-1)\ket{i}\bra{i} \right)
\end{align*}
where $i,j \in \{0,.. n-1\}$. 

We consider a system where we have $N$-sites each with $n$ local degrees of freedom. 
Since, the overall Hilbert space is constructed from the direct product of the local Hilbert spaces, the unitary algebra for the larger system can also be constructed 
by taking the direct product of representations of the individual unitary algebras. To this end we define the matrices:
\begin{equation}
\label{Generallambda}
\Lambda_{Aa} = (\mathbb{I}^{\otimes <A}) \otimes \lambda_a \otimes (\mathbb{I}^{\otimes >A}), \quad \forall\, \lambda_a \in \mathfrak{su}(n).
\end{equation}
For example in this notation one has 
\begin{align}
\Lambda_{1a} = \lambda_a \otimes \I &\otimes \cdots \otimes \I  \label{oldnotation}\\
\Lambda_{2a} = \I \otimes \lambda_a &\otimes \I \otimes \cdots \otimes \I\\
\vdots & \nonumber\\
\Lambda_{Na} = \I \otimes \I & \otimes \cdots \otimes \lambda_{a}
\end{align}
and so on. Note that in these equations $a$ takes values from $1$ to $n^2-1$. 

We can see explicitly that $\Lambda_{A,a}$ form $N$ commuting $\mf{su}(n)$ subalgebras is given as:

\begin{align*}
\Tr(\Lambda_{Aa}\Lambda_{Bb}) &= 2 \delta_{AB} \delta_{ab},\\
[\Lambda_{Aa},\Lambda_{Bb}] &= i \delta_{AB} f_{ab}{}^{c} \Lambda_{Ac},\\
\{\Lambda_{Aa},\Lambda_{Bb}\} &= \delta_{AB} \left( \frac{4}{\dR} \delta_{ab} + d_{ab}{}^c \Lambda_{Ac} \right) \notag \\ &+ 2 \xi_{AB} \Lambda_{Aa} \Lambda_{Bb}
\end{align*}
where $\xi_{AB} = 1-\delta_{AB}$ to ensure that $A\neq B$. We can span the entire space of the $\mf{u}(n^N)$ algebra by taking the products of $\Lambda_{Aa}$ and including the identity operator. In other words the full $\mf{u}(n^N)$ algebra can be obtained by $\mf{u}(n^N)\cong \left\{ \prod\limits_{A\in \mc{P}(\mc{A})} \Lambda_{A a_{A}}\right\}=\{\I, \Lambda_{1a_1},...,\Lambda_{1a_1}\Lambda_{2a_2}...\}$ where $\mc{A}\equiv\{1,..., N\} $ and $\mc{P}(\mc{A})$ is the power set\protect\footnote{The power set is the set of all subsets of a set, including the empty set and the set itself, e.g.
$\mathcal{P}(\{1,2\}) =
\{\emptyset,\{1\},\{2\},\{1,2\}\}$.} of $\mc{A}$. { For example, for $\mf{u}(2^2)$ we write the generators as
\begin{align*}
\Lambda_{1 a} = \sigma_a \otimes 1 \\
\Lambda_{2 a} = 1 \otimes \sigma_a \\
\Lambda_{1 a} \Lambda_{2b} = \sigma_{a}\otimes \sigma_{b}  \\
1 \otimes 1
\end{align*}
Hamiltonian with one-body, two-body, ... interactions can then be written as 
\begin{equation}
    H = B^{Aa}\Lambda_{Aa} + J^{AaBb}\Lambda_{Aa}\Lambda_{Bb} + ...
    \label{BEq}
\end{equation}
To ensure that the Hamiltonian is Hermitian, the coefficients should be real.

In many quantum many-body systems, the Hamiltonian truncates at the quadratic level, keeping only one- and two-body interactions}. Examples include collective neutrino oscillations (which we consider here in the forward-scattering approximation), the Fermi-Hubbard model, ånuclear models, many-electron atomic and molecular Hamiltonians, and quantum spin models such as the Heisenberg, XXZ, XY, and SYK models, Error Correcting Hamiltonians, among many others\cite{Rosenhaus:2018dtp, Kitaev:2005hzj}. The ubiquity of this structure suggests that these seemingly disparate systems should not be viewed as isolated models, but rather as members of a common algebraic class. Motivated by this observation, we now turn to a general treatment of quadratic dynamics in $n$-level many-body systems and develop the framework that will underlie the novel results of this work.

\subsection{Product Structure of $\mf{u}(n^N)$}

We next introduce a convenient compact notation $\lambda_{\mu} = (\I, \lambda_a)\in \mf{u}(n)$. The basis set, $\{\lambda_{\mu}\}$, is now closed under both the commutation and anti-commutation and so it is also be closed under multiplication as $\lambda_{\mu}\lambda_{\nu}  = \frac{1}{2}[\lambda_{\mu},\lambda_{\nu} ] + \frac{1}{2}\{\lambda_{\mu},\lambda_{\nu} \}$. We can then write the generalization of Equation \ref{Generallambda} as 
\begin{equation}
    \lambda_{\mu}\lambda_{\nu} = g_{\mu\nu}{}^{\xi}\lambda_{\xi}
\end{equation}
As stated above since $\lambda_{\mu}\lambda_{\nu}  = \frac{1}{2}[\lambda_{\mu},\lambda_{\nu} ] + \frac{1}{2}\{\lambda_{\mu},\lambda_{\nu} \}$, we can calculate $g$ explicitly in-terms of the Lie-Algebra's structure constants:
\begin{multline}
    g_{\mu\nu}{}^{\xi} = \beta^{\xi} Tr(\lambda_{\mu}\lambda_{\nu}\lambda^{\xi}) = \inParen{\delta_{\mu}{}^{ 0}\delta_{\nu}{}^{ 0} + \delta_{\mu}{}^{ a}\delta_{\nu}{}^{b}\frac{2}{n} \delta_{ab}} \\ \times\delta_{0}{}^{ \xi}+  \inParen{2\delta_{(\mu}{}^{0}\delta_{\nu)}{}^{c} + \delta_{\mu}{}^{a}\delta_{\nu}{}^{b} \frac{1}{2}(d_{ab}{}^{c}+ if_{ab}{}^{c}) }\delta_{c}{}^{\xi}
    \label{g-form}
\end{multline}
where $[\beta^{\xi}]^{-1} = \Tr(\lambda^{\xi}\lambda^{\xi})\Rightarrow \beta^{\xi} = \delta_{0}{}^{\xi}/\dR+\delta_{a}{}^{\xi}/2$. From Equation  \ref{g-form} one sees that $g_{\mu 0}{}^{\xi} = \delta_{\mu}{}{}^{\xi}$ and $g_{0 \mu}{}^{\xi} = \delta_{\mu}{}{}^{\xi}$.

We can now generalize Equation \ref{Generallambda}
by defining the tensor product $\Lambda_{A \mu}$ where $\mu$ is either $0$ or $a=1, \cdots N^2-1$. 
In this notation one can write 
\begin{equation}
    \Lambda_{10} = \Lambda_{20} = \cdots = \Lambda_{N0} = \I \otimes \I \otimes \cdots \otimes \I 
\end{equation}
where  $N$ identity matrices are tensor multiplied. 
In this notation, for example Equation \ref{oldnotation} can be written as 
\begin{equation}
\Lambda_{1a} = \lambda_a \otimes \lambda_0 \otimes \cdots \otimes \lambda_0     
\end{equation}
It is clear that the most general element of $u(n^N)$ can be written as 
\begin{equation}
\label{mostgeneral}
\Lambda_{A \mu} = \lambda_{\mu_1} \otimes \lambda_{\mu_2} \otimes \cdots \otimes \lambda_{\mu_N} .
\end{equation}
The most general element of the algebra then ban be written as a product of appropriate $\Lambda_{A \mu}'$s.

Indeed, these elements of the $u(n^N)$ algebra satisfy the following commutation and anticommutation relations:
\begin{multline}
        \inSqr{\prod_{A\in \mc{A}}\Lambda_{A\mu_A}, \prod_{A\in\mc{A}}\Lambda_{A\nu_A}}_{\pm}   \\= \prod_{A\in\mc{A}}g_{\mu_A\nu_A}{}^{\xi_A}\Lambda_{A\xi_A} \pm \prod_{A\in\mc{A}}g^*_{\mu_A\nu_A}{}^{\xi_A}\Lambda_{A\xi_A}.
        \label{Commutator}
\end{multline}
These products represent the fundamental representation matrices for $u(n^N)$.}
\subsubsection{Operator Decomposition, Completeness, Operator Transformations, and Operator Action \label{Decomp}}
{ Since all possible products of $\Lambda_{A\mu_{A}}$'s are Hermitian matrices of the fundamental representation of the $u(n^N)$ algebra (for an explicit illustration see Ref. \cite{Balantekin:2024pwc}) we can decompose any arbitrary $n^N \times n^N$ matrix in terms of them: 
\begin{multline*}
\mc{O} = \frac{1}{\dR^{N}}\Tr(\mc{O}) + \frac{1}{\dR^{N-1}\TR}\Tr(\mc{O}\Lambda^{Aa})\Lambda_{Aa} + ... \\= \underbrace{\Tr(\mc{O}\beta^{\mu_{1}}\Lambda^{1 \mu_1}...\beta^{\mu_{N}}\Lambda^{N \mu_N})}_{\tilde{o}^{\mu_1...\mu_N}}\Lambda_{1\mu_1}...\Lambda_{N\mu_N}
\end{multline*}
It is important to note Hermitian matrices will have $\tilde{o}^{\mu_{1}...\mu_{N}}\in \mathbb{R}$, where as a generic matrix must be $\tilde{o}^{\mu_{1}...\mu_{N}}\in \mathbb{C}$. In particular the density matrix can be written as 
\begin{equation}
\label{densitymatrix}
\rho = \frac{1}{n^N}\Lambda_{10} + \frac{\langle \Lambda^{Aa} \rangle}{2 n^{N-1}} \Lambda_{Aa} + \frac{\langle \Lambda^{Aa} \Lambda^{Bb} \rangle  }{2^2n^{N-2}}\Lambda_{Aa} \Lambda_{Bb} + \cdots
\end{equation}
where repeated indices are summed over and we defined 
\begin{equation}
    \langle {\cal O} \rangle = \Tr (\rho {\cal O} ). 
\end{equation}

}

\subsection{Quantum Information using product structures}
{
\subsubsection{R\'{e}nyi Entropy \& R\'{e}nyi Mutual Information\label{sec:RenyiInfo}}

In many-body physics, the entanglement structure is often the primary lens through which we examine the behavior of large $N$-systems. To examine the entanglement structure, we introduce the R\'{e}nyi Entropy:
\begin{equation}
    S_{\alpha,\{\mc{A}\}} = \frac{1}{1-\alpha}\ln(\Tr(\rho_{\{\mc{A}\}}^{\alpha}))
\end{equation}
For $\alpha=1$ R\'{e}nyi Entropy reduces to the von Neumann entropy. For $\alpha =2$, Using Equation \ref{densitymatrix}, we get
\begin{multline}
S_{2,\{\mc{A}\}} = 
-\ln\left(\right. \\ \left.\frac{1}{\dR^N}  +\frac{\expval{\Lambda_{Aa}}\expval{\Lambda^{Aa}}}{\dR^{N-1}\TR} + \frac{\expval{\Lambda_{Aa}\Lambda_{Bb}}\expval{\Lambda^{Aa}\Lambda^{Bb}}}{\dR^{N-2}\TR^{2}}\right. \\ \left. 
+\frac{\expval{\Lambda_{Aa}\Lambda_{Bb} \Lambda_{Cc}}\expval{\Lambda^{Aa}\Lambda^{Bb} \Lambda^{Cc}}}{2^3 n^{N-3}} + \cdots \right)
\label{2ENT}
\end{multline}  
Thus, we have arrived at a generic representation of the N-body R\'{e}nyi entropy in terms of the expectation values of $N$-body operators ($2N$-point functions). Building on this, we can define a standard estimator for the two-body mutual information\cite{Alhambra:2022gpw, Scalet:2021jaq}:
\begin{equation}
\mathcal{I}^{\text{Naive}}_{\alpha}(A:B) = S_{\alpha, A} + S_{\alpha, B} - S_{\alpha, AB}
\label{MutInf}
\end{equation}

It is important to emphasize, however, that $S_{\alpha}$ satisfies the sub-additivity condition only for $\alpha=1$. Consequently, this expression does not coincide with the formal information-theoretic R\'{e}nyi mutual information, which is defined via the Petz-R\'{e}nyi relative entropy~\cite{Alhambra:2022gpw}:
\begin{equation}
D_{\alpha}(\rho =\rho^{AB}|| \sigma=\rho^{A}\otimes\rho^{B}) = \frac{1}{\alpha-1} \ln \text{Tr} \left( \rho^{\alpha} \sigma^{1-\alpha} \right)
\end{equation}

While $\mc{I}^{\text{Naive}}_{\alpha} $ requires careful interpretation for $\alpha \neq 1$ as it is not guaranteed to be positive-definite, it remains computationally accessible. Hence it is widely used as a probe which effectively captures the universal scaling and entanglement structure of the many-body systems ~\cite{Scalet:2021jaq}. Furthermore, in the context of many-body systems an often ignored but important quantity is the three-body (or tripartite) mutual information \cite{Seshadri:2018yya}:
\begin{multline}
    \mc{I}^{\text{Naive}}_{\alpha(A:B:C)} = \mc{I}_{\alpha (A:B)}^{\text{Naive}}+\mc{I}_{\alpha(A:C)}^{\text{Naive}} - \mc{I}_{(A:BC)}^{\text{Naive}} = \\ S_{\alpha(A)}+S_{\alpha(B)}+S_{\alpha(C)} - S_{\alpha(AB)} - S_{\alpha(BC)} \\ - S_{\alpha (AC)} + S_{\alpha(ABC)}
    \label{MutInf3}
\end{multline}
This quantity helps characterize weather or not we observe information scrambling as it being negative implies that the joint $BC$ system contains more information about $A$ then $B$ and $C$ separately, indicating non-trivial information is contained in the many-body operators in Eq.~\eqref{densitymatrix} \cite{Seshadri:2018yya}.

\subsubsection{Discrete Phase Space and Mana \label{sec:Mana}}
Another fundamental characteristic of these many-body systems is their representation in phase space via the Wigner function. The Wigner function provides a quasi probability distribution that maps the many-body density matrix onto a discrete phase space, offering a direct visual and mathematical diagnostic of quantum non-classicality and non-stabilizerness~\cite{Chernyshev:2024pqy, Wigner}. To construct this framework, we define a generalized Heisenberg-Weyl(clock) algebra generated by the shift ($X$) and clock ($Z$) operators, which satisfy:

    \begin{align*}
    ZX = \omega XZ\\
    X^{\ell}\ket{i} = \ket{(i+\ell)\,\text{mod}\,\dR}\\
    Z^{\ell}\ket{i} = \omega^{\ell i}\ket{i}
\end{align*}
where $\omega = e^{\frac{2\pi i}{\dR}}$. (For another application of this algebra in neutrino physics see Ref. \cite{Heimsoth}).Thus, a particular realization of the generators can be given as:
    \begin{align*}
    X =\sum_{i=0 }^{\dR-2} \ket{i}\bra{(i+1)} + \ket{\dR - 1}\bra{0}\\
    Z = \sum_{i=0}^{\dR-1}\omega^{i}\ket{i}\bra{i}
\end{align*}
We can now use this clock algebra to define the discrete phase-point operator or Wigner kernel ~\cite{Wigner}.
\begin{equation}
    \Delta_{(q,p)} = \frac{1}{\dR}\sum_{k,\ell}\omega^{p\ell - k q}\omega^{-\frac{k\ell}{2}}Z^{\ell }X^{k}
\end{equation}
Since the operator above satisfies a completeness relation:
\begin{equation}
\Tr(\Delta^{\dagger}_{(q,p)}\Delta_{(\ell,k)}) = \dR\delta_{pk}\delta_{q \ell}
\end{equation}
We can decompose any one body operator $\mc{O}$ into the Wigner-Weyl basis by trace decomposition.
\begin{equation}
    \mc{O}= \sum_{q,p} \frac{\Tr(\mc{O} \Delta_{(q,p)}^{\dagger})}{\dR} \Delta_{(q,p)} =\sum_{q,p} \msc{W}_{\mc{O}(q,p)} \Delta_{(q,p)}
\end{equation}
where  $\msc{W}_{\mc{O}(q,p)}$ is the Weyl Symbol. Additionally, since we have already established the trace decomposition of any operator on the basis of $\mf{u}(n^N)$ we can define a coordinate transform in operator space as: 
\begin{equation}
   \mc{J}_{(\Delta,\lambda)}^{(q,p)}{}_{\mu}=\frac{1}{\dR}\Tr(\Delta^{\dagger(q,p)} \lambda_{\mu})
   \label{WignerJacobian}
\end{equation}
And the inverse transformation is given as:
\begin{equation}
    \mc{J}_{(\lambda,\Delta)}^{(q,p)}{}_{\mu}  = \beta_{\mu}  \Tr(\Delta^{(q,p)}\lambda_{\mu}) = n\beta_{\mu}\mc{J}_{(\Delta,\lambda)}^{* (q,p)}{}_{\mu}
\end{equation}
Thus, the Weyl Symbol of any one-body operator ${\cal O}$, can be found by simple coordinate transformation:
\begin{equation}
 \msc{W}_{\mc{O}, (q,p)} = \mc{J}_{(\Delta,\lambda)}^{(q,p)}{}_{\mu} \tilde{o}^{\mu}
\end{equation}
Now that we have established how one body operators transform we can generalize such a property to the $N$-body system. Due to the completeness relationship for the phase-point operator $\Delta$ we can tensor together these operators at each lattice site, and thus the Weyl Symbol for any $N$-body operator $\mc{O}$ is given as:
\begin{equation}
    \msc{W}_{\mc{O},(Q,P)} = \prod_{A}\mc{J}_{(\Delta,\lambda)}^{(q_{A},p_{A})}{}_{\mu_{A}}\tilde{o}^{\mu_{1}...\mu_{N}}
    \label{WeylSymbol}
\end{equation}
Where $(Q,P)$ are the set of all coordinate pairs  $(q_{A},p_{A})\,\forall\, A\in \{{1,...N}\}$. Weyl symbol of the density operator, known as the Wigner Function, is 
\begin{subequations}
\begin{align}
        \msc{W}_{\rho_A;(q,p)} = & \frac{\mc{J}_{(\Delta,\lambda)}^{(p,q)}{}_{0}}{\dR} + \frac{\mc{J}_{(\Delta,\lambda)}^{(p,q)}{}_{a}\expval{\Lambda_{A}{}^{a}}}{\TR}
        \label{WignerFunction}\\
         \msc{W}_{\rho_{\{\mc{A}\}};(Q,P)}=&\prod_{A}\mc{J}_{(\Delta,\lambda)}^{(q_A,p_A)}{}_{\mu_{A}}\tilde{\rho}^{\mu_1...\mu_N}
         \label{WignerFunction}
\end{align}
\end{subequations}
  As stated above the Wigner function is a quasi-distribution with $\sum_{Q,P}(\msc{W}_{\rho_{\{\mc{A}\}};(Q,P)}) = 1$, but is not positive definite~\cite{Wigner}. The more quantum a state the more negative its corresponding Wigner function~\cite{Wigner}. Thus, one particular measure of ``quantumness" is Mana which measures the total negativity of the Wigner function:
\begin{equation}
    \mc{M}_{\{\mc{A}\}} = \log(\sum_{(Q,P)} |\msc{W}_{\rho_{\{\mc{A}\}};(Q,P)}|)
    \label{Mana}
\end{equation}
 For states which are explicitly more quantum one would expect the Mana of a system to be non-zero.

 \subsubsection{Mana to Magic\label{sec:Magic}}
  The Gottesman-Knill theorem states that quantum systems that utilize only stabilizer resources, i.e. circuits that only consist of gates from the Pauli or Clifford groups, can be efficiently simulated in polynomial time on a probabilistic classical computer \cite{Gottesman:1998hu}. Thus, it is important to have a measure for the amount of non-stabilizer resources in a computation \cite{Chernyshev:2024pqy, Leone:2021rzd}. The generalized Pauli group$(\mf{P})$ is spanned by the clock algebra, $\mf{P}\equiv \{\omega^{k}Z^{q}X^{p}\,\forall\, k,p,q \in\{0, \dR-1\}\}$. Thus we can define an indexing scheme like that of the Wigner kernel phase-space with $(q,p)$ coordinate system thus $\sigma_{(q,p)}\in \{Z^{q}X^{p}|q,p\in\{0,\cdots \dR -1\}\}$ giving us the completeness relationship:
  \begin{equation}
      \Tr(\sigma_{(q,p)}^{\dagger}\sigma_{(\ell,k)})=\dR \delta_{q\ell}\delta_{pk}
  \end{equation}
Via the completeness relationship any operator can be decomposed into the Pauli basis via:
\begin{equation}
    \mc{O} = \sum_{Q,P}\msc{P}_{\mc{O}, (Q,P)}\prod_{q_{A},p_{A}}\sigma^{\dagger}_{q_{A},p_{A}}
\end{equation}
where $(Q,P) = (q_1,p_1...q_N,p_N)$ and we define a ``Pauli Symbol":
\begin{equation}
     \msc{P} _{\mc{O}, (Q,P)}=\frac{1}{n^N} \Tr(\mc{O}\prod_{A\in\{\mc{A}\}} \sigma^{\dagger}_{(q_A,p_A)})
\end{equation}
This leads to a Jacobian like object:
\begin{equation}
    \mc{J}_{(\sigma,\lambda)}^{(q,p)}{}_{\mu} = \frac{1}{\dR}\Tr(\sigma^{\dagger(q,p)}\lambda_{\mu}) .
\end{equation}
Note that we can also transform to the Pauli basis from the Wigner-Weyl basis via the Jacobian like object:
\begin{equation}
     \mc{J}_{(\sigma,\Delta)}^{(q',p')}{}_{(q,p)} = \frac{1}{\dR}\Tr(\sigma^{\dagger(q',p')}\Delta_{(q,p)}). 
\end{equation}
In the Gell-Mann basis the Pauli symbol becomes:
\begin{equation}
   \msc{P}_{\mc{O};(q,p)} = \prod_{A\in\mc{A}}\mc{J}_{(\sigma,\lambda)}^{(q_A,p_A)}{}_{\mu_A}\tilde{o}^{\mu_1\cdots \mu_{N}}
\end{equation}
It can be shown that
\begin{equation}
    \Tr(\rho_{\{\mc{A}\}}^2) =  \sum_{Q,P}n^N|\msc{P}_{\rho_{\{\mc{A}\}}, (Q,P)}|^2
\end{equation}
Then 
\[ 
\frac{n^N|\msc{P}_{\rho_{\{\mc{A}\}}, (Q,P)}|^2}{\Tr(\rho^2)}
\]
is a probability distribution\cite{Leone:2021rzd}. For stabilizer states, the value of $|\msc{P}_{\rho_{\{\mc{A}\}}, (Q,P)}|^2\in\{0,1\}$. Consequently, if we define a R\'{e}nyi Entropy for the distribution function, the stabilizer states will have zero entropy. Thus, a relevant quantity that characterizes the non-stabilizerness of the system is:
 \begin{multline}
     \msc{M}_{\alpha,\{\mc{A}\}} = \frac{1}{1-\alpha} \ln(\dR^{\alpha N}\sum_{Q,P} |\msc{P}_{\rho_{\{\mc{A}\}},(Q,P)}|^{2\alpha}) \\- \frac{\alpha}{\alpha-1} S_{2,\{\mc{A}\}}
     \label{Magic}
 \end{multline} 
 This quantity is known as Magic.

\section{Cumulants, BBGKY, \& Time Evolution with the Product Structure \label{Cumulant}}
\subsection{Hamiltonian} 
We consider the forward scattering limit in which the Hamiltonian resembles a generic quadratic $\mf{su}(n)$ Hamiltonian as seen in previous studies~\cite{Balantekin:2006tg,Balantekin:2018mpq,Balantekin:2024pwc}. 
\begin{equation}
H = B^{Aa}\Lambda_{Aa} + \frac{\mu_{(t)}}{2N}\Pi^{AB}\delta^{ab}\Lambda_{Aa}\Lambda_{Bb}
\label{Hamiltonian}
\end{equation}
As elaborated in Section~\ref{Algebra} any local $n$-level system with $1$ and $2$-body interactions that is unitary must be able to be reduced to a Quadratic   $\mf{su}(n^N)$ Model. Thus it is beneficial to consider the full $\mf{su}(n)$ type algebra. For the factor $\Pi^{AB}$, which accounts for the relative orientation of neutrino momenta, we take $\Pi^{AB} = \frac{1}{\TR}\inParen{1 - \cos\!\left(\theta_{\mathbf{p}_A \mathbf{p}_B}\right)}$. $\mu_{(t)} $ sets the overall interaction strength and is related to the neutrino density:
\begin{equation}
\mu_{(t)} = \frac{\sqrt{2}G_F N}{V}
\end{equation}
In the mass basis the one-body part of the Hamiltonian in the Gell-Mann realization is
\begin{equation}
B_{A,(k^2-1)} = \frac{1}{\TR} \sqrt{\frac{2}{k(k-1)}} \left( \sum_{j<k} E_{\mathbf{p}_A,j} - (k-1) E_{\mathbf{p}_A,k} \right)
\end{equation}
where $E_{\bp_A, k} = \sqrt{\bp_A^2 + m_k ^2}$ and recall we are using the convention $i,j,k,\cdots\in \{0,\cdots, n-1\}$ and $a,b,c,\cdots \in \{1, \cdots, n^2-1\}$. This gives 
\begin{equation}
B_{Aa} = \frac{1}{2} (E_{\mathbf{p}_A,0} - E_{\mathbf{p}_A,1}) \delta_3{}^a \approx \frac{\Delta_{01}}{4 |\mathbf{p}_A|} \delta_3{}^a
\end{equation}
for the two flavor case and 
\begin{align}
B_{Aa} &= \frac{\delta_3{}^a}{2}(E_{\mathbf{p}_A,0} - E_{\mathbf{p}_A,1}) + \frac{\delta_8{}^a}{2 \sqrt{3}} (E_{\mathbf{p}_A,0}+E_{\mathbf{p}_A,1} \nonumber \\& - 2 E_{\mathbf{p}_A,2}) \approx \frac{\Delta_{01}}{4|\mathbf{p}_A|} \delta_3{}^a + \frac{\Delta_{02} + \Delta_{12}}{4\sqrt{3} |\mathbf{p}_A|} \delta_8{}^a
\end{align}
in the three flavor case. In these equations $\Delta_{ij} = m_i^2 - m_j^2$. 

\subsection{Evolution and Hierarchy Truncation}
Instead of evolving the entire density matrix of the system 
we can evolve the coefficients ($n$-point operators) of the density matrix by solving the Heisenberg equation of motion. 
It is convenient to define: 
\begin{subequations}
\begin{align}
    \Phi_{Aa}& = \expval{\Lambda_{Aa}} \label{Phidef}\\
    \Gamma_{AaBb}&= \expval{\Lambda_{Aa}\Lambda_{Bb}} \label{Gammadef}\\
    \dL_{Aa} &= \Lambda_{Aa}-\Phi_{Aa} \label{dLdef}
 \end{align}
\end{subequations}
Note that if $A=B$ we must have $\Gamma_{AaAb} = g_{ab}{}^{\xi}\Phi_{A\xi}$ and thus we can restrict $\Gamma_{AaBb}$ to be defined only for $A\neq B$. Since $[\Lambda_{Aa},\Lambda_{Bb}]\propto \delta_{AB}$,  $\Gamma_{AaBb}$ must be symmetric under $(A,a)\leftrightarrow(Bb)$ interchange and thus $\dim(\Gamma ) \propto \begin{pmatrix}
     N\\ 2
 \end{pmatrix}$. $\dL$ is the central moment generating operator, and up to third order products of $\dL$ are the cumulants. Thus we can explicitly evaluate $1^{\text{st}},\;2^{\text{nd}},\; \& \;3^{\text{rd}}$-order cumulants finding:
\begin{subequations}
\begin{align}
\expval{\dL_{Aa}} &= \expval{\Lambda_{Aa}} - \Phi_{Aa} \equiv 0 \\
\expval{\dL_{Aa}\dL_{Bb}} &= \Gamma_{AaBb} - \Phi_{Aa}\Phi_{Bb} \\
\expval{\dL_{Aa}\dL_{Bb}\dL_{Cc}} & = \expval{\Lambda_{Aa}\Lambda_{Bb}\Lambda_{Cc}}\label{mom2}\\&
 - 3 \Phi_{(Aa} \Gamma_{BbCc)} + 2 \Phi_{Aa}\Phi_{Bb}\Phi_{Cc}
 \label{mom3}
\end{align}
\end{subequations}
where $A\neq B \neq C$   and $T_{(A_1a_1...A_Na_N)} = \frac{1}{N!}\sum\limits_{\sigma\in \mc{S}_N} T_{\sigma(1)... \sigma(N)}$ with $\mc{S}_{N}$ being the set of all permutations of  the set of indices $ \{(A_1,a_{1}),...(A_N,a_{N})\}$.

We find the time evolution of some $n$-body operator using Eq.~\eqref{Commutator} as
    \begin{multline}
     d\expval{\prod_A \Lambda_{A \nu_A}} = \delta_{\nu_{B}}{}^{a} f_{ab}{}^{c} \left( B^{Bb}\expval{\Lambda_{Bc}\prod_{A\neq B}\Lambda_{A\nu_A} } \right.  \\+ \left. \frac{\Pi^{BC}\mu_{(t)}}{N}\bar{g}^{b}{}_{\nu_C}{} ^{\xi_{C}}\expval{\Lambda_{Bc}\Lambda_{C\xi_C}\prod_{A\not\in\{B,C\}}\Lambda_{A\nu_{A}}}\right)dt .
     \label{NExpEvolve}
\end{multline}
Using Eq. (\ref{NExpEvolve}) we can find the evolution of the evolution of the one-body operator as:
\begin{equation}
        d\Phi_{Aa} = B_{Ab}f_{a}{}^{bc}\Phi_{Ac} dt+ \frac{\mu_{(t)}\Pi_{A}{}^{B}}{N}f_{a}{}^{bc}\Gamma_{BbAc}dt .
        \label{dPhi}
\end{equation}
Here the BBGKY hierarchy is evident as the time-evolution of the $N$-body operators is dependent on the $N+1$-body operators (for a complete description of the BBGKY hierarchy in neutrino physics see Ref. \cite{Volpe:2023met}). Solving Eq.~\eqref{mom2} for $\Gamma_{AaBb}$ we can rewrite Eq.~\eqref{dPhi} as:
\begin{multline}
      d\Phi_{Aa} = B_{Ab}f_{a}{}^{bc}\Phi_{Aa} dt+ \frac{\mu_{(t)}\Pi_{A}{}^{B}}{N}\\\times f_{a}{}^{bc}\inParen{\Phi_{Bb}\Phi_{Ac}+\expval{\dL_{Bb}\dL_{Ac}}}dt 
        \label{dPhi2}
\end{multline}
The simplest and often used prescription to solve this equation is to truncate the hierarchy at first order in the $N$-body operators (mean-field level), meaning that we simply enforce $\expval{\dL\dL} = 0 $  and thus $\Gamma_{AaBb}\to \Phi_{Aa}\Phi_{Bb}+\expval{\dL_{Aa}\dL_{Bb}}\approx \Phi_{Aa}\Phi_{Bb}$. However, from Eq.~\eqref{2ENT} we can find how the 1-body entropy changes as a result:
\begin{equation}
    dS_{2,A} = \frac{2\Phi_{A}{}^{a}}{\frac{1}{\dR}+\frac{1}{\TR}\Phi_{Aa}\Phi_{A}{}^{a}} \frac{f_{a}{}^{bc}\Pi_{A}{}^{B}\mu_{(t)}}{N} \Gamma_{BbAc}
\end{equation}
\\
Since $f_{a}{}^{bc}\Phi_{Bb}\Phi_{Ac} \equiv 0$ by symmetry implying $dS_{2A}\propto f_{a}{}^{bc}\expval{\dL_{Bb}\dL_{Ac}}\stackrel{\text{MFT}}{\longrightarrow} 0 $ , and thus at the mean field level we will see no one body entropy production. At the very least, we need a better estimator for the two-body operator evolution that preserves the antisymmetry induced by the cumulants evolution. 

Using Eq.~\eqref{NExpEvolve} again we can find the explicit evolution of the 2-body  operator $\Gamma_{AaBb}$. This evolution will depend on the value of the 3-body operator. If we wish to impose closure of higher moments for the 2-body evolution at second order, we need to solve Eq.~\eqref{mom3} for $\expval{\Lambda_{Aa}\Lambda_{Bb}\Lambda_{Cc}}$ to obtain 
\begin{widetext}
\begin{multline}
 d\Gamma_{AaBb} = \inParen{f_{a}{}^{cd}B_{Ac}\Gamma_{AdBb} +f_{b}{}^{cd}B_{Bc}\Gamma_{AaBd}}dt  + \frac{\mu_{(t)}}{N}\Pi_{AB}\inParen{\frac{\TR f_{ab}{}^{d}}{\dR}(\Phi_{Ad}-\Phi_{Bd})+f_{a}{}^{cd}\bar{g}_{cb}{}^{e}\Gamma_{AdBe} } dt\\+  \frac{\mu_{(t)}}{N}\Pi_{AB}f_{b}{}^{cd}\bar{g}_{ac}{}^{e}\Gamma_{AeBd} dt+ \frac{\mu_{(t)}}{N} \Pi_{A}{}^{C}\xi_{AB}\xi_{BC}f_{a}{}^{cd}\Big(3 \Phi_{(Ad}\Gamma_{CcBb)} - 2 \Phi_{Ad}\Phi_{Cc}\Phi_{Bb}  + \expval{\dL_{Ad}\dL_{Cc}\dL_{Bb}}\Big)dt\\+ \frac{\mu_{(t)}}{N}\Pi_{B}{}^{C}\xi_{AB}\xi_{AC}f_{b}{}^{cd}\Big(3\Phi_{(Aa}\Gamma_{CcBd)} - 2 \Phi_{Aa}\Phi_{Cc}\Phi_{Bd}  + \expval{\dL_{Aa}\dL_{Cc}\dL_{Bd}}\Big)dt .
 \label{LongAhhdG}
\end{multline}
\end{widetext} 
If we impose $\expval{\dL_{Aa}\dL_{Bb}\dL_{Cc}}\to 0$ we decouple $\Phi_{Aa} \,\&\,\Gamma_{AaBb}$ from the rest of the BBGKY Hierarchy. Note that this second order Truncation results in an inherent growth of the asymmetric term which controls how the entropy grows. Since we can write $\Gamma_{AaBb} = \Gamma_{A(a|B|b)} + \Gamma_{A[a|B|b]}\Rightarrow f_{c}{}^{ab}\Gamma_{AaBb} = f_{c}{}^{ab}\Gamma_{A[a|B|b]}$, the kinetic part of the evolution is canceled and the growth of entropy is entirely dependent on the anti-symmetry of the 2-body operators. To elucidate this procedure a little more we can write the Hamiltonian Eq.~\eqref{Hamiltonian} in the fermionic coherent state basis as seen in~\cite{Salasnich:2026qmr}. Where $\Lambda_{Aa} \to  \delta_{A}{}^{B}[\lambda_{a}]^{ij}a^{\dagger}_{Ai}a_{Bj} = \delta_{A}{}^{B}[\lambda_{a}]^{ij}\psi^{\dagger}_{Ai}\psi_{Bj}$ and the Hamiltonian becomes:
\begin{multline*}
	H = B^{Aa}[\lambda_{a}]^{ij}\psi^{\dagger}_{Ai}\psi_{Aj} +\frac{\mu_{(t)}\Pi^{AC}}{2N}\times\\ \delta^{AC}\delta^{BD}\delta^{j\ell}[\lambda_{a}]^{ij}\psi^{\dagger}_{Ai}\psi_{Cj}[\lambda^{a}]^{\ell k}\psi^{\dagger}_{B\ell}\psi_{Dk}
\end{multline*} 
The fermionic coherent state path integral is given by~\cite{Salasnich:2026qmr}:
\begin{multline*}
	Z[\eta^{\dagger},\eta] = \int \mc{D}[\psi^{\dagger}\psi]\\\times e^{i\int dt \sum_{A}i\psi^{\dagger}_{Ai}\delta^{ij}\partial_{t}\psi_{Ai}- H[\psi^{\dagger},\psi] + \eta^{\dagger}_{Ai}\psi^{Ai} + \psi^{\dagger}_{Ai}\eta^{Ai}}
\end{multline*}
In probability theory the moment generating functional for some variable X is given as $\expval{e^{Xt}} $ \cite{Kubo:1962dyl}, thus the moment generating functional for the quantum system given as the partition function $\frac{1}{Z[0,0]}Z[\eta^{\dagger},\eta]$. Additionally the cumulant generating functional in classical probability is given as $\ln(\expval{e^{Xt}})$ and thus the cumulant generating functional is the connected generating functional $W[\eta^{\dagger},\eta]= \ln(Z[\eta^{\dagger},\eta])$ ~\cite{Kubo:1962dyl}. Thus, the cumulant operators give exactly the fully connected $2N$-connected Green's functions.
\begin{multline*}
G^{(c)}_{p_1p'_1..p_N p'_N} \sim \kappa_{n}(\Lambda) \\\sim\prod_{A\in\mc{A}}\delta_{\eta^{Ai}}[\lambda_{a_A}]\delta_{\eta^{\dagger,Aj}} W[\eta^{\dagger},\eta]
\end{multline*}
Where $\kappa_{n}(\Lambda)$ is the $n^{\text{th}}$ order cumulant. Thus such a prescription admits a cluster expansion, where we consider only terms up to some order in the $2n$-point connected greens functions. At the mean-field level one models the evolution of the system by considering only the $2$-point connected Green's function: 
\begin{equation}[\lambda_{a}]^{bc}\expval{\bpsi_{Ab}\psi_{Ac}}=
\begin{tikzpicture}[baseline=(i.base), scale=.75]
    \begin{feynman}
        \vertex(a) at (-1,0){};
        \vertex(b) at (1,0){};
        \vertex (i) at (0,-.1);
        \diagram*{
            (a)--[fermion](b)
        };
    \end{feynman}
\end{tikzpicture}
\sim \expval{\Lambda_{Aa}}
\end{equation}
Note that at the mean-field level the 4-point connected Green's function is set to zero:
\begin{equation}G^{(c)}_{p_1...p_4} \sim \expval{\dL_{Aa}\dL_{Bb}}\sim\begin{tikzpicture}[baseline=(i.base), scale=0.5]
\begin{feynman}
    \vertex (a) at (-1,-1);
    \vertex (b) at (-1,1);
    \vertex (c) at (1,1);
    \vertex (d) at (1,-1);
    \vertex (i) at (0,0);
    \diagram*{
        (a) -- [fermion] (i) -- [fermion] (c),
        (b) -- [fermion] (i) -- [fermion] (d),
    };
\end{feynman}
\end{tikzpicture}
        \stackrel{\text{MFT}}{\longrightarrow}0 ,
        \end{equation}
whereas the full $4$-point Green's function is given as 
\begin{equation}
\expval{\Lambda_{Aa}\Lambda_{Bb}}\sim
\begin{tikzpicture}[baseline=(i.base), scale=.5]
    \begin{feynman}
     \vertex (a)  at (-1,-1){};
     \vertex(b) at (-1,1){};
     \vertex(c) at (1,1){};
     \vertex(d) at (1,-1){};
     \vertex(i) at (0,0);
     \diagram*{
         (a)--[fermion](i),
         (b)--[fermion](i),
         (i)--[fermion](c),
         (i)--[fermion](d),
     };
     \end{feynman}
\end{tikzpicture} + 
\begin{tikzpicture}[baseline=(i.base), scale=.5]
    \begin{feynman}
        \vertex (a)  at (-1,-.5){};
        \vertex(b) at (-1,.5){};
        \vertex(c) at (1,.5){};
        \vertex(d) at (1,-.5){};
        \vertex(i) at(0,0);
        \diagram*{
            (a)--[fermion](d),
            (b)--[fermion](c)
        };
    \end{feynman}
\end{tikzpicture}
+ \text{perms}.
\end{equation}
Thus, by truncating the BBGKY-hierarchy at the mean-field level, one neglects one of the fully connected diagrams involved in the evolution of the system. 
There are two fully connected $6$-point diagrams we must consider. In-fact, we can be more generic since each vertex gets $4$-lines, each internal line connects to $2$-vertices and each external line connects to $1$-vertex, the diagram has the topological constraint for the connected diagrams that $E = 4V -2I$, where $E$ is the number of external lines, V is the number of vertices and I is the number of internal lines. One also has $\mc{E} = 2V - I = V - L+1$ where $\mc{E} = E/2$ is the number of external operators and $L$ is the loop order\cite{Peskin:1995ev}. So for our connected $6$-point function we have $2$ vertices and one internal line. Thus, we have three unique diagrams with permutations:
\begin{multline}
\label{Conn3}
    \expval{\dL_{Aa}\dL_{Bb}\dL_{Cc}}\sim 
      \begin{tikzpicture}[baseline=(v1.base), scale=.75]
    \begin{feynman}[scale = 1]
        \vertex (i1)  at (-1,-1){};
        \vertex (i2)  at (-1,0){};
        \vertex (i3)  at (-1,1){};
        \vertex(v1) at (-.5,0);
         \vertex (v2) at (.5,0);
        \vertex (f3)  at (1,-1){};
        \vertex (f2)  at (1,0){};
        \vertex (f1)  at (1,1){};
        \diagram* {
            (i1)--[fermion] (v1),
            (i2)--[fermion] (v1),
            (i3)--[fermion] (v1),
            (v1)--[fermion] (v2),
            (v2)--[fermion] (f1),
            (v2)--[fermion] (f2),
            (v2)--[fermion] (f3)
        };
    \end{feynman} 
\end{tikzpicture}
+ \begin{tikzpicture}[baseline=(v1.base), scale=.75]
    \begin{feynman}
        \vertex (i1) at (-1,1) {};
        \vertex (i2) at (-1,0) {};
        \vertex (i3) at (-1,-1) {};
         \vertex (v1) at (-.35,0);
         \vertex (v2) at (+.35,0);        
        \vertex (f1) at (1,1){};
        \vertex (f2) at (1,0) {};
        \vertex (f3) at (1,-1){};
        \diagram* {
            (i1) -- [fermion] (v2);
            (i2) -- [fermion] (v1);
            (i3) -- [fermion] (v1);
            (v1) -- [fermion] (v2);
            (v2) -- [fermion] (f3);
            (v2) -- [fermion] (f2);
            (v1) -- [fermion] (f1);
        };
    \end{feynman}
\end{tikzpicture}\\
+
\begin{tikzpicture}[baseline=(v1.base), scale=.75]
    \begin{feynman}
        \vertex (i1) at (-1,1) {};
        \vertex (i2) at (-1,0) {};
        \vertex (i3) at (-1,-1) {};
         \vertex (v1) at (-.35,0);
         \vertex (v2) at (+.35,0);        
        \vertex (f1) at (1,1){};
        \vertex (f2) at (1,0) {};
        \vertex (f3) at (1,-1){};
        \diagram* {
            (i1) -- [fermion] (v2);
            (i2) -- [fermion] (v1);
            (i3) -- [fermion] (v2);
            (v1) -- [fermion] (v2);
            (v1) -- [fermion] (f3);
            (v2) -- [fermion] (f2);
            (v1) -- [fermion] (f1);
        };
    \end{feynman}
\end{tikzpicture}
+ \text{perms}
\end{multline}
Since $\Gamma_{AaBb} = \Phi_{Aa}\Phi_{Bb}+\expval{\dL_{Aa}\dL_{Bb}}$ the full $6$-point fermion function is given as 
\begin{flalign*}  &\expval{\Lambda_{Aa}\Lambda_{Bb}\Lambda_{Cc}}  \sim -2\left( \begin{tikzpicture}[baseline=(i2.base), scale=.65]
    \begin{feynman}[scale = 1]
        \vertex (i1)  at (-1,-.5){};
        \vertex (i2)  at (-1,0){};
        \vertex (i3)  at (-1,.5){};
        \vertex (f1)  at (1,-.5){};
        \vertex (f2)  at (1,0){};
        \vertex (f3)  at (1,.5){};
        \diagram* {
            (i1)--[fermion] (f1);
            (i2)--[fermion] (f2);
            (i3)--[fermion] (f3);
        };
    \end{feynman}
\end{tikzpicture} + \text{perms} \right) \\&+3 \left(
\begin{tikzpicture}[baseline=(i2.base), scale=.65]
    \begin{feynman}[scale = 1]
        \vertex (i1)  at (-1,-.5){};
        \vertex (i2)  at (-1,0){};
        \vertex (i3)  at (-1,1){};
        \vertex(v) at(0,.5);
        \vertex (f1)  at (1,-.5){};
        \vertex (f2)  at (1,0){};
        \vertex (f3)  at (1,1){};
        \diagram* {
            (i3)--[fermion] (v);
            (i2)--[fermion] (v);
            (v)--[fermion] (f3);
            (v)--[fermion] (f2);
            (i1)--[fermion] (f1);
        };
    \end{feynman}
\end{tikzpicture}+\text{perms }
\right)
+
\begin{tikzpicture}[baseline=(v1.base), scale=.65]
    \begin{feynman}[scale = 1]
        \vertex (i1)  at (-1,-1){};
        \vertex (i2)  at (-1,0){};
        \vertex (i3)  at (-1,1){};
        \vertex(v1) at (-.5,0);
         \vertex (v2) at (.5,0);
        \vertex (f3)  at (1,-1){};
        \vertex (f2)  at (1,0){};
        \vertex (f1)  at (1,1){};
        \diagram* {
            (i1)--[fermion] (v1),
            (i2)--[fermion] (v1),
            (i3)--[fermion] (v1),
            (v1)--[fermion] (v2),
            (v2)--[fermion] (f1),
            (v2)--[fermion] (f2),
            (v2)--[fermion] (f3)
        };
    \end{feynman}
\end{tikzpicture} \\ & +\begin{tikzpicture}[baseline=(v1.base), scale=.65]
    \begin{feynman}
        \vertex (i1) at (-1,1) {};
        \vertex (i2) at (-1,0) {};
        \vertex (i3) at (-1,-1) {};
         \vertex (v1) at (-.35,0);
         \vertex (v2) at (+.35,0);        
        \vertex (f1) at (1,1){};
        \vertex (f2) at (1,0) {};
        \vertex (f3) at (1,-1){};
        \diagram* {
            (i1) -- [fermion] (v2);
            (i2) -- [fermion] (v1);
            (i3) -- [fermion] (v1);
            (v1) -- [fermion] (v2);
            (v2) -- [fermion] (f3);
            (v2) -- [fermion] (f2);
            (v1) -- [fermion] (f1);
        };
    \end{feynman}
\end{tikzpicture}
+
\begin{tikzpicture}[baseline=(v1.base), scale=.65]
    \begin{feynman}
        \vertex (i1) at (-1,1) {};
        \vertex (i2) at (-1,0) {};
        \vertex (i3) at (-1,-1) {};
         \vertex (v1) at (-.35,0);
         \vertex (v2) at (+.35,0);        
        \vertex (f1) at (1,1){};
        \vertex (f2) at (1,0) {};
        \vertex (f3) at (1,-1){};
        \diagram* {
            (i1) -- [fermion] (v2);
            (i2) -- [fermion] (v1);
            (i3) -- [fermion] (v2);
            (v1) -- [fermion] (v2);
            (v1) -- [fermion] (f3);
            (v2) -- [fermion] (f2);
            (v1) -- [fermion] (f1);
        };
    \end{feynman}
\end{tikzpicture}
+ \text{perms}
\end{flalign*}
Enforcing third order truncation neglects the last term in the expansion of the $3$-point operator, but includes the contribution of the fully connected $2$-point operator.  From the action we can write down the bare-vertex is given as:
\begin{multline}
    \mc{V}_{(t)}^{ABCD,ij\ell k} = 
    \frac{\Pi^{AB}\delta^{ab}\mu_{(t)}}{2N}\delta^{\bp_{A}-\bp_{C}}\delta^{\bp_{B}-\bp_{D}}[\lambda_{a}]^{ij}[\lambda_{b}]^{\ell k}\\\delta^{\bp_{A}+\bp_{B}-\bp_{C}-\bp_{D}}
\end{multline}
The bare propagator can be found particularly easily in the mass basis by inverting the kinetic term giving us:
\begin{multline}
    G_{(\delta t)}^{AB;ij} = \int \frac{d\omega}{2\pi}\sum_{i}\frac{i\delta^{\bp_A-\bp_B}\ketbra{i}e^{i\omega \delta t}}{\omega -\mc{B}_{Ai} } \\=  \delta_{\bp_A-\bp_B}\delta_{ij}e^{ i\mc{B}_{Ai} \delta t}
\end{multline}
Where we define $\mc{B}_{Ai}=\sum_{j>i}\sqrt{\frac{2}{j(j-1)}} B_{A(j^2-1)} + \sqrt{\frac{2 (i-1)}{i}}B_{A(i^2-1)}$. From the form of the bare vertex it is clear that the first term goes to $0$ 
\begin{align}
\label{zerozero}
   \begin{tikzpicture}[baseline=(v1.base), scale=1]
    \begin{feynman}[scale = 1]
        \vertex (i1)  at (-1,-1){$A$};
        \vertex (i2)  at (-1,0){$B$};
        \vertex (i3)  at (-1,1){$C$};
        \vertex(v1) at (-.5,0);
        \vertex (v2) at (.5,0);
        \vertex (f3)  at (1,-1){$A'$};
        \vertex (f2)  at (1,0){$B'$};
        \vertex (f1)  at (1,1){$C'$};
        \diagram* {
            (i1)--[fermion] (v1);
            (i2)--[fermion] (v1);
            (i3)--[fermion] (v1);
            (v1)--[fermion, momentum=$p_D$] (v2);
            (v2)--[fermion] (f1);
            (v2)--[fermion] (f2);
            (v2)--[fermion] (f3);
        };
    \end{feynman}
\end{tikzpicture}   \propto \delta^{\bp_A-\bp_C}\delta^{\bp_B-\bp_D} &\delta^{}_{\bp_{A}+\bp_{C}+\bp_{B}-\bp_{D}}\notag\\ &\stackrel{{\scriptsize\begin{matrix}{}
    \text{Forward}\\
    \text{Scattering}
\end{matrix}}}{\longrightarrow} 0 .
\end{align}
Since this is due to the conservation of momentum, corrections to the diagram in Eq. (\ref{zerozero}) must be 0 to all orders. However, note that this diagram is zero in the forward scattering approximation. If one were to include contributions from the beyond forward scattering terms in the Hamiltonian this diagram would have non-zero contributions; this agrees with the recent results by Cirigliano et al. in ~\cite{Cirigliano:2024pnm} and Cervia in ~\cite{Cervia:2025pfg}. We only need to consider the other two connected diagrams depicted in Eq.~\eqref{Conn3} as all the other diagrams can be gotten by permuting the momentum state connections
\begin{multline}
\begin{tikzpicture}[baseline=(v1.base), scale=1]
    \begin{feynman}
        \vertex (i1) at (-1,1) {$Ai$};
        \vertex (i2) at (-1,0) {$Bj$};
        \vertex (i3) at (-1,-1) {$Ck$};
         \vertex (v1) at (-.25,0);
         \vertex (v2) at (+.25,0);        
        \vertex (f1) at (1,1){$A'i'$};
        \vertex (f2) at (1,0) {$B'j'$};
        \vertex (f3) at (1,-1){$C'k'$};
        \diagram* {
            (i1) -- [fermion] (v2);
            (i2) -- [fermion] (v1);
            (i3) -- [fermion] (v1);
            (v1) -- [fermion] (v2);
            (v2) -- [fermion] (f3);
            (v2) -- [fermion] (f2);
            (v1) -- [fermion] (f1);
        };
    \end{feynman}
\end{tikzpicture} \sim \frac{\Pi^{BD}\Pi^{AB'}}{4N^2} \delta^{\bp_B-\bp_{A'}}\delta^{\bp_{C}-\bp_{D}} \\ \times  [\lambda_{a}]^{i'j}[\lambda^{a}]^{mk} \delta_{mn}[\lambda_{a'}]^{j'n}[\lambda^{a'}]^{k'i}\delta_{\bp_{D}\bp_{E}}\\ \times \delta^{\bp_{E}-\bp_{B'}}\delta^{\bp_{A}-\bp_{C'}}\delta_{\bp_{B}+\bp_{C}-\bp_{A'}-\bp_{D}} \\\times \delta_{\bp_{A}+\bp_{D}-\bp_{C'}-\bp_{B'}} \int dt\int dt' \mu_{(t)}\mu_{(t')} e^{i\mc{B}_{Dm}(t-t')} ,
\end{multline}
and 
\begin{multline}
\begin{tikzpicture}[baseline=(v1.base), scale=1]
    \begin{feynman}
        \vertex (i1) at (-1,1) {$Ai$};
        \vertex (i2) at (-1,0) {$Bj$};
        \vertex (i3) at (-1,-1) {$Ck$};
         \vertex (v1) at (-.25,0);
         \vertex (v2) at (+.25,0);        
        \vertex (f1) at (1,1){$A'i'$};
        \vertex (f2) at (1,0) {$B'j'$};
        \vertex (f3) at (1,-1){$C'k'$};
        \diagram* {
            (i1) -- [fermion] (v2);
            (i2) -- [fermion] (v1);
            (i3) -- [fermion] (v2);
            (v1) -- [fermion] (v2);
            (v1) -- [fermion] (f3);
            (v2) -- [fermion] (f2);
            (v1) -- [fermion] (f1);
        };
    \end{feynman}
\end{tikzpicture} \sim \frac{\Pi^{BD}\Pi^{AB'}}{4N^2} \delta^{\bp_B-\bp_{A'}}\delta^{\bp_{C'}+\bp_{D}} \\ \times  [\lambda_{a}]^{i'j}[\lambda^{a}]^{mk} \delta_{mn}[\lambda_{a'}]^{j'n}[\lambda^{a'}]^{k'i}\delta_{\bp_{D}-\bp_{E}} \\ \times \delta^{\bp_{C}-\bp_{B'}}\delta^{\bp_{A}+\bp_{E}}\delta_{\bp_{B'}-\bp_{C}-\bp_{A}-\bp_{E}} \\\times \delta_{\bp_{B}-\bp_{D}-\bp_{C'}-\bp_{A'}} \int dt\int dt' \mu_{(t)}\mu_{(t')} e^{i\mc{B}_{Dm}(t-t')} .
\end{multline}
Note that these diagrams are $\mc{O} (\mu_{(t)}^2/N^2)$ and thus they are exceedingly small for 
$N\sim 10^{57}$. Additionally the number of disconnected diagrams will grow much faster than the fully-connected diagrams and thus dominate the evolution both due to being lower order in the coupling and their combinatorial growth.  
In this work we are going beyond improving the dynamical mean field approximation using a cluster expansion by enforcing Truncation from the rest of the hierarchy at the order of the $2$-body operators. Of course one can systematically improve this approximation by setting the $(N+1)^{\text{th}}$-order cumulant to zero, and thus obtaining an approximation for the $(N+1)$-body operators in terms of the $1$-body, $2$-body, $\cdots$, and $N$-body operators. One can then insert the approximation for $(N+1)$-body operator in Eq. \eqref{NExpEvolve}. This, procedurally, is analogous to moment closure in a stochastic process~\cite{StochMom2016}, however in the context of operator dynamics and quantum mechanics it is better illustrated by the term hierarchy truncation as the analogy with probability theory ends with the Heisenberg equation of motion. In our approximation, instead of traversing the entire hierarchy, we only have to solve the coupled differential equations for the $1$ and $2$-body operators, depicted in Eq. \eqref{dPhi} and Eq. \eqref{LongAhhdG}.  To do so we simply use an adaptive Runge-Kutta 4 (RK4) procedure for the coupled equations as outlined below in Sec.~\ref{Consist}. 

\section{Numerics and Computational Complexity{\label{Num}}}
\subsection{Conserved Quantities, Consistency, \& Adaptive stepping\label{Consist} }
From Eq. (\ref{densitymatrix}) one can obtain reduced density matrices. For example, the reduced density matrix obtained by taking the trace of all the other qudits, but the qudit corresponding to $A=1$ is
\begin{equation}
    \rho_1 = \frac{1}{n} + \frac{1}{2}\langle \Lambda_{1a} \rangle \Lambda_{1a} .
\end{equation}
One can write similar expressions by tracing over all the other qudits but $A=2$ or $A=3$, and so on.  Note that in the above expression sum over the index $a$ is implied.  Similarly the reduced density matrix obtained by taking trace over all the qudits but qudits $A=1$ and $A=2$ is 
\begin{multline}
    \rho_{12} = \frac{1}{n^2} (1 \otimes 1) + \frac{1}{2n} \left( \langle \Lambda^{1a} \rangle \Lambda_{1a} + \langle \Lambda_{2a} \rangle \Lambda_{2a} \right) \\
    +\frac{1}{4} \langle \Lambda^{1a} \Lambda^{2b} \rangle \Lambda_{1a} \Lambda_{2b} .
\end{multline}
One can again write similar expressions for all pairs of qudits.

Since $\Tr(\rho) = 1$ and $0\leq\Tr(\rho^2)\leq 1$, using these results and definitions from Eqs. (\ref{Phidef}) and (\ref{Gammadef})
we obtain two sets of inequalities. The first set is 
\begin{equation}
    0\leq \Phi_{1a}\Phi^{1a} \leq \TR - \frac{\TR}{\dR}\label{1-res}
\end{equation}
with similar inequalities for qudits 2 or 3 and so on . The second set is 
\begin{align}
    \frac{-\TR}{\dR}\sum_{C\in\{1,2\},a}\Phi_{Ca}\Phi_{C}{}^{a}  \leq \Gamma&_{1}{}^{a}{}_{2}{}^{b}\Gamma_{1a2b}\notag \\\leq 4 - \frac{4}{\dR^2 } -\frac{\TR}{\dR}&\sum_{C\in\{1,2\},a}\Phi_{Ca}\Phi_{C}{}^{a}\label{2-res} .
\end{align}
Again similar inequalities with other pairs of qudits follow.

At each time-step we check if Eqs.~\eqref{1-res} and ~\eqref{2-res} are satisfied, if not we adjust the $n$-point function by setting it equal to the closer bound. A pure state is at the upper bounds of Eqs.~\eqref{1-res} and ~\eqref{2-res}, whereas entangled/mixed states are below the upper bound. 
Since the coupling is large at the beginning, we initially need smaller time steps. To achieve such an adoptive time step, we take $\delta t_{(t)} = \operatorname{min}(\epsilon\mu_{(t)}^{-1}, \delta t_{\text{max}})$. This enables us to avoid extra evaluations of RK45 and instead utilize the four  evaluations of RK4 with guaranteed acceptance. Additionally we know that the total diagonal elements of the Lie algebra are conserved, i.e.,$d(\sum_{A}\Phi_{A,{d}}) = 0\; \forall \;{d} \in \msc{D}$ where $\msc{D} = \{i^2-1 | i\in\{2, \cdots, \dR\}\}$ or the set of diagonal indices. We can enforce this by adding a constraint to the update equation:
\begin{equation}
   \lim_{\delta t\to 0} \Phi_{A,a (t+\delta t)} - \Phi_{A,a(t)} = d\Phi_{A,a(t)} - \sum_{A,{d}\in\msc{D}}\delta_{a,{d}} \frac{d\Phi_{A ,{d}(t)}}{N}
\end{equation}
We can compute $d\Phi_{A,a(t)}$ directly using RK4 or any method and then impose the Lagrange-multiplier constraint to ensure that the conserved quantity is actually conserved. 

\subsection{Big $\mc{O}$ and Index contractions and $n_{\text{flavors}}$} 
We have $\dim(\Phi) = N(n^2-1) = \mf{D}$ and, since $\Gamma_{AaBb}$ is symmetric under $(A,a)\leftrightarrow(B,b) $ interchange, $\dim(\Gamma) = \frac{1}{2}\mf{D}^2\inParen{1-\frac{1}{N}}$. We indicate four indices of $\Gamma_{AaBb}\Rightarrow \Gamma_{\mc{A}}$ as $\varphi(\mc{A}) = (A,a,B,b)$ keeping the $(A,a)\leftrightarrow (B,b)$ symmetry of the two-point functions. Since $f$ and $g$ of Eq. (\ref{g-form}) are sparse tensors, we pre-compute all nonzero index contractions in advance, eliminating vanishing terms and thereby optimizing the time step. In practice, this amounts to storing only the nonzero contraction channels that appear in equations \eqref{dPhi} and \eqref{LongAhhdG}.

\vspace{-1em}
\begin{figure}[h]
    \centering
    \captionsetup{width=1\linewidth}
    \includegraphics[width=1\linewidth]{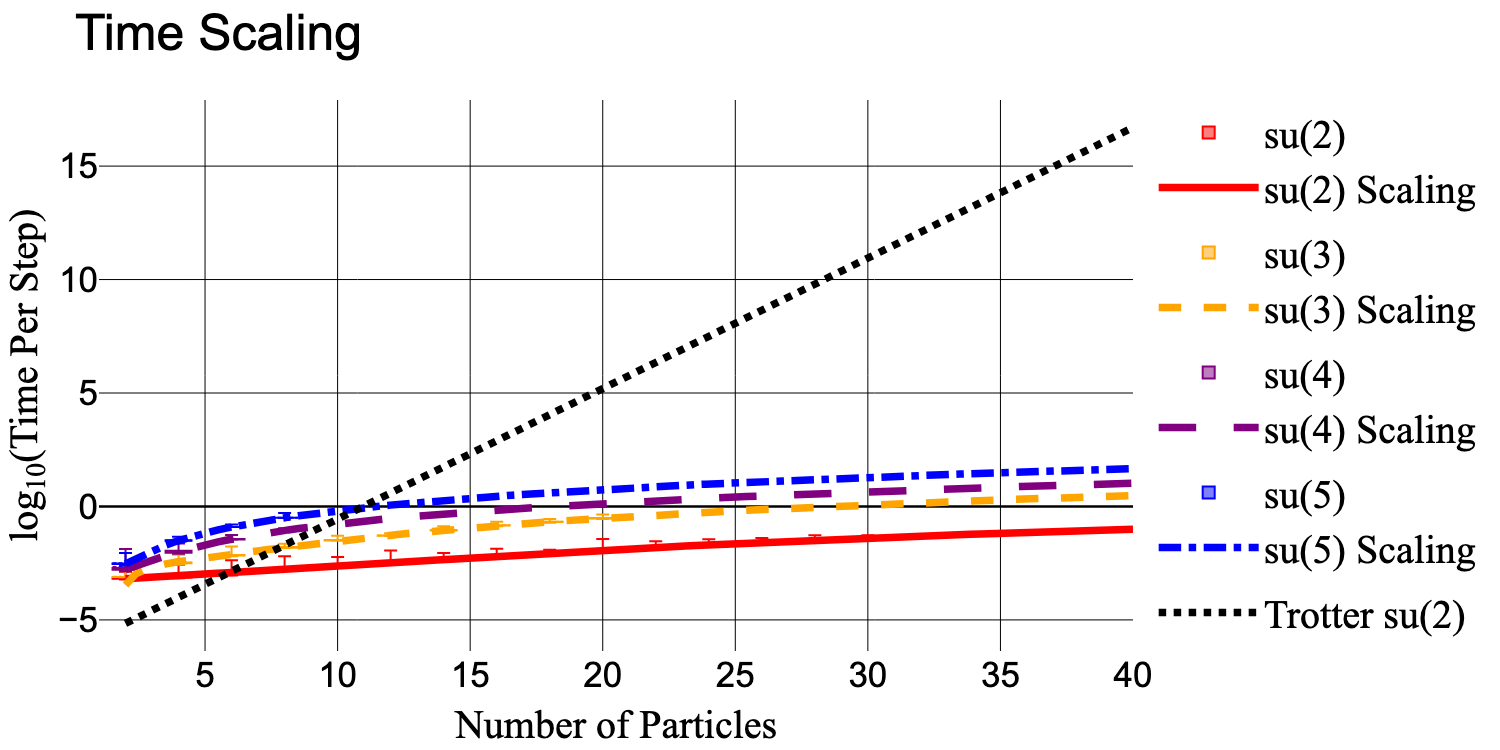}
    
    
    \caption{$\log_{10}(\Delta t_{\text{step}})$ for RK4 forward integration vs Trotter Expansion. Simulations performed on Apple MacBook Air (M3, 8-core CPU, 16 GB RAM), where we solve Eq. \eqref{dPhi} and Eq. \eqref{LongAhhdG} together by direct Trotter expansion. {Note directly the sub-exponential scaling of the Hierarchy truncation and that at second-order Truncation scales as $\mathcal{O}(N^3 n^6)$}. The lines were fit to the mean time per step of using the equation $\bar{\delta t}_{\text{step}}(N,n) = \sum\limits_{i=0}^3\beta_{i(n)}(N)^{i} $ via OLS.}
    \label{fig:timescale}
\end{figure}

We can now examine how the evolution of the system scales with changing $N$ and $n$ as seen in Fig~\ref{fig:timescale}. The raw evolution of the two-point function, without exploiting sparsity, scales as $\mathcal{O}(n^6 N^3)$. As shown in Fig.~\ref{fig:timescale}, this represents a logarithmic improvement over pure Trotter evolution $U \approx \prod_{t} e^{-i H(t)\,\delta t_{(t)}}$, which, even in the case of $\mathfrak{su}(2)$, scales exponentially with system size $\sim \mc{O}(2^N)$. (For $\mathfrak{su}(n)$, classical Trotter evaluation scales as $\mathcal{O}(n^N)$). 
Furthermore, the generalization of this method to the larger values of $n_{\text{flavors}}$ is straightforward. One simply constructs a realization of the $\mf{u}(n)$ algebra, such as the generalized Gell-Mann basis~\cite{GellRepr}, computes the associated structure constants $f$, $d$, and $g$ for the chosen realization (the dimensions of which are $\dim(f) = \dim(d) = (n^2 - 1) \times (n^2 - 1) \times (n^2 - 1)$ and $\dim(g) = n^2 \times n^2 \times n^2$, respectively), and then solves the coupled ODEs depicted in Eqs.~\eqref{dPhi} and~\eqref{LongAhhdG} in the associated realization.

\section{Results and Discussion\label{Results}}
\subsection{Parameterization and angular dependence}
In our simulations, we can adopt the phenomenological neutrino bulb model~\cite{Duan:2006an} :
\begin{equation}
{\mu}_{(t)} = \mu_0 \inParen{1 + \sqrt{1-\frac{R_\nu^2}{(r_0 + t)^2}}}^2 .
\end{equation}
In addition, we introduce an energy scale $\omega_{0}$ so that $t\to \omega_0 \tau$. For physical supernovae $\mu_0\sim 10^4-10^5$ thus we can choose $r_0$ such that $\bar{\mu}_{(0)}\propto \omega_0$. Similarly $B^{Aa}\to B^{Aa}/\omega_{0}$ gets scaled by the energy scale. We can take $\omega_0 = \frac{\Delta_{01}}{4 E_{0}}$ with $E_{0}\sim 10MeV$.   for simplicity we will initialize our state with all electron neutrinos, ie $\ket{\psi_0 } = \bigotimes_{A} \ket{\nu_e}$ and utilize the mixing angles $\theta_{12} = \theta_{1x} = 33.7^{\circ}$, $\theta_{23} = 44.72^{\circ}$, $\theta_{13} = 8.62^{\circ}$, and $\delta_{cp} = 0$. It is important to note that in $d>1$ dimensions we have a continuum of angles that determine our states as $\Lambda_{(|\bp_{A}|, \Omega_{\bp_A}),a}$. We wish to integrate out the angular dependence of this continuum of states to capture the dependence of the neutrino energy. Note that in $d+1$-dimensions each momentum state has $d-1$ angular degrees of freedom, hence we should integrate out all $(d-1)N$ angles. One way to do such an integral in a high dimensional space is to use a Monte-Carlo integrator for the angular degrees of freedom. 

In $3+1$-dimensions each momenta state $\bp_{A}$ is dependent on the magnitude and two angular degrees of freedom:
\begin{equation}
    \bp_{A} = \inParen{|\bp_0| + (A-1)\delta |\bp|}\begin{pmatrix}\sin(\theta_{\bp_{A}})\cos(\phi_{\bp_{A}})\\ \sin(\theta_{\bp_{A}})\sin(\phi_{\bp_{A}})\\
    \cos(\theta_{\bp_A})
    \end{pmatrix}.
\end{equation}
Here we have discretized the magnitude of the momenta with some step size $\delta |\bp|$, and we sample the angles $\{(\theta_{\bp_{1}},\phi_{\bp_{1}}),...(\theta_{\bp_{N}}, \phi_{\bp_{N}})\}$ uniformly over the $2$-sphere. Since the Jacobian for the $2$-sphere is given as $\sin \theta$, $\phi_{\bp_A}$ should be uniformly distributed in the range $[0,2\pi]$ or $\phi_{\bp_A} \sim \mc{U}(0,2\pi)$. Hence the probability distribution function of $\phi$ is given as $f(\phi) = \frac{1}{\int_0^{2\pi}d\phi }$. Additionally, since the Jacobian for $\theta$ is not a constant, $\theta_{\bp_{A}}$ is not simply uniformly distributed from $[0,\pi]$, instead its distribution is determined by the the Jacobian, i.e. $f(\theta) = \frac{1}{\int_0^{\pi}d\theta \sin(\theta)}\sin(\theta)$ and thus we can generate $\theta_{\bp_{A}}$ by sampling some variable $x$  uniformly in the range $[0,1]$(ie $x \sim \mc{U}(0,1)$) and applying the inverse Cumulative Distribution Function $\theta_{\bp_{A}}= \cos^{-1}(2x_{\bp_{A}}-1)$.
Then for each sampling we calculate $\Pi^{AB} = \frac{1}{\TR}\inParen{1-\hat{\bp}^{A}\cdot\hat{\bq}^{B}}$ and compute the time evolution by solving Eqs.~\eqref{dPhi} and Eq.~\eqref{LongAhhdG} simultaneously for this particular sampling of $\Pi^{AB}$. We repeat this procedure a number of times and then average $\Phi_{Aa}$ and $\Gamma_{AaBb}$ over all the runs to get the final result.

\subsection{Comparison to Trotter expansion}
\begin{figure}[btp]
    \centering
    \includegraphics[width=1\linewidth]{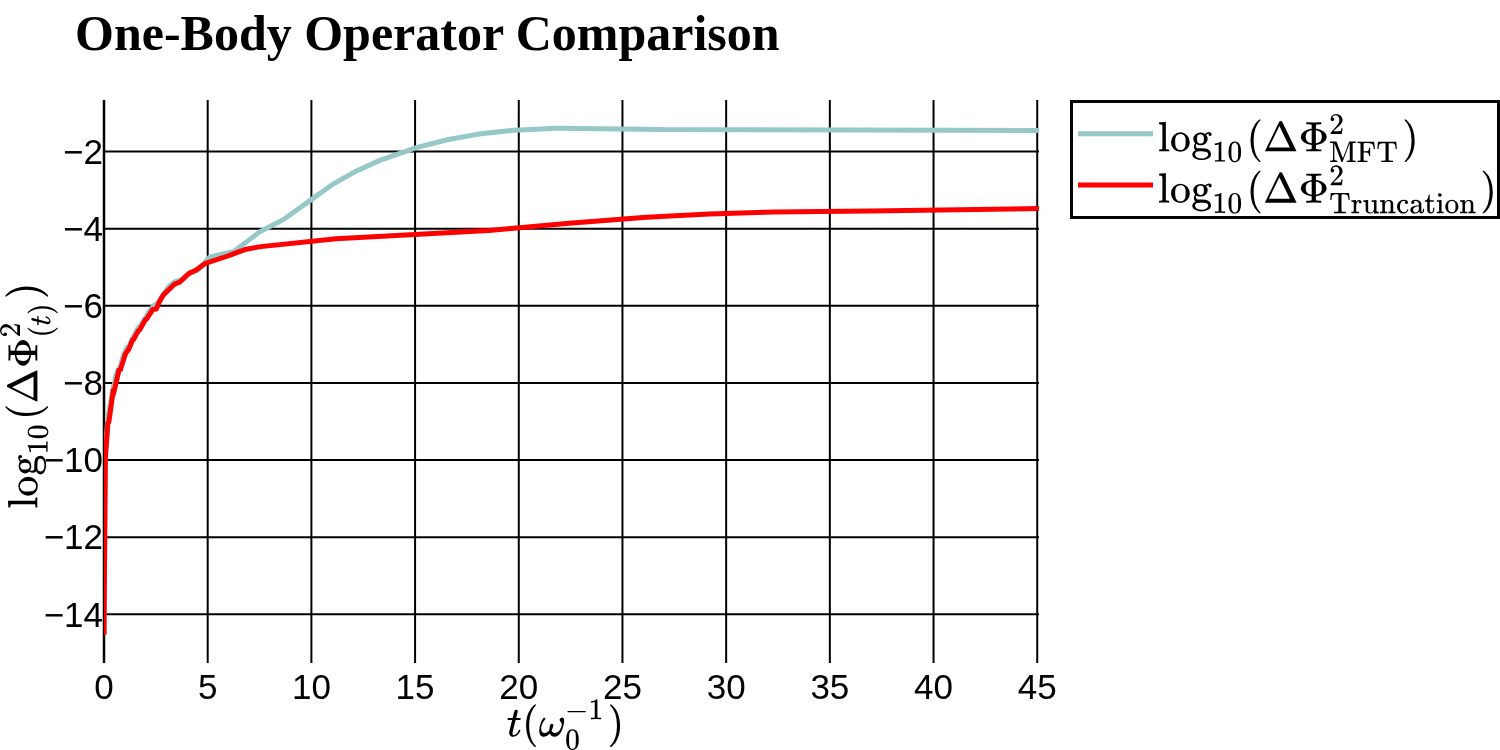}
    \includegraphics[width=1\linewidth]{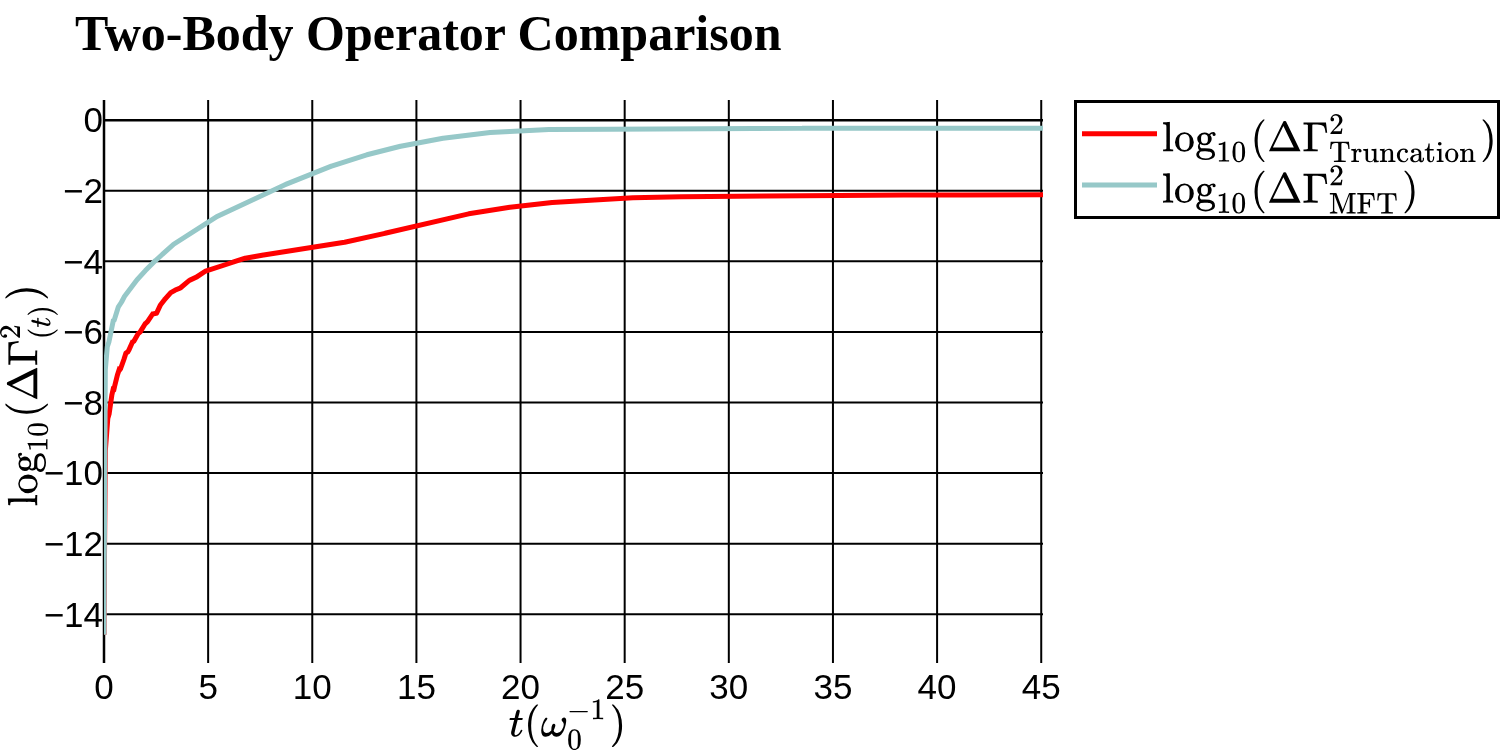}
    \caption{ Comparison of the mean-field, Trotter expansion, and hierarchy truncation for operator expectation values. For hierarchy truncation three-body cumulants are set to zero in Eq. \ref{LongAhhdG}. 
    We  use the definitions $\Delta\Phi^2=\sum_{A,a}({\Phi_{Aa}^{\text{Trotter}}}-\Phi_{Aa}^{\text{Estimator}})^2$,  $\Delta\Gamma^2 = \sum_{A,a, B,b}({\Gamma_{AaBb}^{\text{Trotter}}}-\Gamma_{AaBb}^{\text{Estimator}})^2$, and the mean field estimator for $\Gamma_{AaBb}^{\text{MFT}} = \Phi_{Aa}\Phi_{Bb}$. Note that the hierarchy truncation method matches the operator expectation values about two orders of magnitude better than that of the mean-field. }
    \label{fig:CompareOpers}
\end{figure}
\begin{figure}[tph]
    \centering
    \includegraphics[width=1\linewidth]{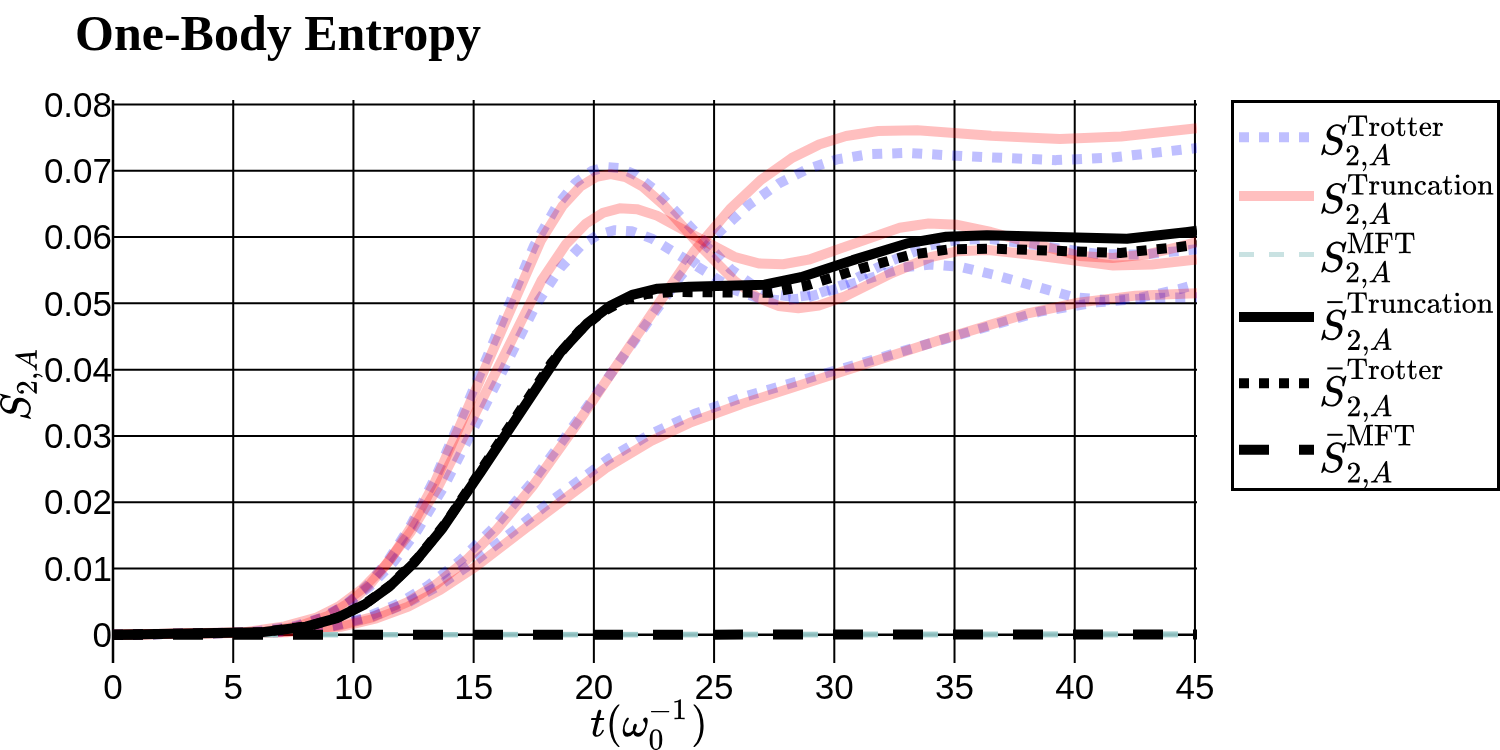}
    \includegraphics[width=1\linewidth]{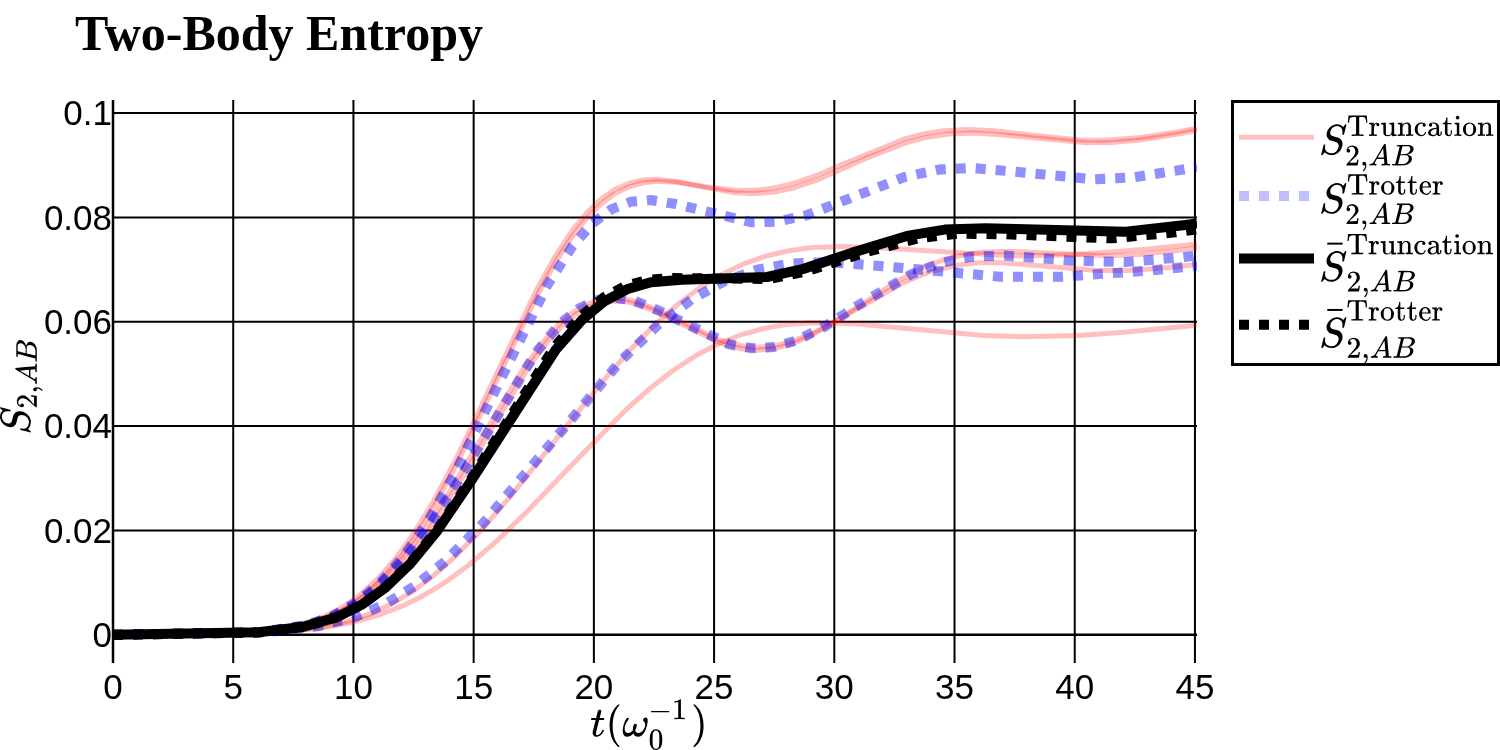}
    \caption{ Comparison of the mean-field, Trotter expansion, and hierarchy truncation for entropy values. For hierarchy truncation three-body cumulants are set to zero at Eq. \ref{LongAhhdG}. 
     The one- and two-body entropies are defined in Eq.~\eqref{2ENT}. Notice that the entropies for truncation matches almost exactly that of the Trotter expansion. The mean field, as expected, has zero entropy.}
    \label{fig:CompareEnt}
\end{figure}
\begin{figure}
    \centering
    \includegraphics[width=1\linewidth]{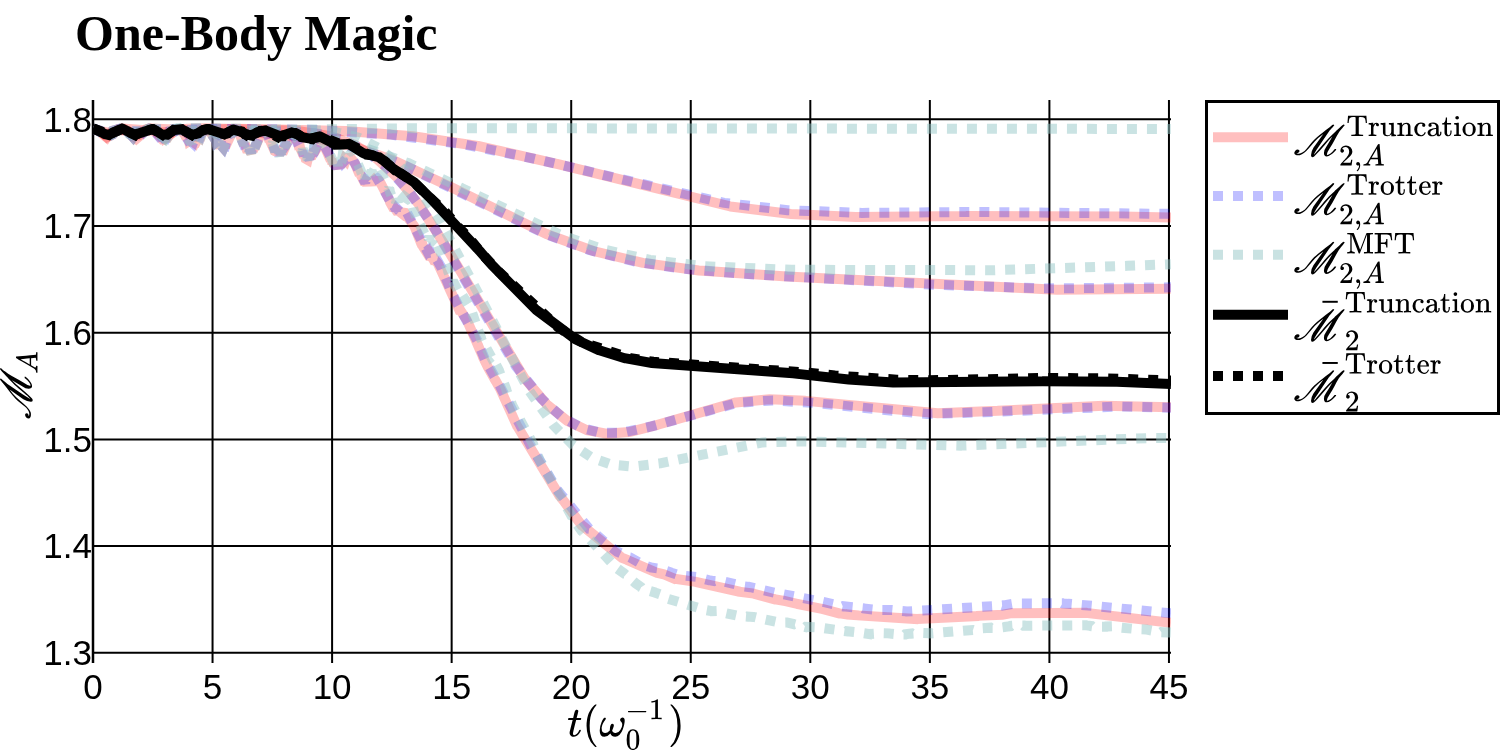}
    \includegraphics[width=1\linewidth]{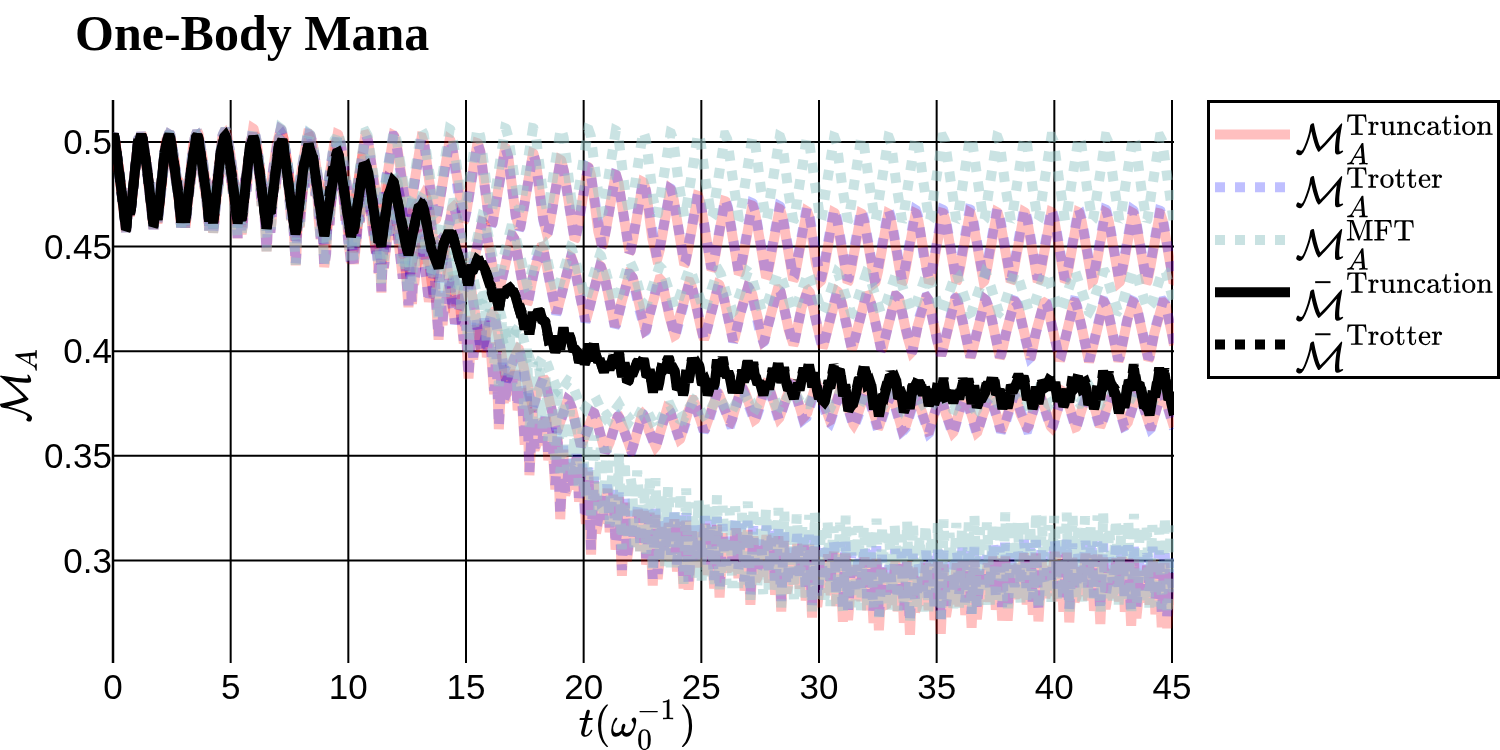}
    \caption{ Comparison of the mean-field, Trotter expansion, and hierarchy truncation for magic and mana values. For hierarchy truncation three-body cumulants are set to zero at Eq. \ref{LongAhhdG}. In this figure we compare the non-stabilizer measures of magic and mana, defined in  Eqs.~\eqref{Magic} and \eqref{Mana}, respectively. Notice that hierarchy truncation matches almost exactly that of the Trotter expansion, whereas the mean field diverges in its estimate of both magic and mana.
    }
    \label{fig:CompareMagic}
\end{figure}

First, as a consistency check in the, we compare the results of the standard Trotterized evolution (i.e.,  $U\approx \prod_{\delta t} e^{-iH(t)\delta t_{(t)}}$) of the Hamiltonian in Eq.~\eqref{Hamiltonian} with those obtained from the mean-field theory (MFT), solved by direct integration of Eq. \eqref{dPhi} with $\Gamma_{AaBb}\to\Phi_{Aa}\Phi_{Bb}$, and  the hierarchy truncation, solved by direct integration of equations \eqref{dPhi} and \eqref{LongAhhdG} simultaneously. To make the Trotterized evolution computationally tractable and for simplicity of comparison, we only consider a single case with $N=4,n=3, \mu_{(0)} = 5$. 

The results of these simulations are depicted Figs.~\ref{fig:CompareOpers}, ~\ref{fig:CompareEnt}, and  ~\ref{fig:CompareMagic}. These figures illustrate that the hierarchy truncation is approximately two orders of magnitude closer to the Trotterized solution for the one- and two-point functions than the mean field. 

Results obtained for the one and two-body R\'{e}nyi entropy as well as for the magic and mana with truncation match those obtained from the Trotter expansion with only small deviations. In contrast, mean-field calculations have larger deviations in magic and mana when compared to the hierarchy truncation.
The discrepancies in the entropy function and expectation values are predominantly controlled by numerical integration errors of $\mc{O}(\delta t^4)$.

More generally, if we do not truncate the hierarchy one would need to generate $m$-body cumulants.
Increasing the truncation order systematically would improve the approximation, but at the cost of increased computational complexity. Tructating at order $m$ requires a cost that scales as $
\mathcal{O}\!\left(
\binom{N}{m}(n^2 - 1)^{m}
\right).
$  If one were to traverse the entire hierarchy, the total cost would scale as $
\mathcal{O}(n^{N})$ corresponding to the full operator algebra dimension $
\mathfrak{u}(n^N)$, i.e., the complete operator basis of the Hilbert space.

We should point out that in this work we are mainly calculating the R\'enyi entropy $(\alpha=2)$. For reference a comparison of von Neumann $(\alpha = 1)$ and R\'enyi entropies is given in Fig. \ref{fig:Von} .

\subsection{Large N Behavior Results}
\subsubsection{$\mf{su}(2)$ in the Large $N$ limit}
{
Now that we demonstrated the utility of the hierarchy truncation, we can probe the large $N$ behavior of the system. First, we explore the two-flavor system with $n=2, \;N=100$. This will allow us to explore some of the statistical properties that would otherwise be computationally untractable. To do so we randomly sample $100$ independent $\Pi^{AB}{}\text{s}$ with uniform probabilities and average over the different simulations.
\begin{figure}[tp]
    \centering
    \includegraphics[width=\linewidth]{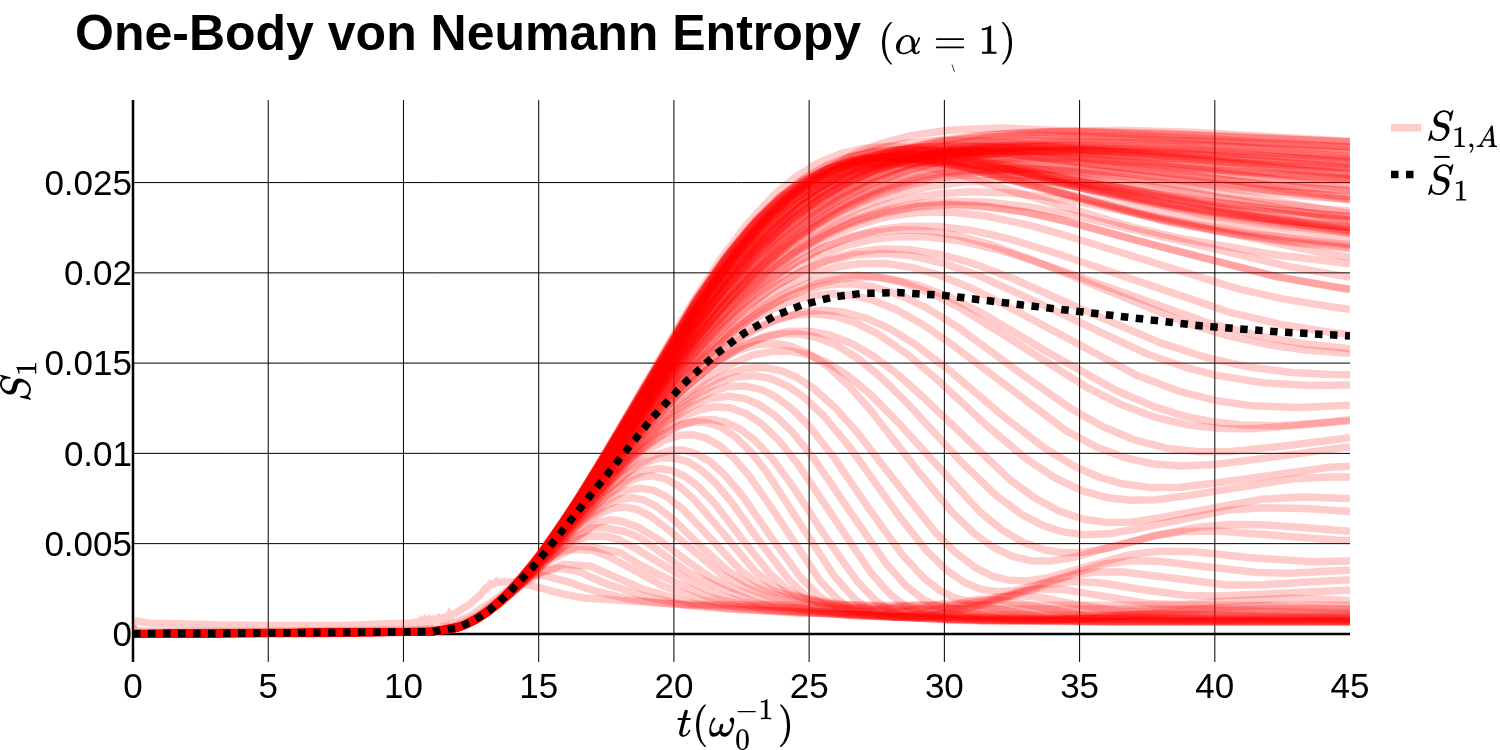}
    \includegraphics[width=\linewidth]{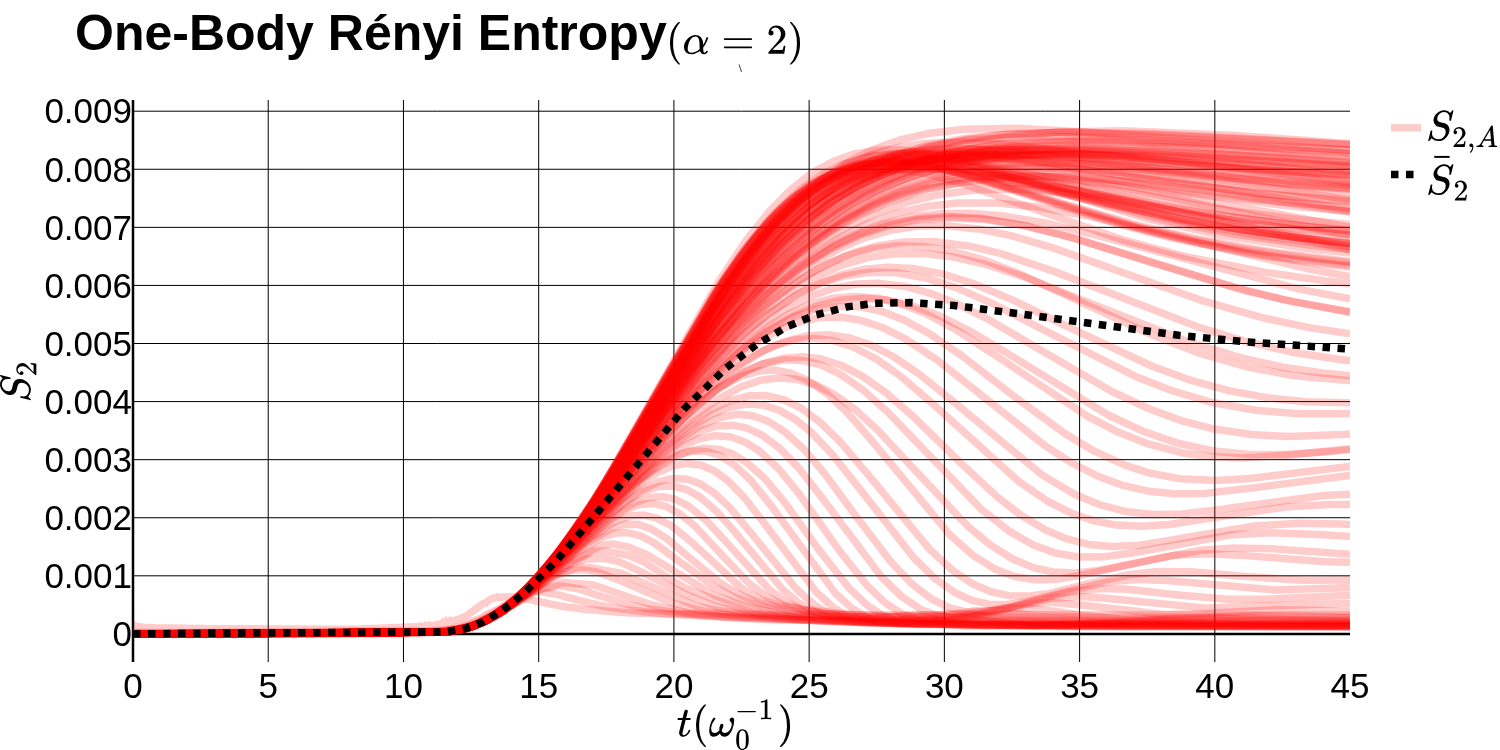}
    \caption{Evolution of the von Neumann Entropy vs that of the $\alpha=2$ R\'enyi Entropy. The evolution is found by solving Eqs.~\eqref{dPhi} and \eqref{LongAhhdG} simultaneously with $N = 100, n = 2,$ and $\mu_{(0)} = 5$. To compute the von Neumann Entropy we construct the one-body reduced density matrix with $\rho_{A}= \frac{1}{n} \I+ \frac{1}{2}\Phi_{Aa}\lambda^{a} $, we then compute the von Neumann entropy numerically for the case of the $n$ flavor system. Of course in the $2$-flavor system this can be computed analytically because the eigenvalues of the reduced density matrix $\rho_{A}$ are trivially given as $\frac{1}{2}(1\pm \sqrt{\Phi_{Aa}\Phi_{A}{}^{a}})$. Additionally the R\'enyi Entropy is calculated using Eq.~\eqref{2ENT}. Note that the R\'{e}nyi entropy is bounded by the von Neumann entropy, $S_{2,A}\leq S_{1,A}$}
    \label{fig:Von}
\end{figure}

One interesting probe would be defining an order parameter $\bar{\Phi}_{a} = \frac{1}{N}\sum_{A}\Phi_{Aa}$(analogous to that of the site-averaged magnetization seen in condensed matter literature \cite{Magnetization}), characterizing flavor coherence for the system. As seen in Fig.~\ref{fig:KwDual} around the crossover point where $\mu_{(t)} \sim |\bar{\bm{B}}|$ ($\bm{B}$ of Eq.~\eqref{BEq}) we see that the order parameter changes rapidly and the magnitude of the order parameter is asymptotically transitioning to some small value, indicating a phase transition around the crossover point.

\begin{figure}[btp]
    \centering
    \includegraphics[width=\linewidth]{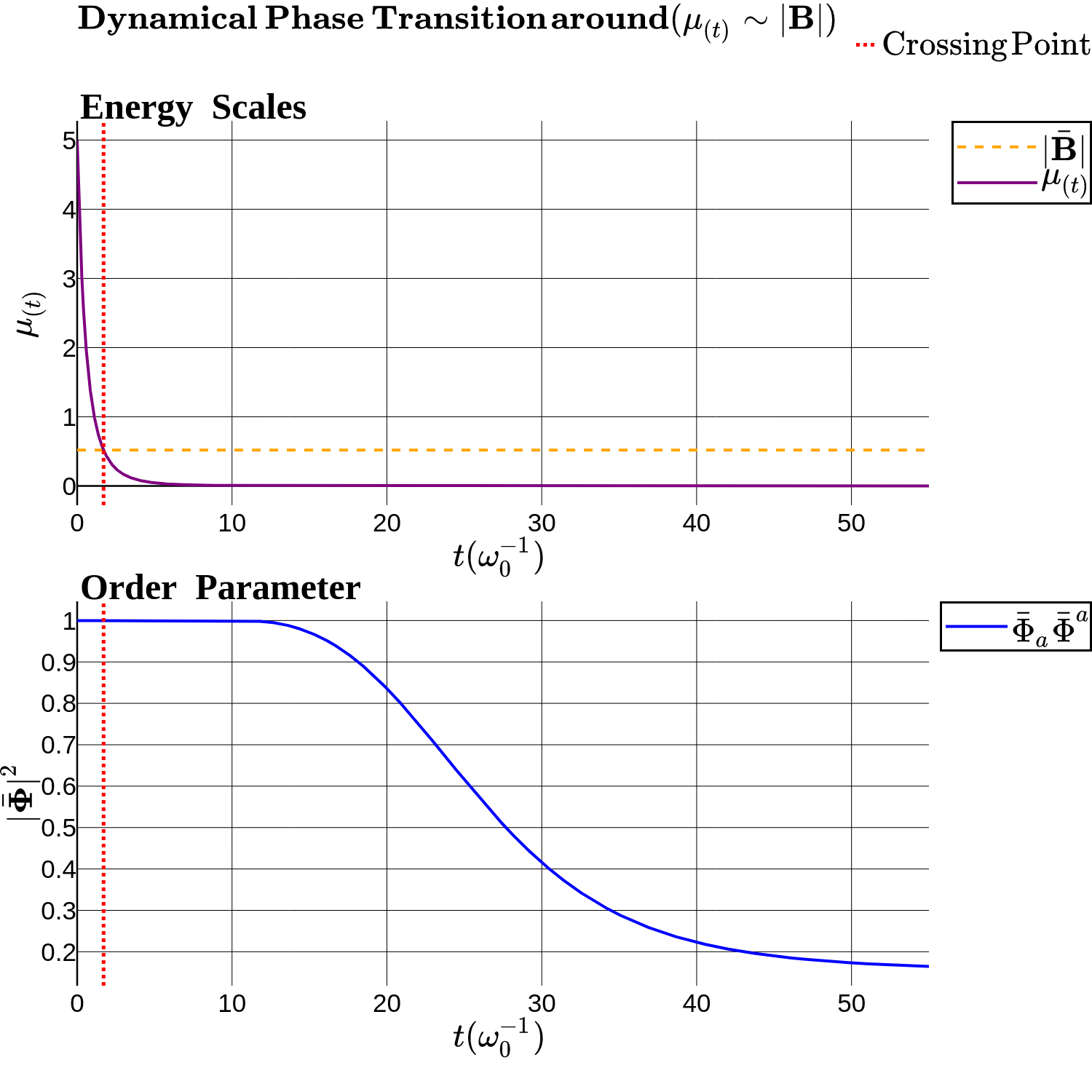}
    \caption{Evolution found by solving Eqs.~\eqref{dPhi} and \eqref{LongAhhdG} simultaneously with N = 100, n = 2, and $\mu_{(0)} = 5$. This figure depicts a dynamical phase transition after the crossover point $\mu(t) \sim |\bar{\bm{B}}|$. Order parameter is $\bar{\Phi}_{a} =\frac{1}{N}\sum_{A}\Phi_{Aa}$. Vertical dotted line corresponds to the critical value of $ | \bar{\bm{B}}|$.   }
    \label{fig:KwDual} 
\end{figure}
\begin{figure}[tp]
    \centering
    \includegraphics[width=\linewidth]{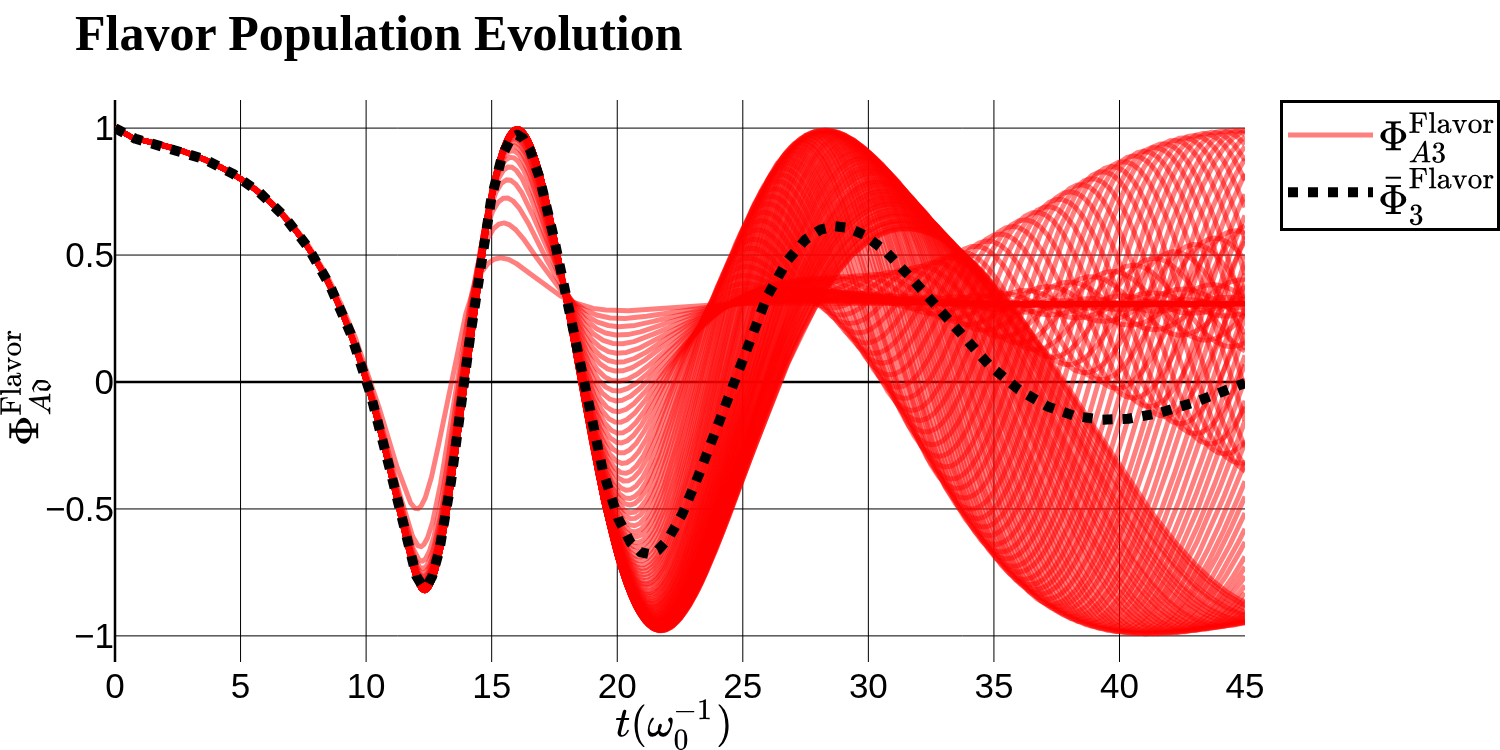}
    \includegraphics[width=\linewidth]{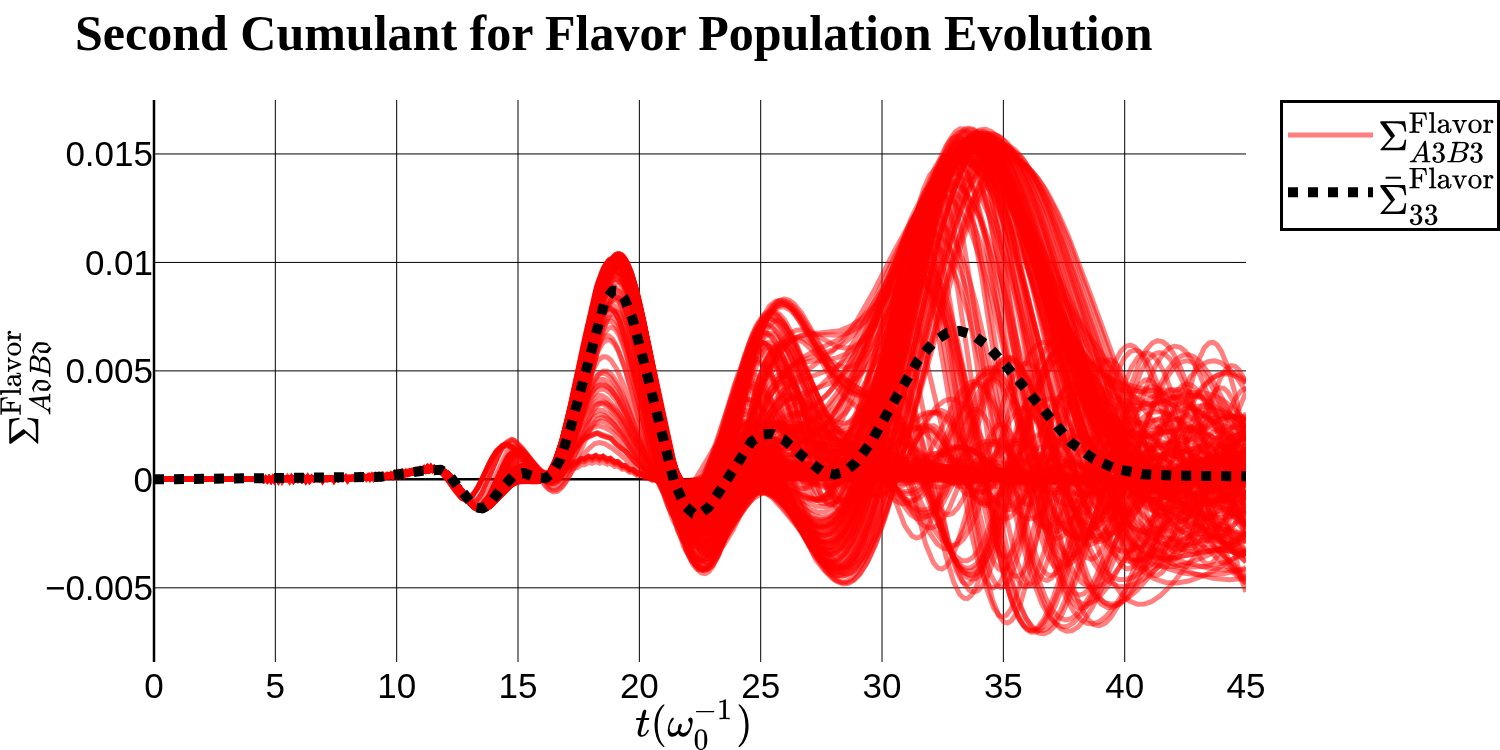}
    \caption{Evolution of the flavor operators for two flavors. The evolution is found by solving Eqs.~\eqref{dPhi} and \eqref{LongAhhdG} together with $N = 100, n = 2,$ and $\mu_{(0)} = 5$. One-body relative populations are defined as $\Phi_{A3} = \expval{n_{A\nu_e}} - \expval{n_{A,\nu_{X}}}$ and two-body flavor cumulant as $\Sigma_{A3B3} = \Gamma_{A3B3} -\Phi_{A3}\Phi_{B3}$. After the transition from coherent oscillations ($\mu(t) > |\bar{\bm{B}}|$) to decoherent vacuum oscillations as $|\bar{\bm{B}}|>>\mu_{(t)}$ two-body cumulant deviates from its zero value.}
    \label{fig:Pz}
\end{figure}

\begin{figure}[btp]
    \centering
    \includegraphics[width=\linewidth]{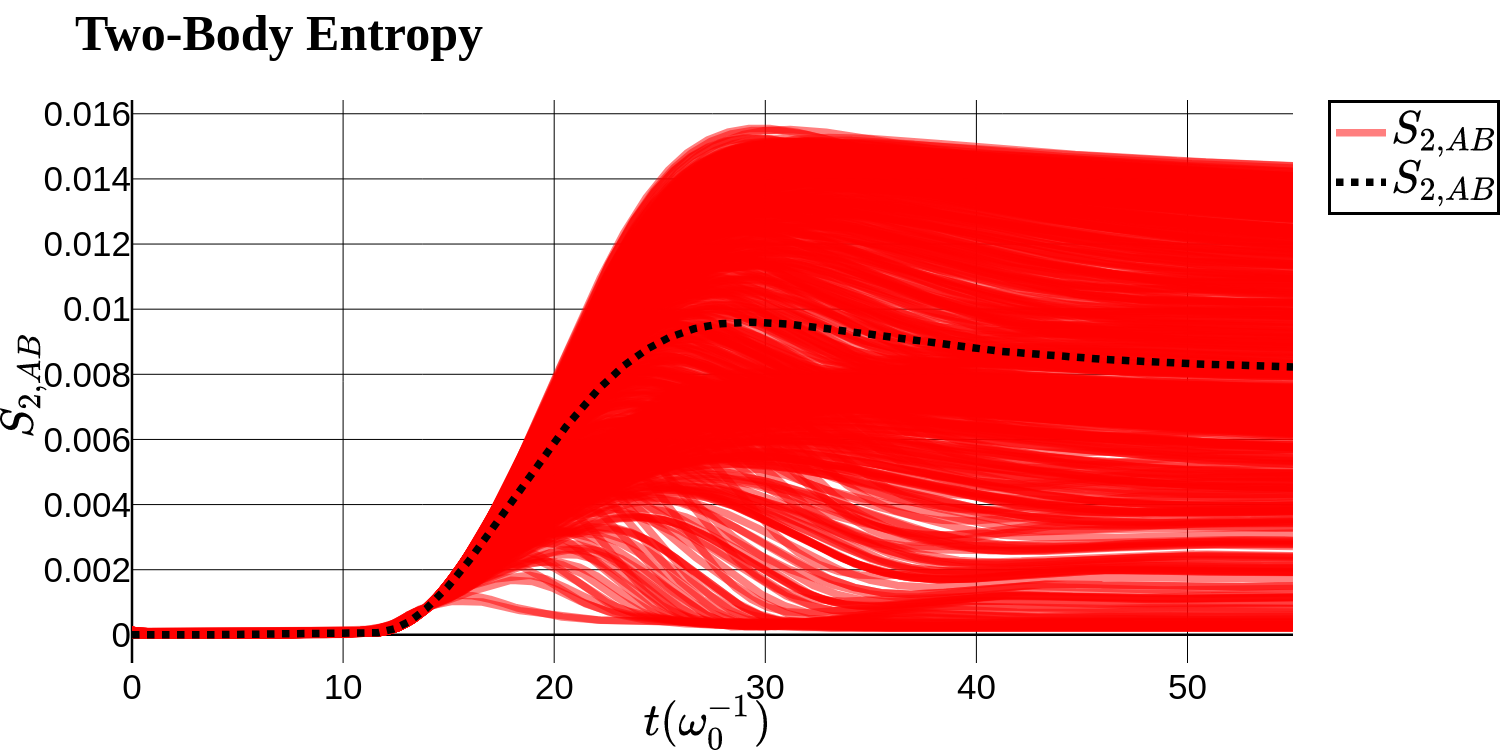}
    \includegraphics[width=\linewidth]{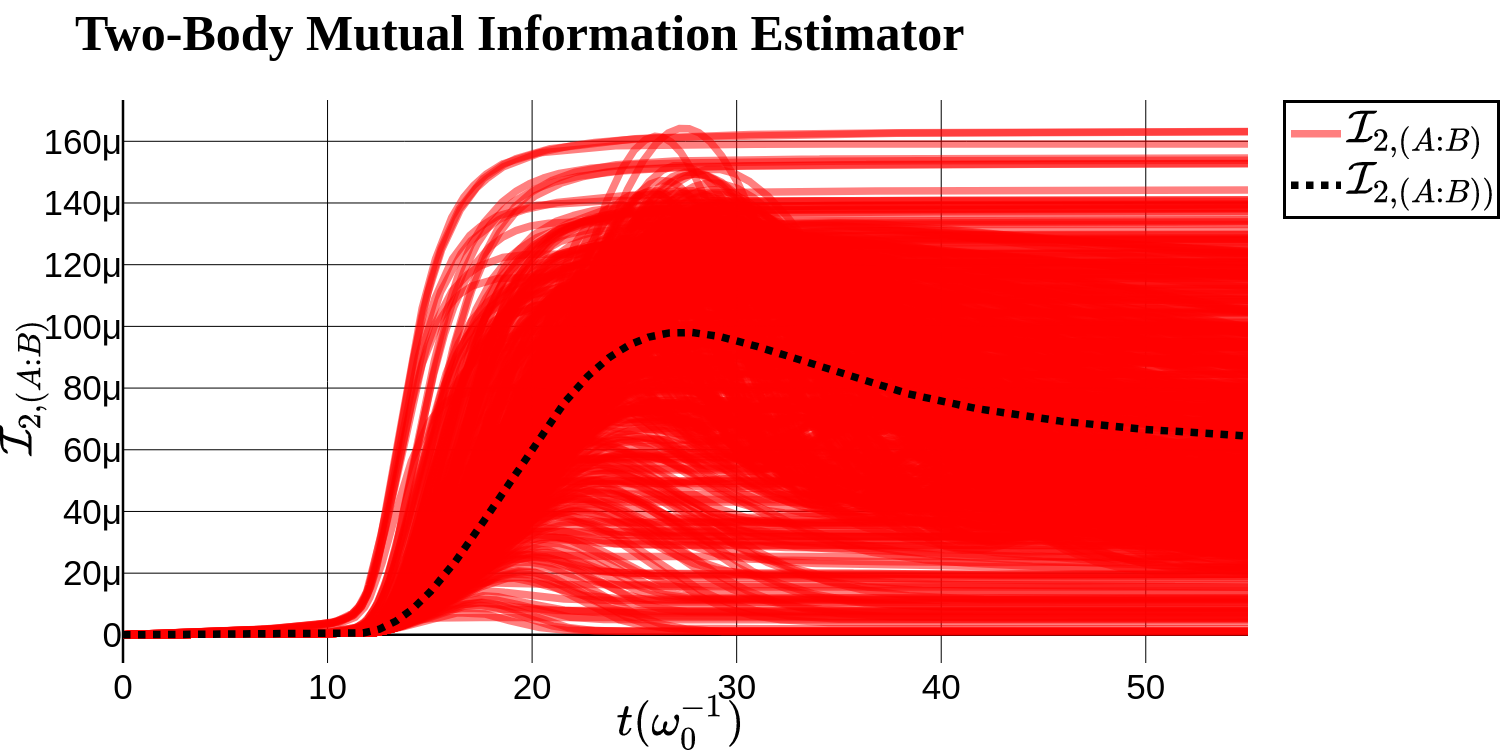}
    \caption{Evolution of the two-body entropy and mutual information estimator for two flavors. The evolution of the system is found by solving Eqs.~\eqref{dPhi} and \eqref{LongAhhdG} simultaneously with $N=100, n=2,$ and $\mu_{(0)} = 5$. The two-body entropy is  calculated using Eq.~\eqref{2ENT} and the two-body mutual information estimator is calculated using Eq.~\eqref{MutInf} with $\alpha = 2$. }
    \label{fig:twoBodyTimesu2}
\end{figure}

If we go to the flavor basis and plot the third component of the flavor polarization (relative populations $\Phi_{A,3} = \expval{n_{\nu_e}}- \expval{n_{\nu_{X}}}$) and the corresponding two-body cumulant ($\Sigma_{A3B3} = \Gamma_{A3B3}-\Phi_{A3}\Phi_{B3}$) in Fig.~\ref{fig:Pz}. We see that increased entanglement results in non-zero values of the two-body cumulant.

Since an apparent dynamical phase transition is identified it is important that we examine some of the quantum properties of the evolution of the density matrix. Particularly it is interesting to see how information is distributed in the system. We start by computing the two-body entropy and the mutual information estimator (Eqs.~\eqref{2ENT} and ~\eqref{MutInf}, respectively). The results for the time evolution can be found in Fig.~\ref{fig:twoBodyTimesu2} and corresponding heat maps for the final time step in the simulation in Fig.~\ref{fig:twoBodysu2Heat}. These heat maps depict the entropy and mutual information correlation functions over the discretized momenta $|\bp_{A}|$ and $|\bp_{B}|$(i.e., $\mc{I}_{2,(A:B)} = \mc{I}_{2,(|\bp_{A}|:|\bp_{B}|)}$ and $S_{2,AB} = S_{2,|\bp_{A}||\bp_{B}|}$, respectively). From these figures we see that information is being distributed in the momentum space. The latter figure shows an increase of entropy when two momenta are equal and sufficiently large.
Also, information from the higher momentum states is distributed into the states with smaller momentum states. This behavior suggests that information is being delocalized in the momentum space, i.e., one cannot reconstruct all of the information in the system from only one-body operators since they are labeled by a single momentum. 

\begin{figure}[btp]
    \centering
    \includegraphics[width=.96\linewidth]{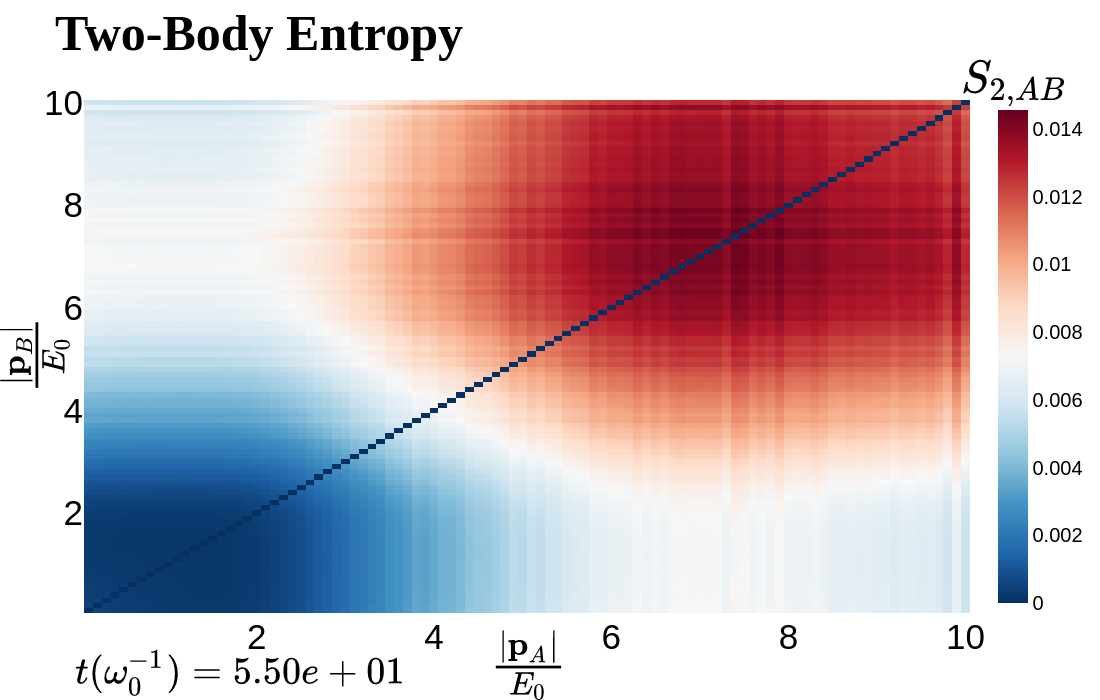}
    \includegraphics[width=.96\linewidth]{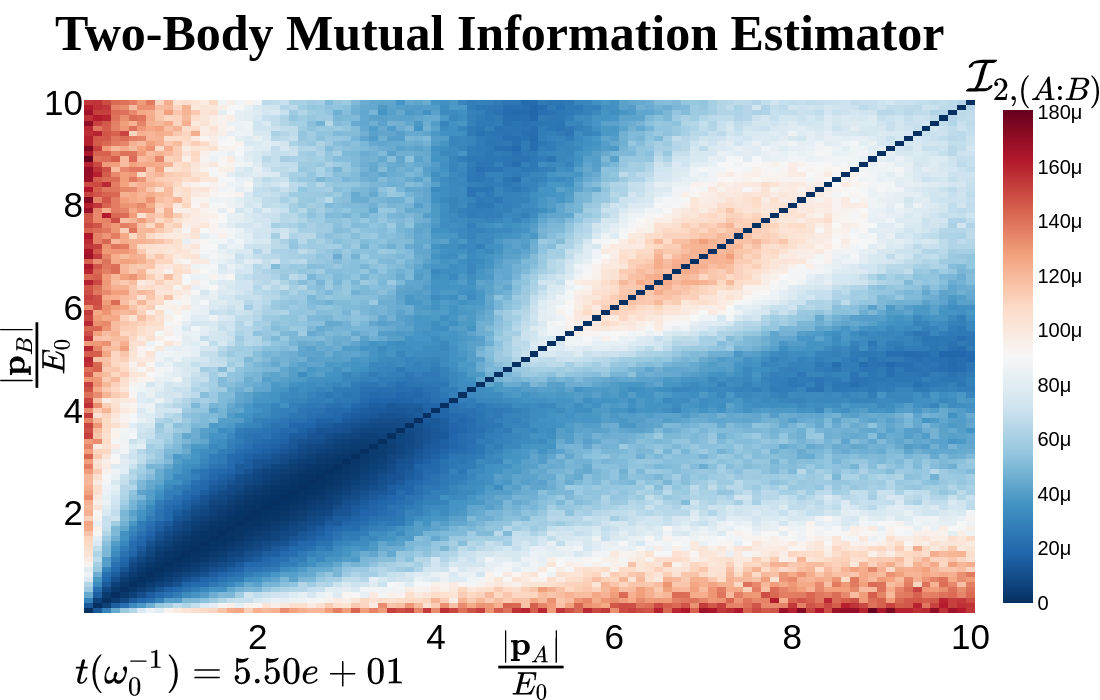}
    \caption{A constant time slice ($t\inParen{\omega_{0}^{-1}} \approx 54.1$) heat map of the distribution of two-body entropy (Eq.~\eqref{2ENT}) and mutual information (Eq.~\eqref{MutInf}) over momentum space. We first calculate the evolution of the system by simultaneously solving Eqs.~\eqref{dPhi} and  \eqref{LongAhhdG} with N = 100, n = 2, $\mu_{(0)} = 5$, then focus only on the $t\inParen{\omega_{0}^{-1}} = 45$ time slice to find the mutual information estimator. Resulting mutual information estimator indicates that the information is being distributed in the momentum space.}
    \label{fig:twoBodysu2Heat}
\end{figure}

 The way in which the information is distributed in the 
 system can be further explored by computing two- and three-body mutual information and the entropy as seen in Figs. ~\ref{fig:twoBodyTimesu2}, ~\ref{fig:twoBodysu2Heat} and~\ref{fig:threeBodyTimesu2}. The rapid growth in the two-body mutual information estimator and entropy are indicators of information distribution into many-body operators. A note of interest is the evolution of the three-body entropy and mutual information. The three-body entropy growth coincides with that of the one and two-body, however more interestingly the three-body mutual information estimator has a richer  structure as it is negative which is a clear indicator of weak information scrambling \cite{Seshadri:2018yya}

 \begin{figure}[btp]
    \centering
    \includegraphics[width=1\linewidth]{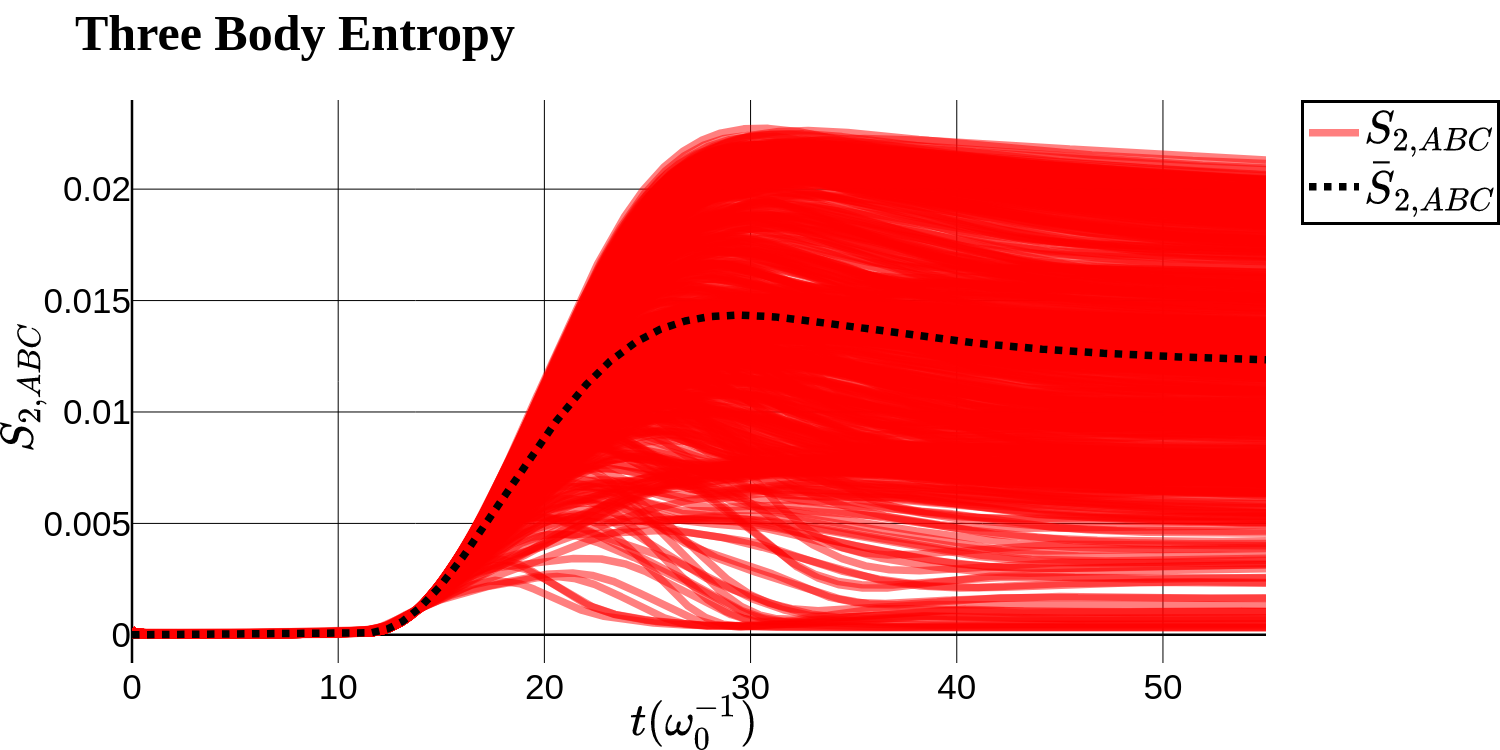}
    \includegraphics[width=1\linewidth]{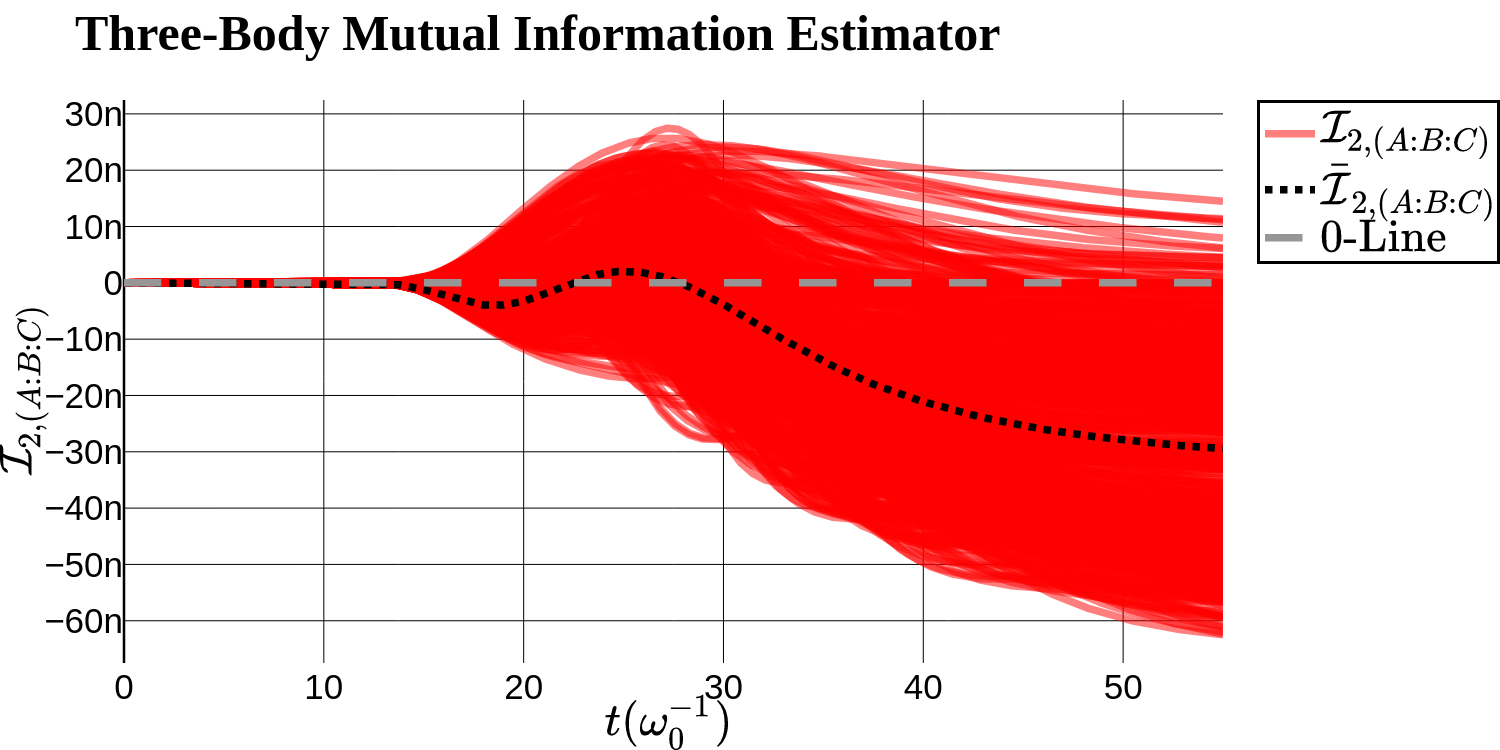}
    \caption{Evolution of the three-body entropy and three-body mutual information estimator. Evolution of the system is found by solving Eqs.~\eqref{dPhi} and  \eqref{LongAhhdG} together using N = 100, n = 2, $\mu_{(0)} = 5$. Entropy and mutual information estimator are calculated using Eqs.~\eqref{2ENT} and ~\eqref{MutInf3} with $\alpha = 2$, respectively. We can see the growth of three-body mutual information and entropy, indicating the growth of the information content of three-body operators. The slight negativity of the estimator indicates the presence of  weak information scrambling~\cite{Seshadri:2018yya}}
    \label{fig:threeBodyTimesu2}
\end{figure}

Once we know that information is being distributed among  many-body operators we can examine non-stabilizer resources inherent in such a  system. Examining Fig.~\ref{fig:MagicMana} 
we can see that the amount of ``non-stabilizerness" (as discussed in sections~\ref{sec:Mana} and \ref{sec:Magic})  such as non-zero magic, in each particle starts coherently and begins to de-cohere, but remains non-zero throughout the entire evolution. This is expected since the states of the current system oscillates around the stabilizer states. }

\begin{figure}[btp]
    \centering
    \includegraphics[width=1\linewidth]{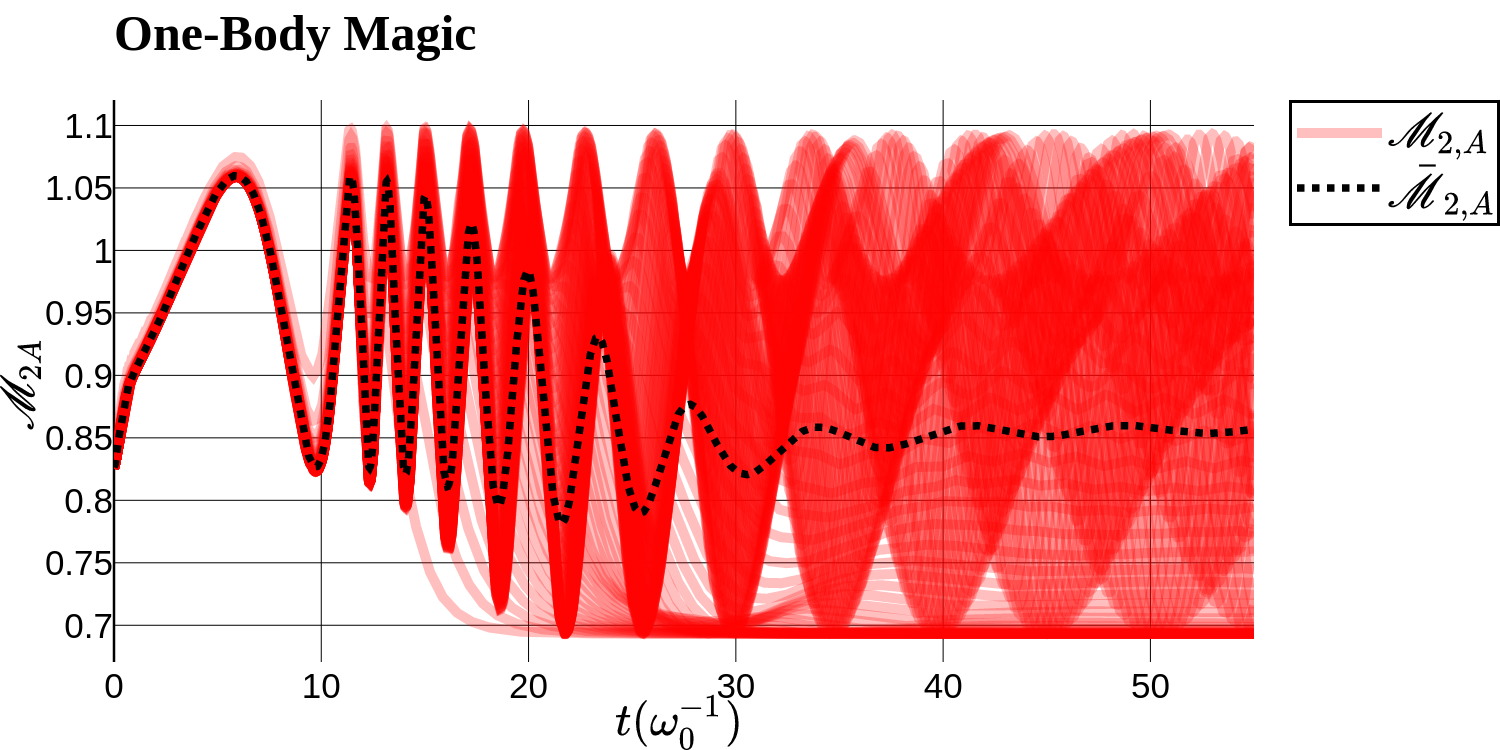}
    \includegraphics[width=1\linewidth]{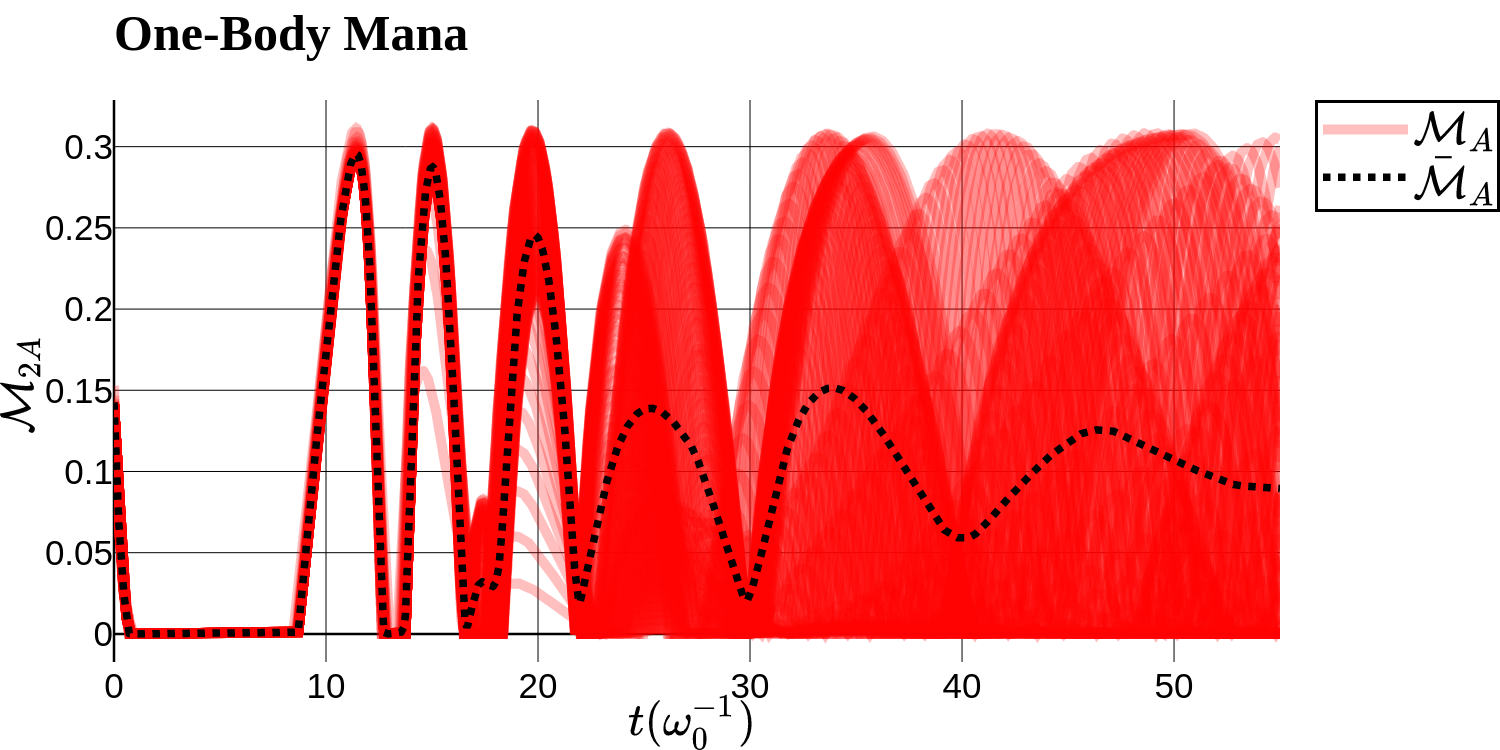}
    \includegraphics[width=1\linewidth]{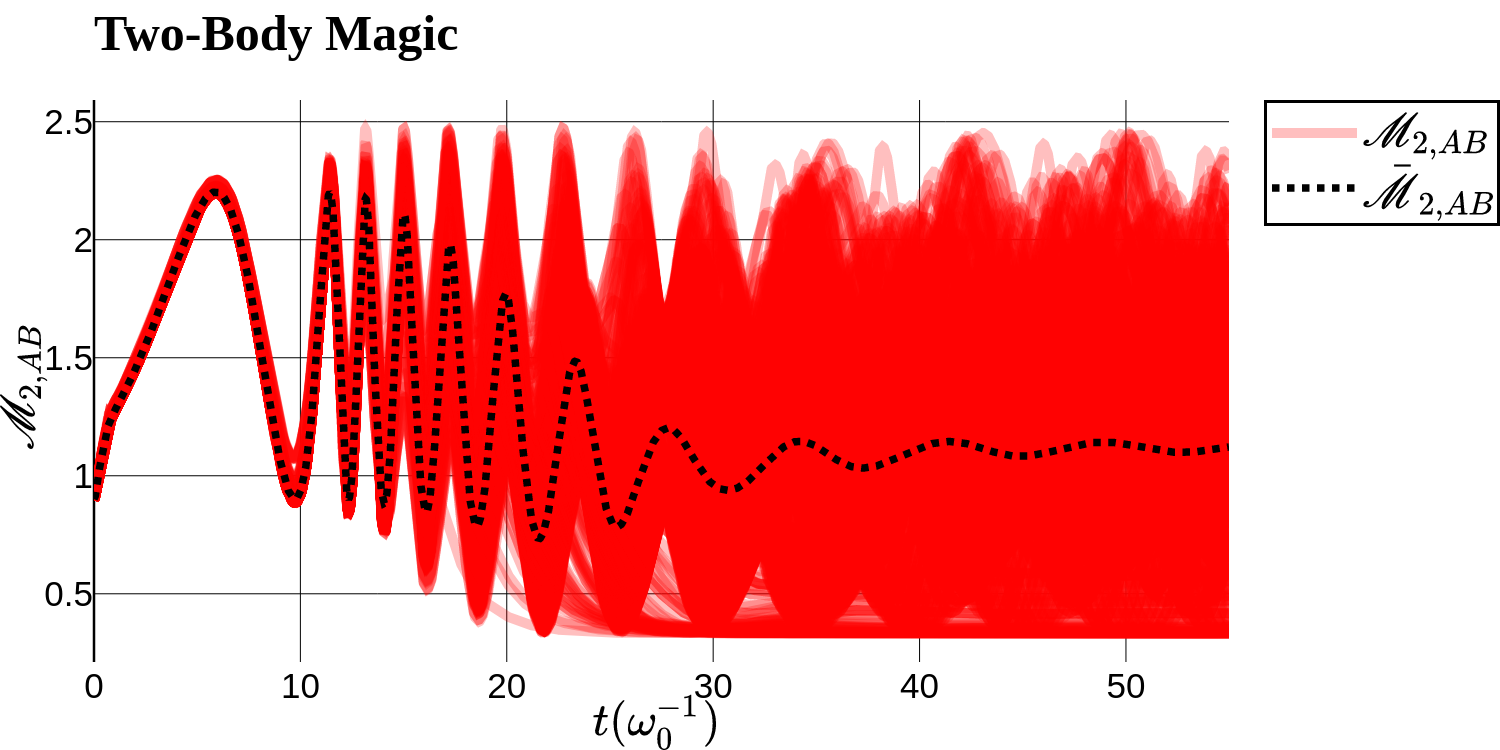}
    \includegraphics[width=1\linewidth]{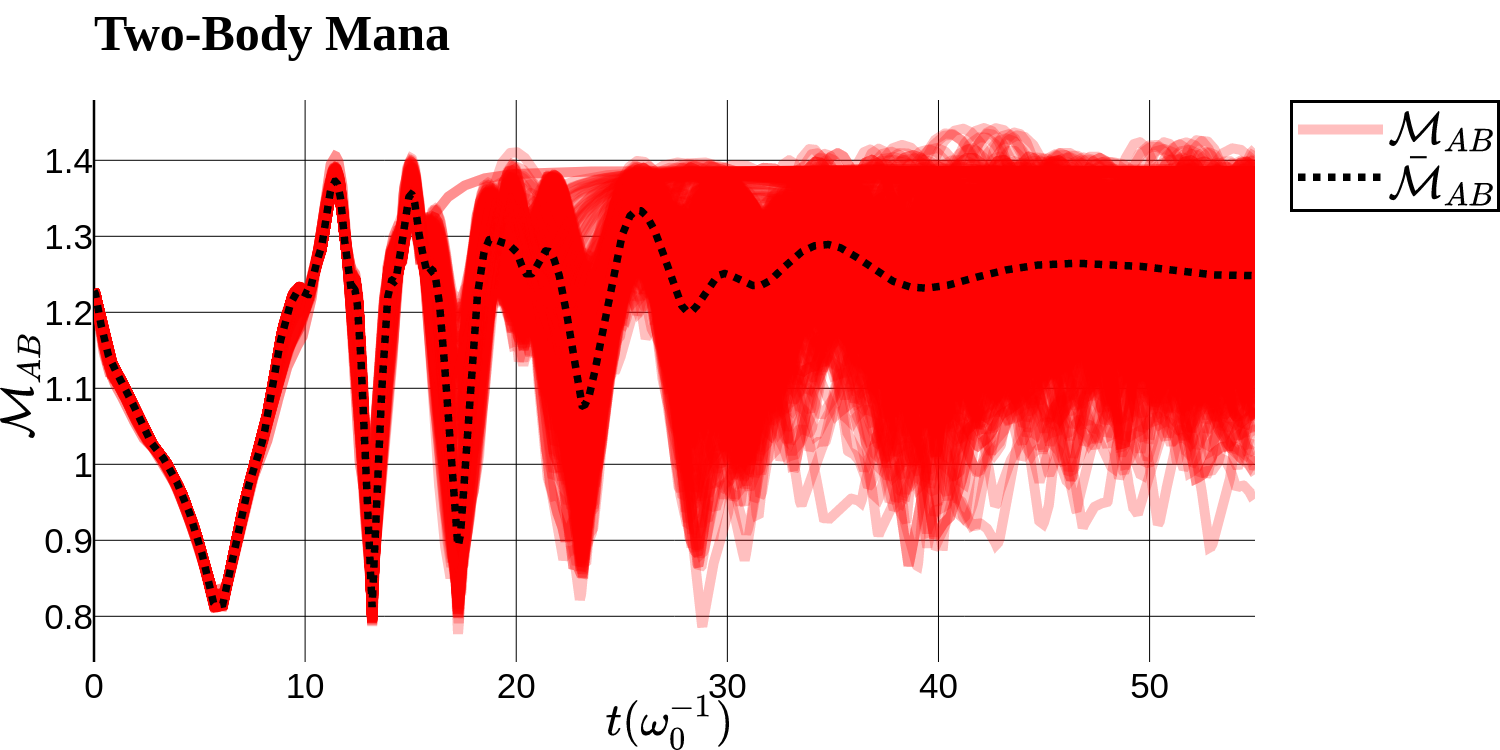}
    \caption{Evolution of the one-body entropy, two-body magic and mana for two flavors. The evolution of the system is found by solving Eqs. ~\eqref{dPhi} and  \eqref{LongAhhdG} together with $N=100, n=2,$ and $\mu_{(0)} = 5$. We calculate magic and mana using Eqs.~\eqref{Magic} and \eqref{Mana}, respectively. Both magic and mana indicate that density matrix contains non-stabilizer states. }
    \label{fig:MagicMana}
\end{figure}

Another note of interest is the one-body mana. We observe a clear hierarchy in the distribution of non-stabilizer resources. While the two-body mana remains consistently positive, the one-body mana frequently hits zero. This suggests that the phase-space negativity is a more fragile resource, susceptible to being 'washed out' by the mixing induced by subsystem entanglement. Persistence of magic in the absence of mana highlights that one-body density matrices are non-stabilizer, yet Wigner-positive. This arises because the density operator becomes mixed, allowing the Wigner negativities of the constituent pure-state components to cancel in the convex combination that defines the mixed-state Wigner function~\cite{Wallman:2012ppe}. Since $\msc{W}_{\rho} = \sum p_{i}\msc{W}_{\rho_i}$ individual $\msc{W}_{\rho_i}$'s can have negativity while the total mixed $\msc{W}_{\rho}$ is non-negative \cite{Wallman:2012ppe}.

\twocolumngrid

\subsection{$\mf{su}(3)$ in the large $N$-limit}

Probing the collective neutrino oscillations within the two-flavor approximation enabled us to consider a relatively large number of neutrinos, as high as $N \sim 100$ and higher. However, as shown in the previous sections computational cost, even with the second order truncation, scales as $\mc{O}(N^3 n^6)$. Although the scaling of $\mf{su}(3)$ and $\mf{su}(2)$ are both $\mc{O}(N^3)$, the time constant in front of $N$ is higher for $\mf{su}(3)$ due to the additional degrees of freedom. Consequently, a simulation with $N = 100$ neutrinos for three flavors would incur a significant computational cost. To keep the simulations manageable we instead perform our $\mf{su}(3)$ simulations with $N = 50$ neutrinos using the parameters $N=50, n=3$ and $\mu_{(0)} = 5$. Following the same procedure as in the $\mf{su}(2)$ case we sample fifty different $\Pi^{AB}$'s a hundred times, evolve every one of them by solving the coupled differential equations \eqref{dPhi} and \eqref{LongAhhdG}; finally average the results to perform the angular integration. Once we obtain the evolution of $\Phi_{Aa}$ and $\Gamma_{AaBb}$ we can compute the order parameter for the three flavor system, $\bar{\Phi}_{a} = \frac{1}{N}\sum_{A}\Phi_{Aa}$. Order parameter behaves almost identically to that of the $\mf{su}(2)$ case: we see a rapid change of it, indicating a dynamical phase transition as seen at Fig.~\ref{fig:su3Phase}. However, it appears that the $\mf{su}(3)$ order parameter changes quicker and reaches its asymptotic value well before that of $\mf{su}(2)$.

Relative populations ($\Phi_{A3} = \expval{n_{A\nu_e}} - \expval{n_{A,\nu_{\mu}}}$ and $\Phi_{A8} = \frac{1}{2\sqrt{3}}\inParen{\expval{n_{A,\nu_{e}}} + \expval{n_{A\nu_{\mu}}}- 2\expval{n_{{A}\nu{\tau}}}}$) begin to de-cohere as seen in Fig. ~\ref{fig:Pzsu3} following the phase transition. Additionally, flavor cumulants ($\Sigma_{AaBb} = \Gamma_{AaBb}-\Phi_{Aa}\Phi_{Bb}$) and the two-body entropy (Fig~\ref{fig:twoBodyTimesu3}) being to grow coinciding with the phase transition just as they did in the two flavor case.

\begin{figure}[btp]
    \centering
    \includegraphics[width=\linewidth]{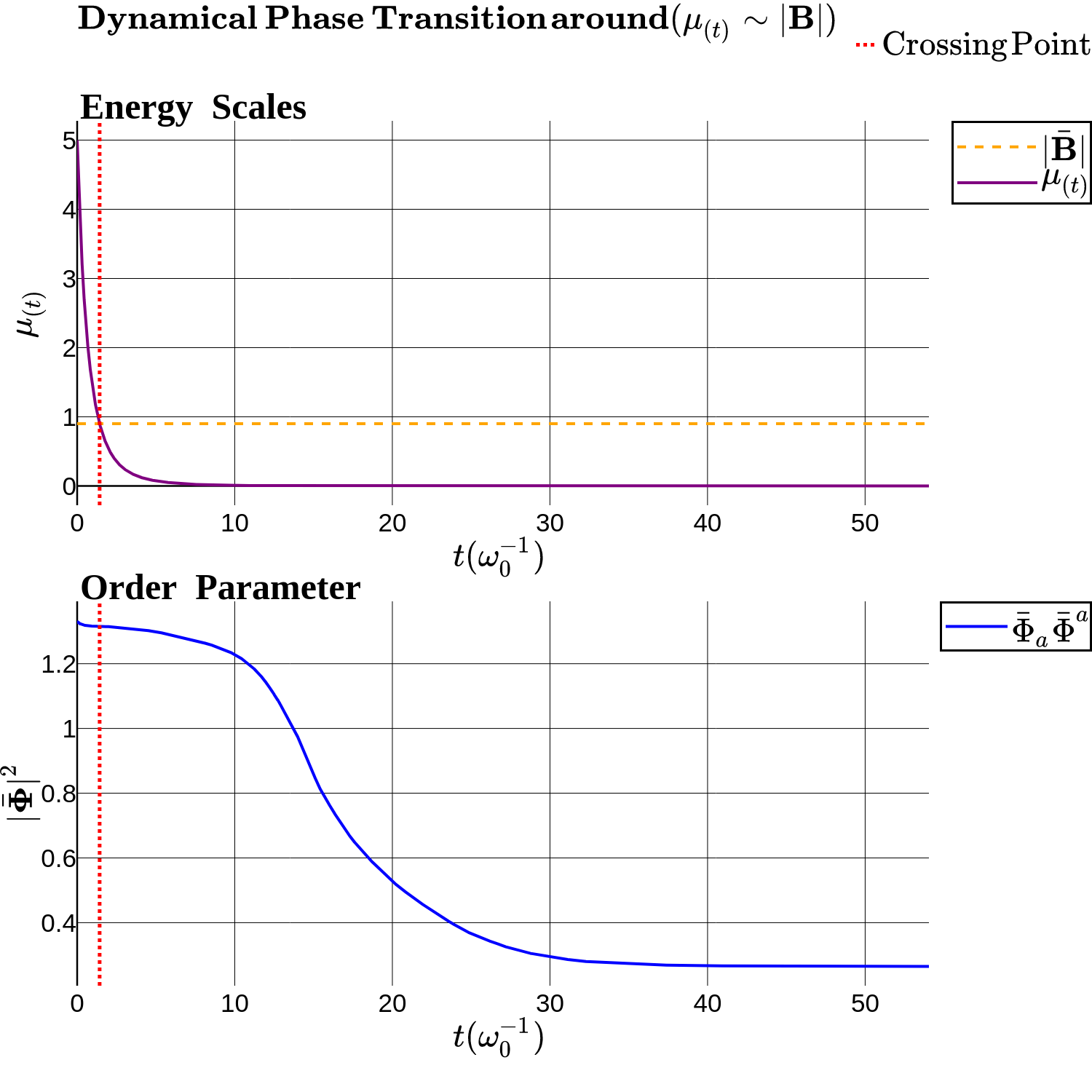}
    \caption{Three-flavor evolution found by solving Eqs.~\eqref{dPhi} and \eqref{LongAhhdG} together  with the parameters $N=50, n=3, {\rm and} \> \mu_{(0)} = 5$. This figure depicts a dynamical phase transition around the crossover point (the dotted line) $\mu(t) \sim |\bar{\bm{B}}|$. The horizantal dashed line is indicates the value of $|\bar{\bm{B}}|$. The order parameter is given by $\bar{\Phi}_{a} =\frac{1}{N}\sum_{A}\Phi_{Aa}$. }
    \label{fig:su3Phase} 
\end{figure}
\begin{figure}[tp]
    \centering
    \includegraphics[width=\linewidth]{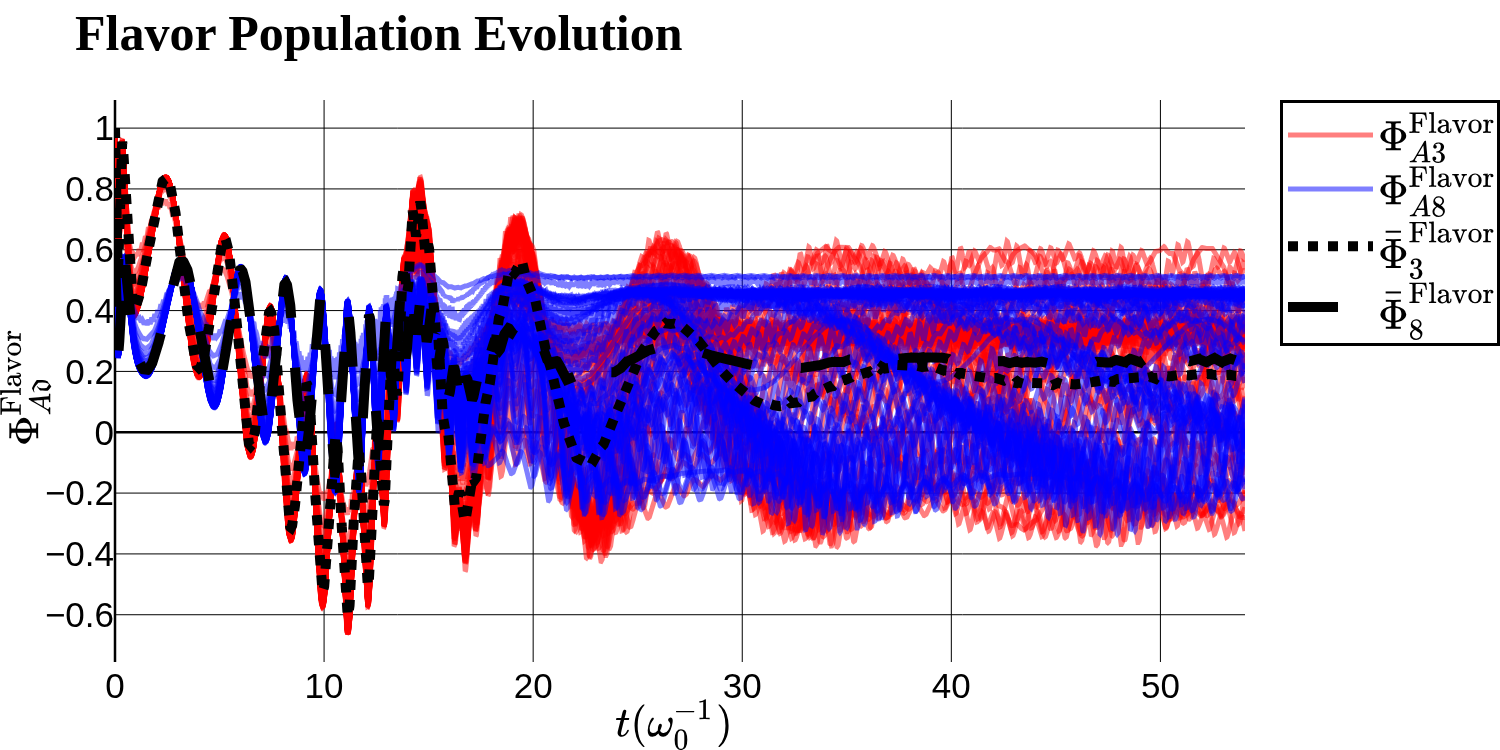}
    \includegraphics[width=\linewidth]{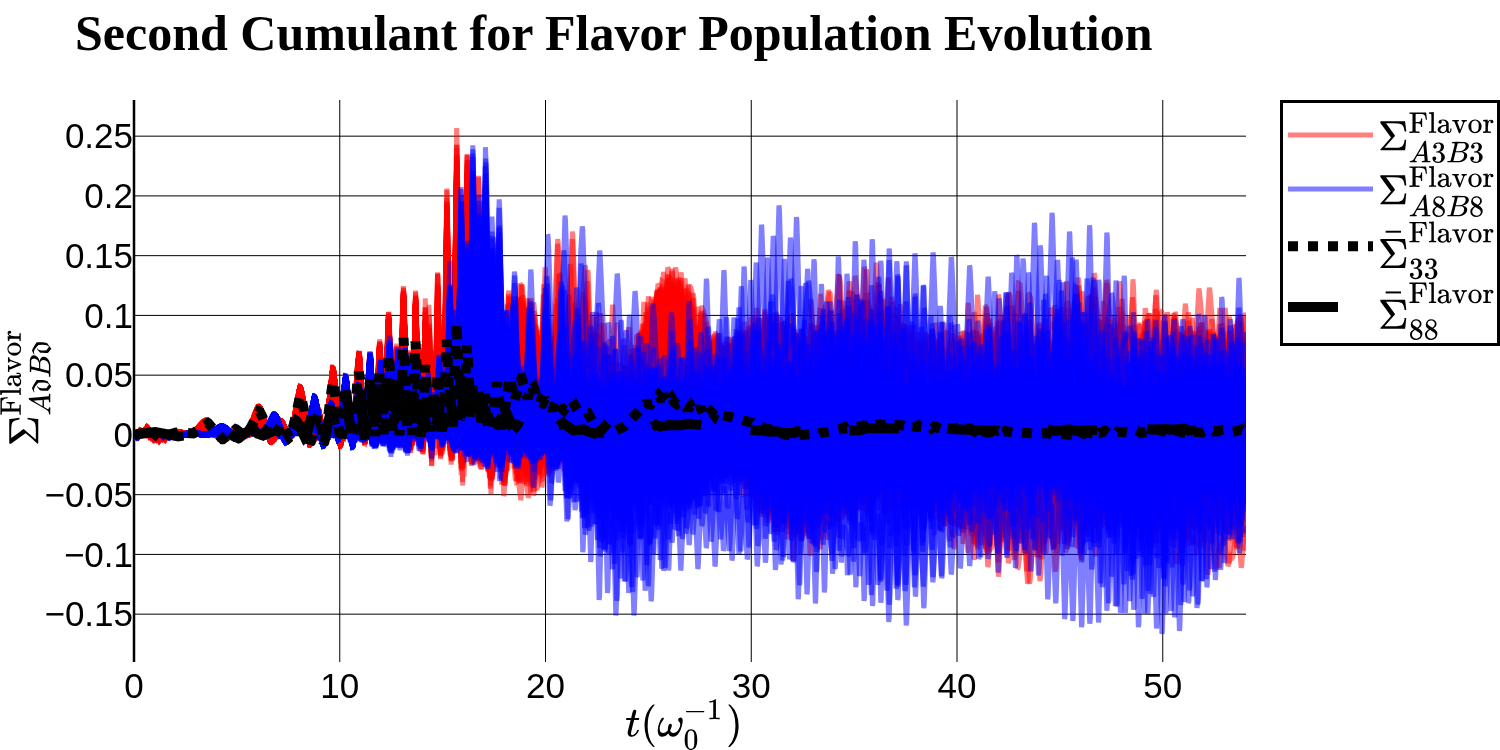}
    \caption{Three-flavor evolution of the one-body relative populations $\Phi_{A3} = \expval{n_{A\nu_e}} - \expval{n_{A,\nu_{\mu}}}$ and $\Phi_{A8} = \frac{1}{2\sqrt{3}}\inParen{\expval{n_{A,\nu_{e}}} + \expval{n_{A\nu_{\mu}}}- 2\expval{n_{{A}\nu{\tau}}}}$ as well as the two-body flavor cumulant $\Sigma_{AaBb} = \Gamma_{AaBb} -\Phi_{Aa}\Phi_{Bb}$. In the calculations leading to this figure we take $N=50, \> n=3, {\rm and} \> \mu_{(0)} = 5$.
    The transition from coherent oscillations ($\mu(t) > |\bar{\bm{B}}|$) to the decoherent vacuum oscillations is seen as the interaction regime shifts to $|\bar{\bm{B}}|>>\mu_{(t)}$. }
    \label{fig:Pzsu3}
\end{figure}
        
Since we are seeing the characteristics of a phase transition, we could examine quantum properties in the evolution of the density matrix. To examine how information is distributed in the system,  we show the two-body entropy and mutual information estimator (Eq.~\eqref{2ENT} and Eq.~\eqref{MutInf} respectively) in Fig~\ref{fig:twoBodyTimesu3}. The two-body entropy grows around the phase transition and later becomes flat much like that of the two-flavor system. Two-body entropies differ quite significantly (by 1-2 orders of magnitude) between two and three-flavor cases, but their overall behavior is characteristically the same. This is likely attributable to the different values of $N$ and $n$. The $\mf{su}(3)$-system has a Hilbert space with a much richer structure, and thus is more readily entangled.

Fig.~\ref{fig:twoBodyTimesu3} shows that 
a sharp peak begins to form in the naive mutual information estimator around the phase transition, indicating that information is being distributed into some of the many body operators.
\begin{figure}[btp]
    \centering
    \includegraphics[width=\linewidth]{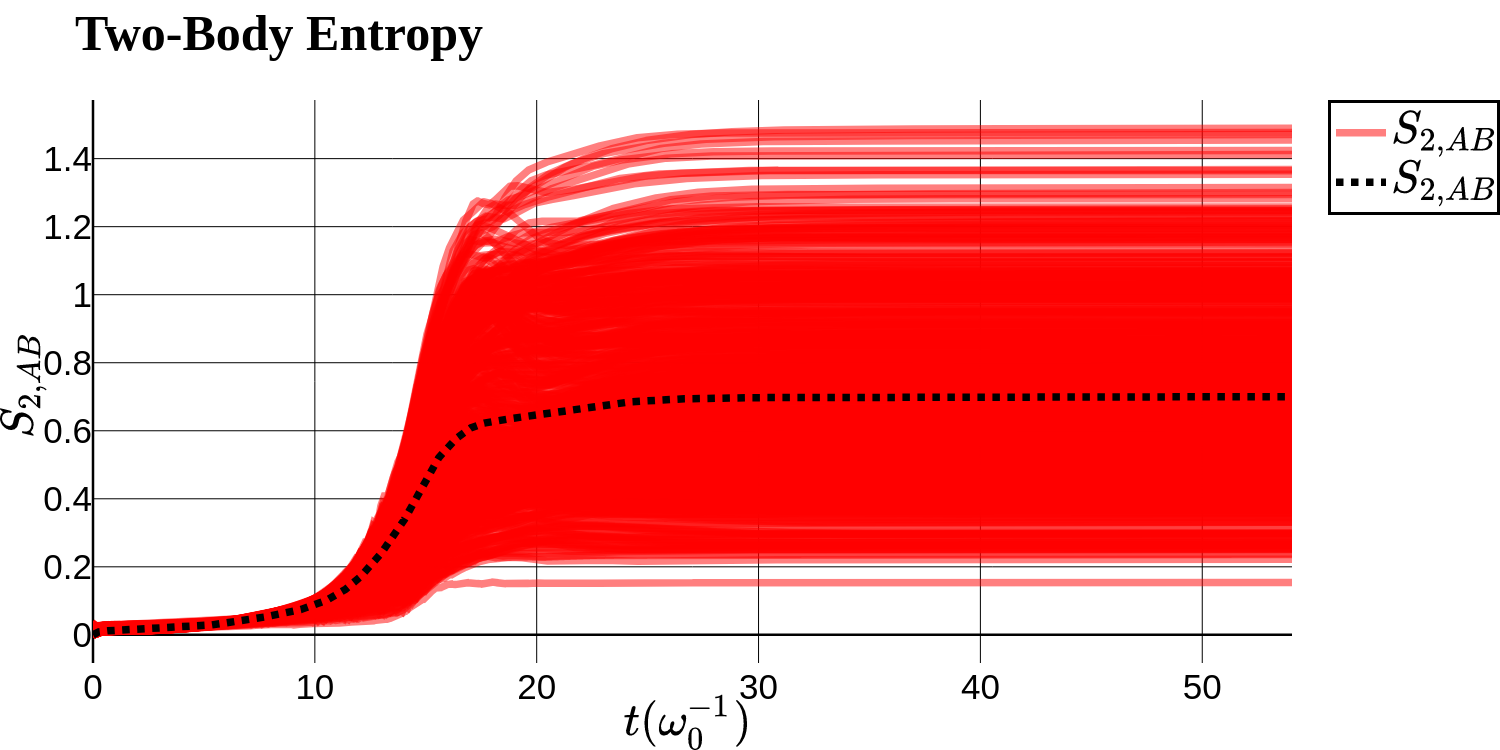}
    \includegraphics[width=\linewidth]{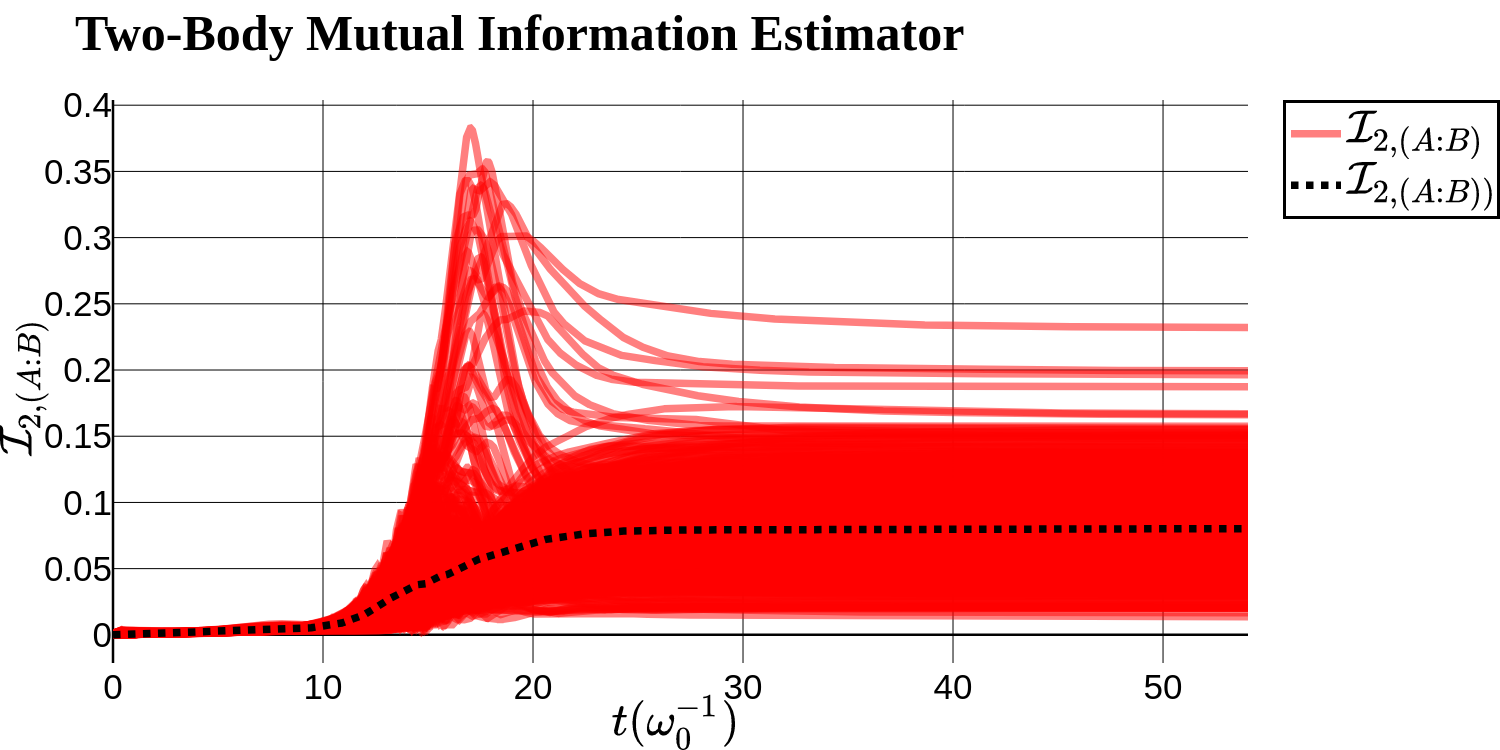}
    \caption{Two-body entropy and mutual information estimator evolution for three flavors found by solving Eqs.~\eqref{dPhi} and \eqref{LongAhhdG} simultaneously with $N=50, n=3,$ and $\mu_{(0)} = 5$. The two-body entropy is then calculated using Eq.~\eqref{2ENT} and the two-body mutual information estimator is calculated using Eq.~\eqref{MutInf} with $\alpha = 2$. Both the two-body entropy and mutual information estimator indicate that information is being distributed in many-body operators.}
    \label{fig:twoBodyTimesu3}
\end{figure}


 The three-flavor two-body entropy and mutual information estimator heat maps shown in Fig.~\ref{fig:twoBodysu3Heat} indicate a similar behavior to that of the two flavor case: A sharp increase around the phase transition and then entering a steady state as $\mu_{(t)}\to 0$. However, we see a sharp peak of the mutual information estimator not seen in the two-flavor case. Additionally, the magnitude of the estimator is much higher in the three-flavor case, meaning that information is more readily being shifted into the many body operators. This is due to the richer structure inherent in the $\mf{su}(3)$ Hilbert Space making it easier for particles to become entangled and share information in the momentum space. 

\begin{figure}[h]
    \centering
    \includegraphics[width=.95\linewidth]{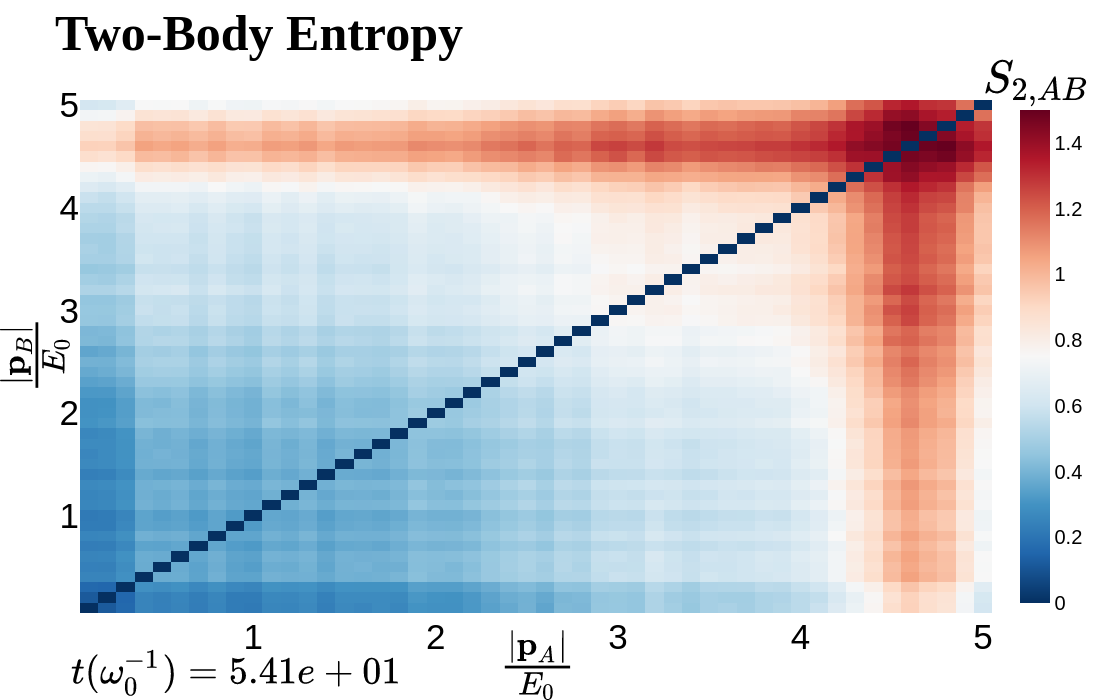}
    \includegraphics[width=.95\linewidth]{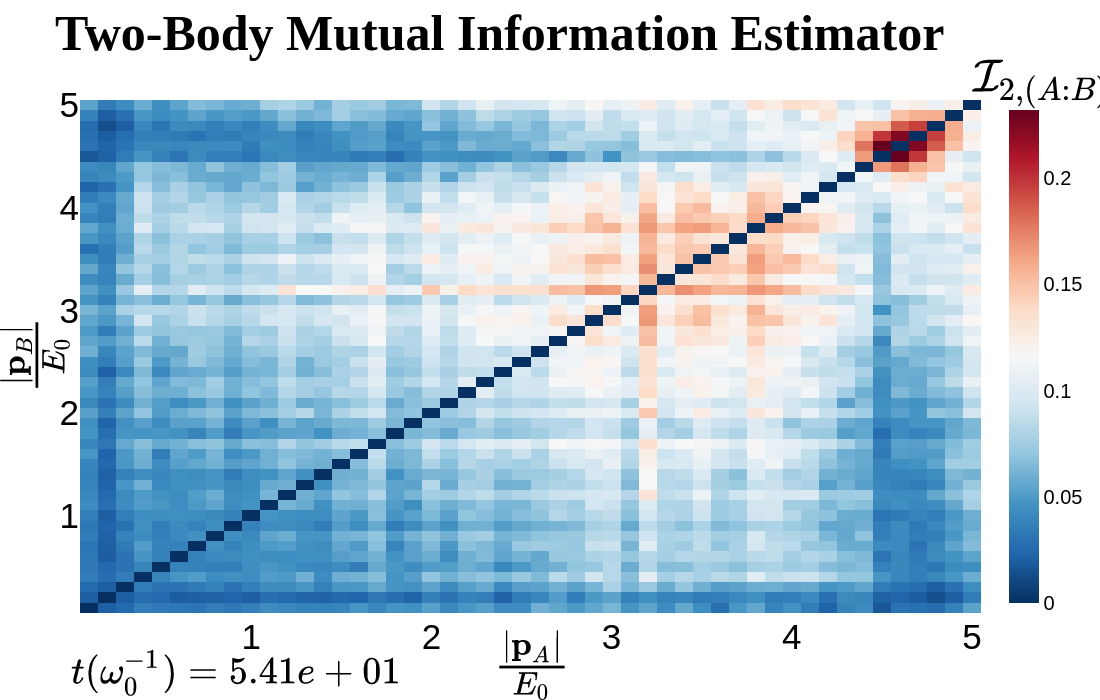}
    \caption{A constant time slice ($t\inParen{\omega_{0}^{-1}} \approx 54.1$) heat map of the distribution of two-body entropy (Eq.~\eqref{2ENT}) and mutual information (Eq.~\eqref{MutInf}) over momentum space for three flavors. In this figure we take $N=50, n=3,$ and $\mu_{(0)} = 5$. This figure indicates that the information is being distributed non-locally between points in momentum space.}
    \label{fig:twoBodysu3Heat}
\end{figure}

To explore further if information is being distributed non-locally, we again compute the three body entropy and mutual information estimator in Fig.~\ref{fig:threeBodyTimesu3}. Compared to $\mf{su}(2)$, the three-body entropy exhibits a higher magnitude but a similar behavior. The three-body mutual information estimator still suggests that information is being weakly scrambled. 
 \begin{figure}[h]
    \centering
    \includegraphics[width=1\linewidth]{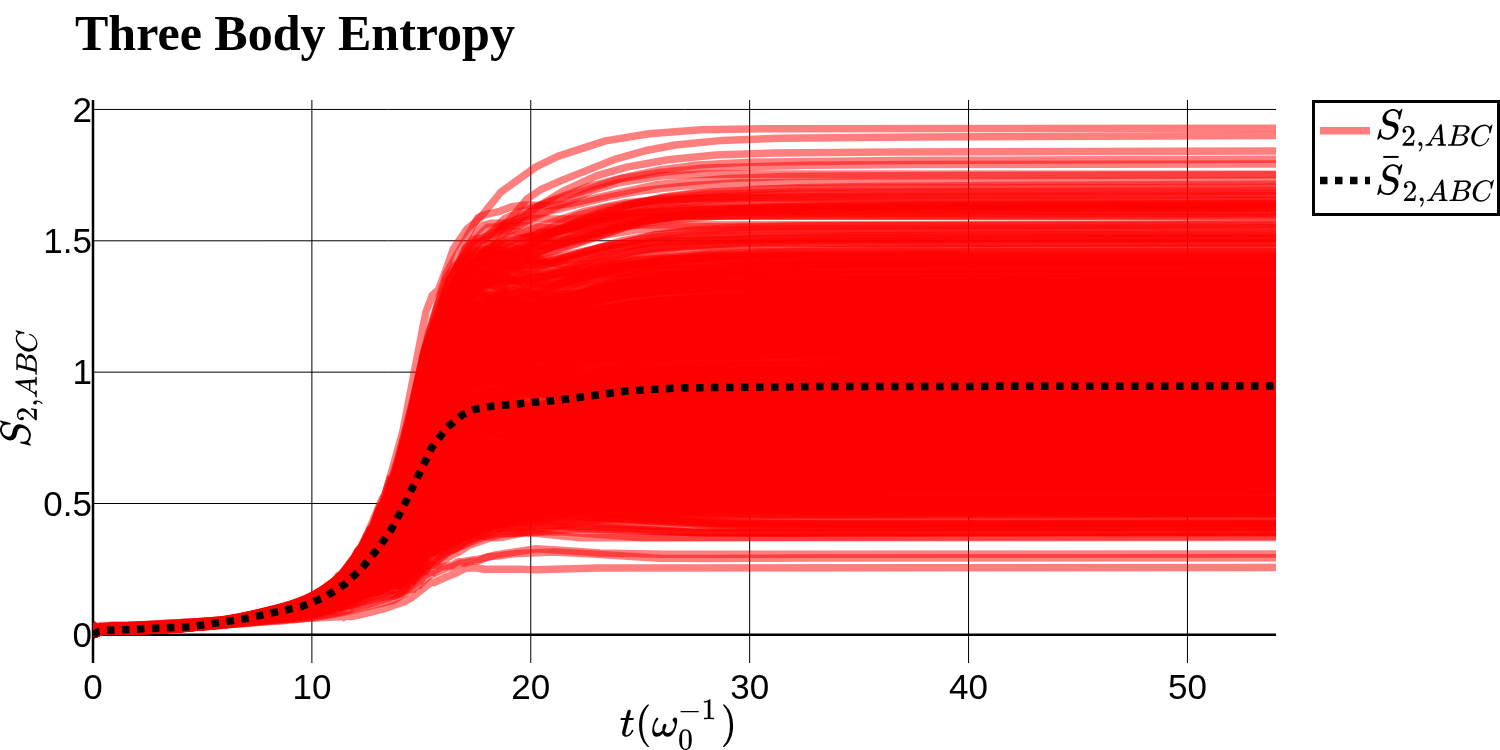}
    \includegraphics[width=1\linewidth]{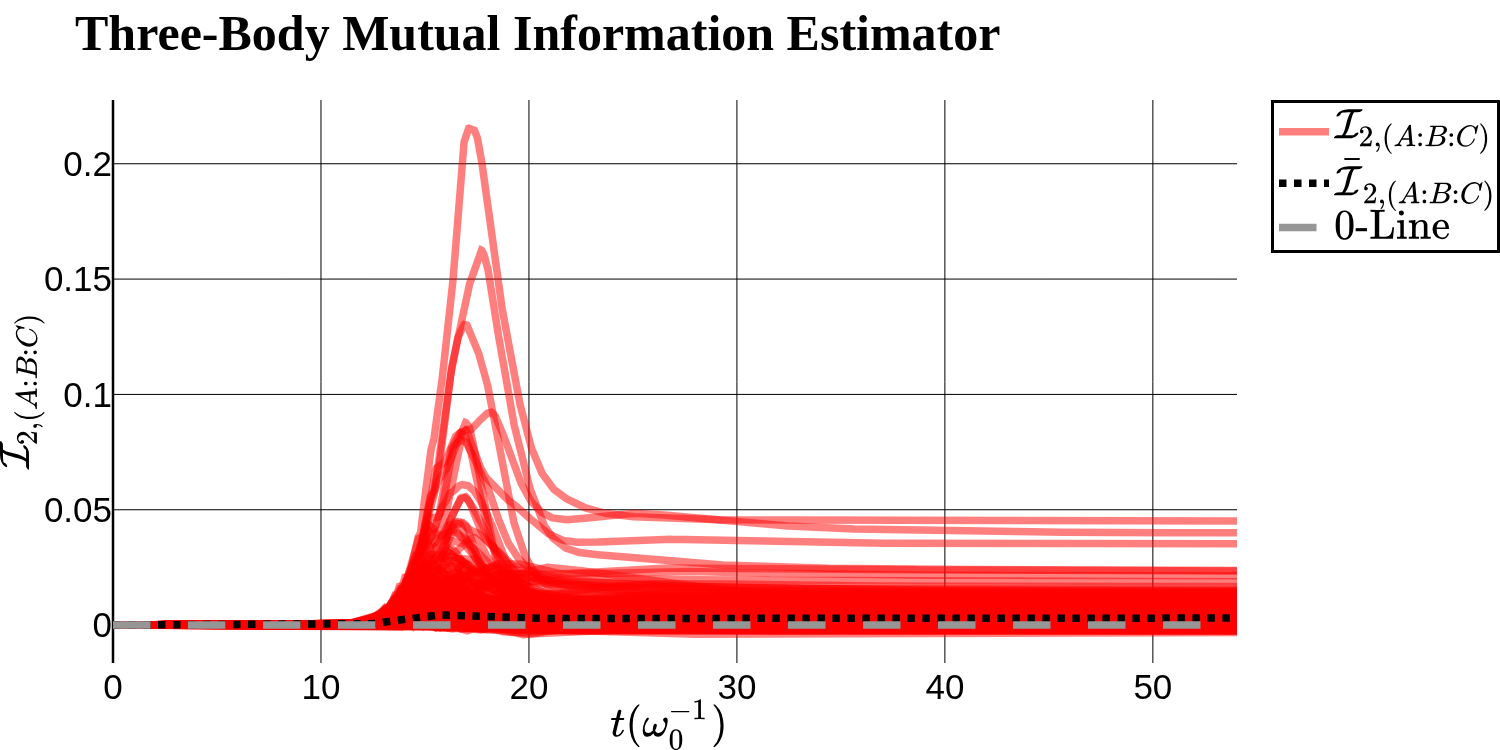}
    \caption{Evolution of the three-flavor three-body entropy and three-body mutual information estimator found by solving Eqs.~\eqref{dPhi} and \eqref{LongAhhdG} together with $N=50, n=3, {\rm and} \> \mu_{(0)} = 5$. Entropy is calculated using Eq.~\eqref{2ENT} and the mutual information estimator using eq.~\eqref{MutInf3} with $\alpha = 2$. The growth of three-body mutual information and entropy with time indicates the growth of the information content of three-body operators. The slight negativity of the estimator suggests that we are seeing information scrambling~\cite{Seshadri:2018yya}}
    \label{fig:threeBodyTimesu3}
\end{figure}

 Since the information is being distributed among many-body operators even more strongly in the $\mathfrak{su}(3)$ case, one should also examine how the non-stabilizer resources vary with time. In Fig.~\ref{fig:su3MagicMana}, we present the temporal evolution of both the one-body and two-body magic and mana.
\begin{figure}[t]
    \centering
    \includegraphics[width=1\linewidth]{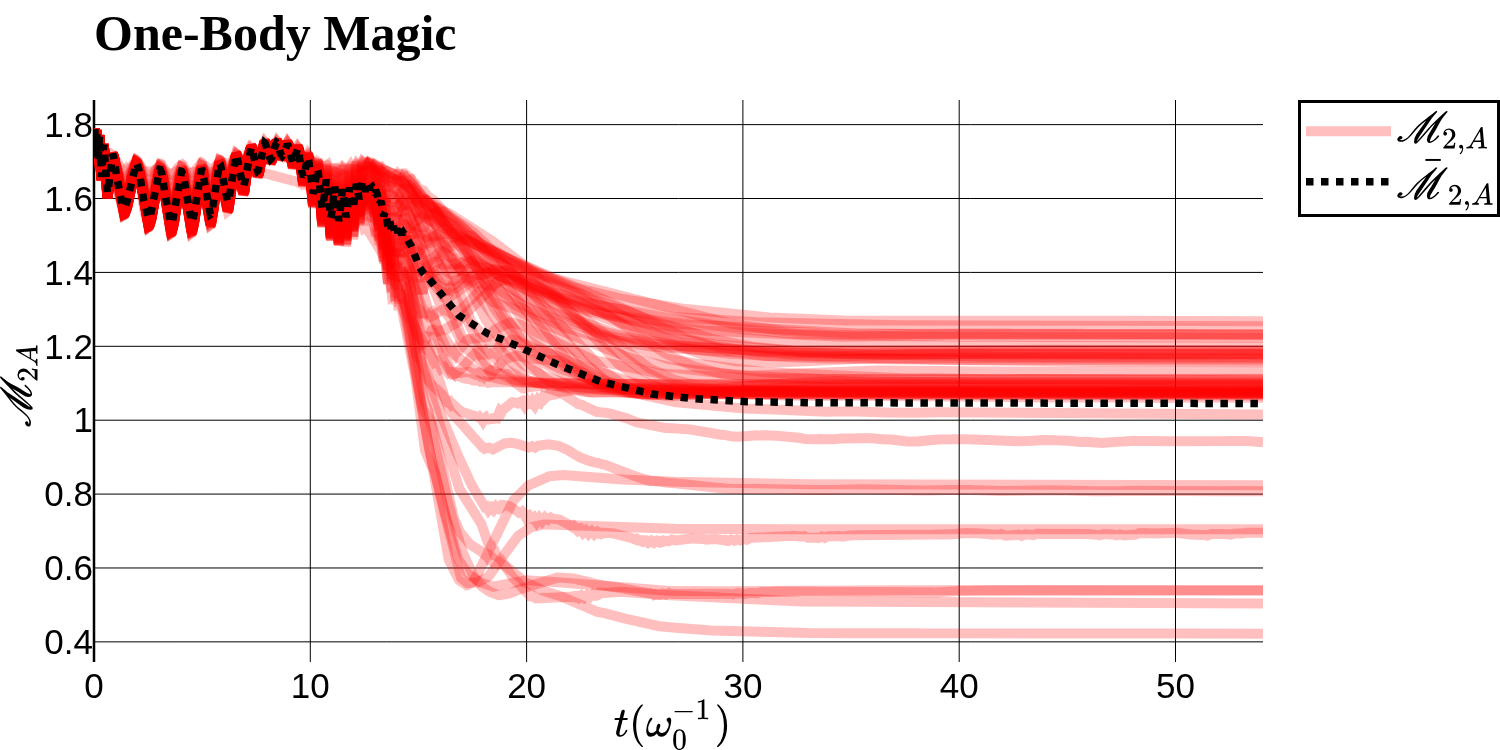}
    \includegraphics[width=1\linewidth]{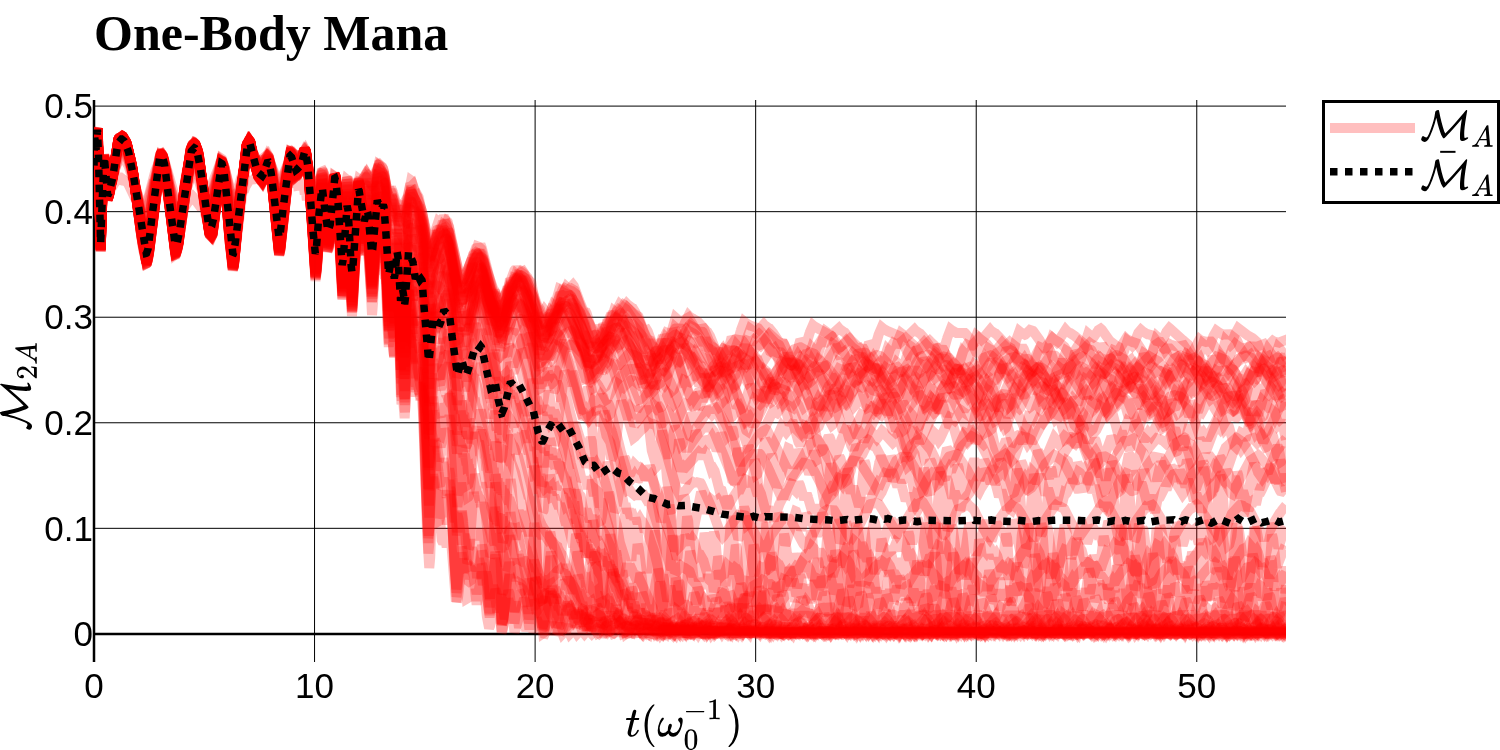}
    \includegraphics[width=1\linewidth]{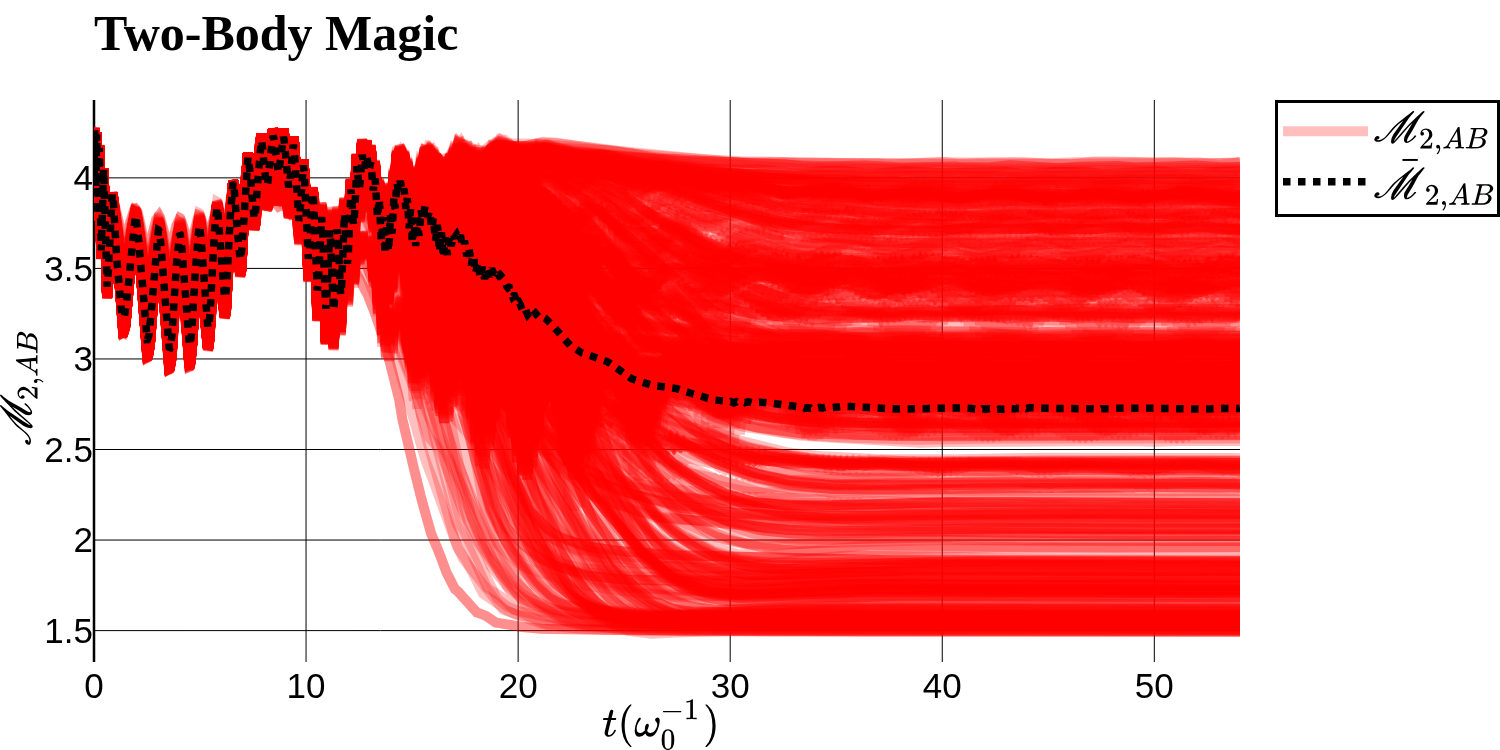}
    \includegraphics[width=1\linewidth]{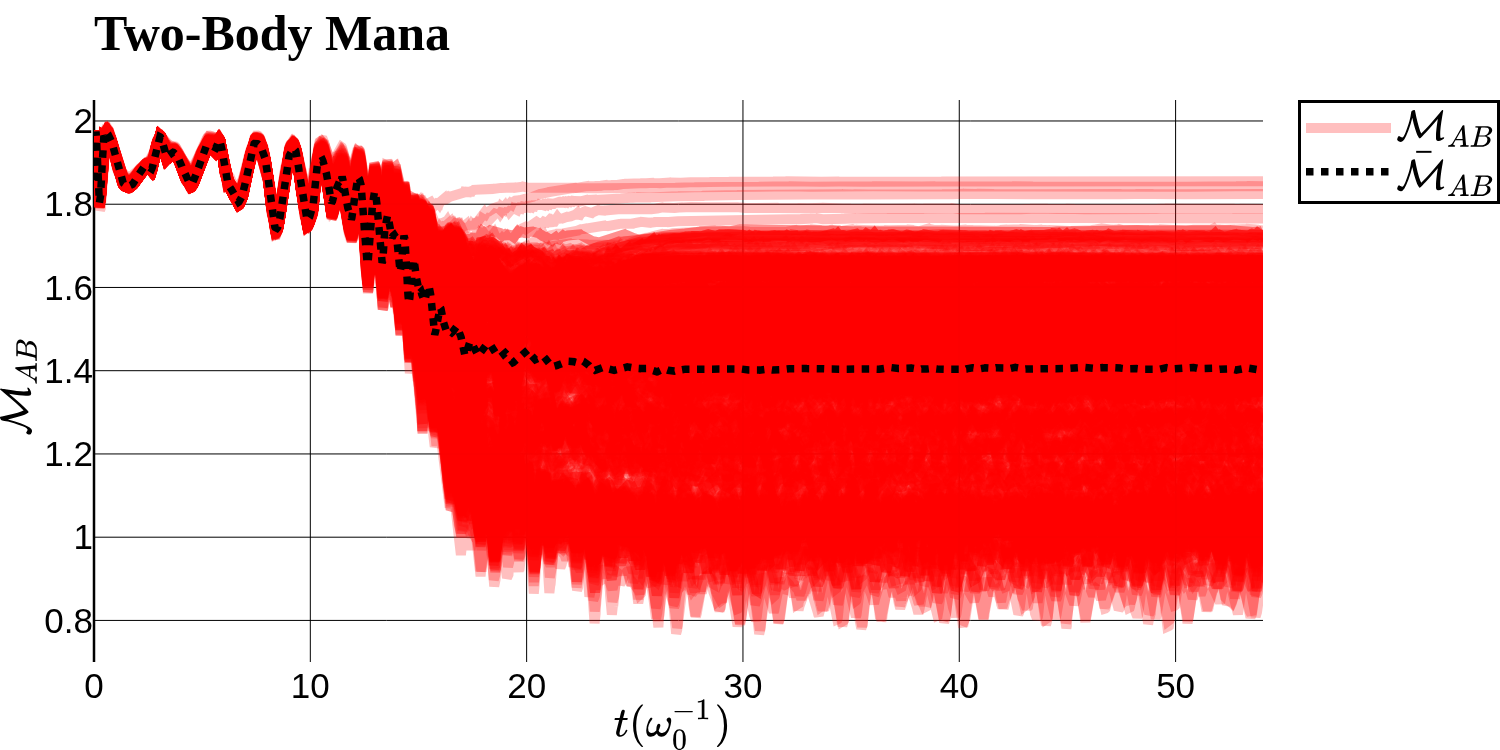}
    \caption{Evolution of the one-body entropy as well as two-body magic and mana for three flavors found by solving Eqs.~\eqref{dPhi} and \eqref{LongAhhdG} together with $N=50, n=3,$ and $\mu_{(0)} = 5$. We calculate magic and mana using Eq.~\eqref{Magic} and Eq.~\eqref{Mana}, respectively. Both magic and mana indicate that the density matrix contains non-stabilizer states. }
    \label{fig:su3MagicMana}
\end{figure}
We see that the non-stabilizerness of the density matrices behaves somewhat similarly to that of the $\mathfrak{su}(2)$ case in the early-time regime ($\mu(t) > |\bar{\mathbf{B}}|$), where these quantities exhibit highly coherent, synchronized oscillations. This behavior reflects the dominant role of the dense neutrino background field, which enforces global collective behavior and suppresses individual state departures from the initial configurations. 

Around the dynamical phase transition the one-body magic and mana, which initially behave coherently, begin to decohere much like for the two-flavor approximation. However, unlike the simpler two-flavor approximation, both the local (one and two-body) magic and mana rapidly decrease for three-flavor case at the phase transition. One observes a sharp peak in the mutual information estimator, and the growth in the one-, two-, and three-body entropies. The rapid growth in the local entropies provides direct evidence that the non-stabilizerness is leaking out of single particles and shifting into the many-body operators. This behavior is a clear signature that the quantum magic is "scrambling" across the entire system \cite{Garcia:2022abt}.

Furthermore, the persistent, non-zero value of the two-body magic and mana in the late-time asymptotic regime is highly significant. If the system completely thermalized into a trivial mixed state, these local magic metrics would drop all the way to zero. Instead, their survival suggests that the generated quantum correlations are not classically simulatable in the Gottesman-Knill sense. Rather than wiping out the system's quantum complexity, the dynamical phase transition permanently seeds the many-body state with genuine non-stabilizer components. This confirms that the collective, scattering-induced scrambling mechanism effectively drives the neutrino gas into a highly non-trivial quantum state that remains fundamentally complicated even at late times.


\section{Conclusions and Summary} 
BBGKY truncation provides a reliable way of improving the mean field approximation. Particularly in problems that require all-to-all connectivity truncation offers an advantage over traditional methods as it grows like $
\mathcal{O}\!\left(\binom{N}{m}(n^2 - 1)^{m}
\right)
$ for an $m^{\text{th}}$ order truncation. This polynomial scaling allows one to approximate the evolution of the neutrino system, including the angular effects, with much larger $N$ then previously believed possible. For the two flavor approximation we solved the system with $N=100$ neutrinos and for the three flavor system with $N=50$ neutrinos. In both cases, we observed that the system is undergoing a dynamical phase transition as the neutrinos become more diffuse. Additionally, we show that the system shows properties of weak information scrambling in the limit we investigated.

We also use these simulations to show the unique properties that the full three flavor system exhibits. For instance, we show that the full three-flavor system's additional complexity allows for more non-trivial correlations (thus is more easily entangled) than that of the two flavor system. 

Finally, we investigated the non-stabilizer resources necessary to simulate the system. We note that the magic for the one- and two-body operators seemingly decays corresponding with a growth in entropy. This is a an indicator that magic is being shifted into higher n-body operators, exhibiting information and magic scrambling.

Our results indicate that many-body treatments of the neutrino problem, even in the approximate regime we examined, contain irreducibly complex quantum mechanical correlations that mean-field, and the standard fast-flavor oscillation treatments do not. 

\begin{acknowledgments}
This work was supported by the U.S. National Science Foundation grant No. PHY-2411495.    
\end{acknowledgments}

\FloatBarrier
    \bibliographystyle{apsrev4-2} 
   
\bibliography{references} 

\end{document}